\documentclass[a4paper,12pt]{article}
\pdfoutput=1
\usepackage[english]{babel}
\usepackage{feynmp-auto,expdlist}
\usepackage{amsmath, amsfonts, amssymb}
\usepackage{graphicx}
\usepackage{enumerate}
\usepackage{hyperref}
\usepackage{latexsym}
\usepackage{dsfont}
\usepackage{hepnicenames}
\usepackage{enumerate}
\usepackage{soul}
\usepackage[normalem]{ulem}
\usepackage{bbold}
\usepackage{appendix}
\usepackage{bm}
\usepackage{epsfig,cite}

\font\tenrsfs=rsfs10 at 12pt
\font\sevenrsfs=rsfs7
\font\fiversfs=rsfs5
\newfam\rsfsfam
\textfont\rsfsfam=\tenrsfs
\scriptfont\rsfsfam=\sevenrsfs
\scriptscriptfont\rsfsfam=\fiversfs

\numberwithin{equation}{section}

\usepackage{mathrsfs}
\usepackage{braket}
\usepackage{titling}
\usepackage{amsmath}
\usepackage{slashed}
\usepackage{amssymb}
\usepackage{epsfig}
\usepackage{graphicx}
\usepackage{multirow}
\usepackage{color}
\usepackage[normal]{subfigure}
\usepackage{rotating}
\usepackage{hyperref}
\usepackage[margin=1.2in]{geometry}
\usepackage[table,xcdraw,dvipsnames]{xcolor}
\usepackage{enumitem}
\usepackage[utf8x]{inputenc}
\usepackage[compress,numbers,sort]{natbib}
\usepackage{authblk}
\usepackage{colortbl}
\usepackage{pdflscape}
\usepackage{color}
\usepackage{mathtools}

\definecolor{nicered}{rgb}{0.7,0.1,0.1}
\definecolor{nicegreen}{rgb}{0.1,0.5,0.1}
\definecolor{red}{rgb}{1.0, 0, 0}
\definecolor{niceblue}{rgb}{0,0,0.8}
\definecolor{red}{rgb}{1.0, 0, 0}
\hypersetup{colorlinks,citecolor= nicegreen,linkcolor= nicered,urlcolor=nicered}

\def\eq#1{{Eq.~(\ref{#1})}}
\def\eqs#1#2{{Eqs.~(\ref{#1})--(\ref{#2})}}
\def\fig#1{{Fig.~\ref{#1}}}

\def\Table#1{{Table~\ref{#1}}}

\def\sect#1{{Sect.~\ref{#1}}}

\def\app#1{{App.~\ref{#1}}}

\def\e{e}

\def\abs#1{\left| #1\right|}

\def\Tr{\mbox{Tr}\,}
\def\tr{\mbox{tr}\,}
\def\det{\mbox{det}\,}
\def\Det{\mbox{Det}\,}
\def\SDet{\mbox{SDet}\,}

\def\v#1{{\textbf#1}}

\renewcommand{\bar}{\overline}

\newcommand{\U}{\,{\rm U}}
\renewcommand{\O}{\,{\rm O}}

\newcommand{\beq}{\begin{equation}}
\newcommand{\eeq}{\end{equation}}
\newcommand{\bea}{\begin{eqnarray}}
\newcommand{\eea}{\end{eqnarray}}

\renewcommand{\[}{\left[}
\renewcommand{\]}{\right]}
\renewcommand{\(}{\left(}
\renewcommand{\)}{\right)}
\definecolor{blus}{cmyk}{1,1,0,0.6}

\begin{document}

\begin{center}  
{\Large
\bf
\color{blus} 
False vacuum decay: an introductory review
} \\
\vspace{1.0cm}

{\bf Federica Devoto$^{a}$, Simone Devoto$^{b}$, Luca Di Luzio$^{c}$, Giovanni Ridolfi$^{d}$}\\[5mm]

{\it $^a$Rudolf Peierls Centre for Theoretical Physics, University of Oxford, \\ Clarendon Laboratory, Parks Road, Oxford OX1 3PU, UK} \\[1mm] 
{\it $^{b}$Dipartimento di Fisica Aldo Pontremoli, Universit\`a di Milano \\ and INFN, Sezione di Milano,
Via Celoria 16, I-20133 Milano, Italy
} \\[1mm] 
{\it $^{c}$Dipartimento di Fisica e Astronomia Galileo Galilei, Universit\`a di Padova \\ and INFN, Sezione di Padova,
Via Marzolo 8, I-35131 Padova, Italy
} \\[1mm] 
{\it $^{d}$Dipartimento di Fisica, Universit\`a di  Genova and INFN, Sezione di Genova, \\ 
Via Dodecaneso 33, I-16146 Genova, Italy
} \\[1mm] 

\vspace{0.5cm}
\begin{quote}

We review the description of tunnelling phenomena in the semi-classical approximation in ordinary quantum mechanics and in quantum field theory. In particular, we describe in detail the calculation, up to the first quantum corrections, of the decay probability per unit time of a metastable ground state. We apply the relevant formalism to the case of the standard model of electroweak interactions, whose ground state is metastable for sufficiently large values of the top quark mass. Finally, we discuss the impact of gravitational interactions on the calculation of the tunnelling rate.

\end{quote}
\thispagestyle{empty}
\end{center}

\bigskip

\clearpage
\tableofcontents


\section{Introduction}

Tunnelling in quantum field theory is a fascinating subject. 
In many respects, it differs substantially from the analogous phenomenon in ordinary quantum mechanics with a finite number of degrees of freedom, and deserves a special treatment. In a couple of famous papers \cite{Coleman:1977py,Callan:1977pt}, 
S.~Coleman and C.~Callan have shown that the semi-classical approximation, in conjunction with path integral techniques, provides 
a suited context to deal with quantum tunnelling with infinite degrees of freedom. The first purpose of this work is to review both the semi-classical approximation in quantum mechanics and its extension to quantum field theory, as illustrated in 
Refs.~\cite{Coleman:1977py,Callan:1977pt}. We will adopt a pedagogical attitude: special attention will be devoted to 
the derivation of the bounce formalism, the definition and calculation of functional determinants, and the role of renormalization. All steps in the derivation of these known results, some of which are omitted in the original literature, are carefully illustrated.

A remarkable feature of tunnelling in quantum field theory is that the transition 
from the false to the true vacuum does not take 
place between two
spatially homogeneous field configurations, 
but rather through the formation of a bubble of true vacuum in a false-vacuum background.
Such field configuration (the bounce) is therefore not spatially homogenous, 
and the gradient term in the potential energy gives a non-zero contribution to the full potential energy. 
This is different from the case of ordinary quantum mechanics, in which the gradient term is absent 
and the tunnelling rate does not depend on the physics beyond the potential barrier. 

An important application of the formalism above is the study of the stability of the electroweak vacuum 
in the case of the standard model of electroweak interactions. 
The accurate measurements of the top quark and Higgs boson masses indicate that the electroweak vacuum, which determines the observed spectrum of physical particles, is not the absolute ground state of the electroweak theory. 
Extrapolating the Higgs effective potential at high energies, 
the electroweak minimum becomes metastable for Higgs field values around $10^{10}$ GeV.  
Many efforts have been devoted to establish whether the lifetime metastable vacuum of the standard model is sufficiently larger that
the age of the Universe, and therefore compatible with observations, or else
if physics beyond the standard model is needed in order to justify the observed electroweak spectrum.

The application of the formalism outlined in  Refs.~\cite{Coleman:1977py,Callan:1977pt} to the standard model is not straightforward. 
At the level of the classical Lagrangian, the electroweak vacuum corresponds to a true minimum of the scalar potential, 
and it is only upon the inclusion of quantum corrections that the instability occurs. 
Furthermore, the tunnelling rate calculation in the standard model 
including the first quantum corrections (originally addressed in Ref.~\cite{Isidori:2001bm})
is complicated by the presence of many degrees of freedom with different spin,
and by the approximate scale invariance of the theory at high-energy scales, 
relevant for the tunnelling process in the standard model.   

In the second part of this work, we will review several aspects of the 
calculation of the electroweak vacuum lifetime. 
In particular, we will carefully discuss the role 
of approximate scale invariance for the determination of the standard model bounce, 
which selects energy scales of order $10^{17}$ GeV. 
Since the latter is very close to the Planck scale, it is a legitimate question 
to ask whether gravitational effects can become relevant in this regime.
After addressing the ultraviolet sensitivity of the standard model vacuum decay rate \cite{Branchina:2013jra}, 
we will take gravitational effects into account within the formalism of 
Coleman and De Luccia \cite{Coleman:1980aw}, 
arguing that as long as gravity can be treated as an effective field theory in 
a perturbative regime \cite{Isidori:2007vm}, 
the standard model calculation does not get drastically affected. 

The review is organized as follows: in \sect{sec:tunnelling_QM} we introduce the phenomenon of quantum tunnelling in quantum mechanics with some explicit examples.
We then generalize this formalism to the case of quantum mechanics with many degrees of freedom, thus providing the tools required for dealing with tunnelling in quantum field theory, which is the main subject of \sect{tunnellingQFT}.
\sect{effatceffpot} is devoted to a review of effective potential methods, a necessary step in order to introduce the problem of the instability of the Higgs potential in the standard model. The calculation of the standard model vacuum decay rate is illustrated in \sect{sec:SMstability},
together with the construction on the so-called standard model phase diagram
and the possible impact of non-standard physics.
\sect{sec:gravity} is devoted to the study of gravitational corrections
to the standard model vacuum decay rate.

Some technical issues are collected in the Appendices, mainly with the purpose of casting well-known results in the notations adopted in the text, for the ease of the reader. \app{app:WKB} is a simple review of the semi-classical approximation in ordinary quantum mechanics, which is applied in the following \app{app:doublewell} to the calculation of the lowest-level energy splitting for a symmetric double-well potential. This is a simple exercise in quantum mechanics, which is included here for the purpose of comparison with the analogous calculation by path-integral methods presented in the text. Small differences between the two results are discussed. The Maupertuis principle is reviewed in \app{app:maup}, while \app{app:LegTra} describes briefly Legendre transformations.
\app{app:FLinstanton} contains a derivation of the Fubini-Lipatov bounce. In \app {app:nummethods} we describe the  techniques employed to obtain bounce solutions by numerical methods. Finally, the derivation of the gravitational field equations for a scalar field minimally coupled to gravity is described in \app{section:graveom}.

\section{Tunnelling in quantum mechanics}
\label{sec:tunnelling_QM}

In this section we review the phenomenon of tunnelling through barriers due to quantum fluctuations. We will start from ordinary quantum mechanics  in one space dimension, extend the results to the case of many degrees of freedom, and finally generalize to quantum field theory. We will follow closely the work by Coleman and collaborators \cite{Coleman:1977py,Callan:1977pt}. 
See also Ref.~\cite{Weinberg:2012pjx} for a comprehensive review.

\subsection{Tunnelling in one dimension}
 
Tunnelling is a typical phenomenon of quantum mechanics. The simplest
example is the case of a beam of particles (or a wave packet) moving
in one space dimension towards a potential energy barrier $U(x)$ whose
maximum value is larger than the particle energy: in quantum mechanics the particles of the beam
have a non-zero probability of being transmitted beyond the potential
barrier, contrary to what happens in classical mechanics.  The relevant quantity to be computed is the transmission coefficient, defined as the ratio between the transmitted flux and the incident flux.

If the potential barrier is sufficiently high and broad, the
transmission coefficient is conveniently computed within the
semi-classical or Wentzel-Kramers-Brillouin (WKB) approximation, which is reviewed in some detail
in~\app{app:WKB}. It turns out that, in the leading semi-classical
approximation, the transmission coefficient $\cal T$ through a potential
barrier is given by \cite{landau2013quantum}
\begin{equation}
{\cal T}=A_T\e^{-B};\qquad B=\frac{2}{\hbar}\int_a^b|p(x)|dx
=\frac{2}{\hbar}\int_a^b\sqrt{2m[U(x)-E]}dx,
\label{transm}
\end{equation}
where $E$ is the energy of the incident particles, and $a,b$ are the
classical turning points, $U(a)=U(b)=E$.  The overall factor $A_T$ is
not fixed in the leading semi-classical approximation.

The result \eq{transm} can also be used to estimate the decay probability
per unit time $\gamma$ of a metastable state, which is the case of interest
to us. Indeed, if the barrier separates a local minimum of the potential energy
from a deeper minimum, then the probability per unit time $\gamma$ that
the particle, initially localized around the local minimum with energy
$E_0$, the fundamental level of the local minimum,
penetrates the barrier, is proportional to $\cal T$. The exact proportionality
coefficient cannot be computed within the lowest-order semi-classical
approximation. However, it must be proportional to the number $N$ of
times the particle hits the barrier per unit time. Approximating
the initial state as an oscillatory state with frequency
$\omega=\frac{E_0}{ \hbar}$, we have simply
$N=\frac{\omega}{2\pi}=\frac{1}{t_{\rm cl}}$, where $t_{\rm cl}$ is
the classical period of the oscillation. Hence \cite{Konishi:2009qva}
\begin{equation}
\gamma\simeq \frac{\cal T}{t_{\rm cl}}\equiv Ae^{-B}.
\end{equation}
It is customary to define a decay width
\begin{equation}
\Gamma=\hbar\gamma,
\end{equation}
with the dimension of an energy, whose physical meaning will be clear in a moment.

Finally, we note that, if the particle has a probability per unit time
$\gamma$ to escape from the local minimum, then the probability $P(t)$
to find it in the vicinity of the local minimum decreases in time
according to
\begin{equation}
\frac{dP(t)}{dt}=-\gamma P(t),
\end{equation}
which has the solution
\begin{equation}
P(t)=P(0)\e^{-\gamma t}=P(0)\e^{-\frac{\Gamma t}{\hbar}}.
\label{prob}
\end{equation}
This suggests that we may interpret the unstable state as an approximate
stationary state with complex energy eigenvalue:
\begin{equation}
E=E_0-i\frac{\Gamma}{2},
\end{equation}
so that the corresponding wave function evolves in time according to
\begin{equation}
\Psi(x,t)=\psi(x)\e^{-i\frac{E_0 t}{\hbar}}\e^{-\frac{\Gamma t}{2\hbar}},
\end{equation}
consistently with \eq{prob}.

\subsection{The double-well potential}
\label{sec:doublewell}
\begin{figure}[ht]
\centering
\includegraphics[width=0.45\textwidth]{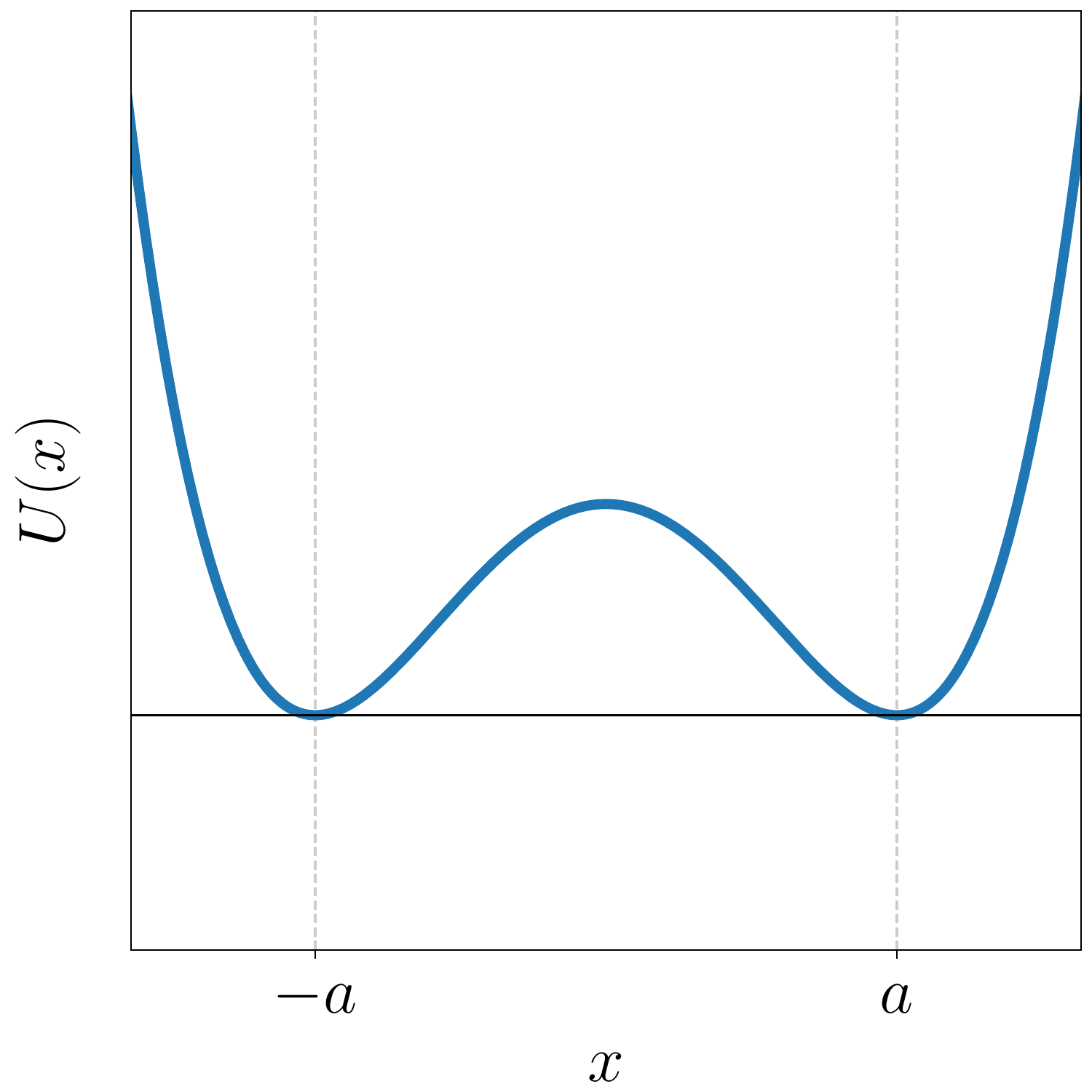} \qquad
\includegraphics[width=0.45\textwidth]{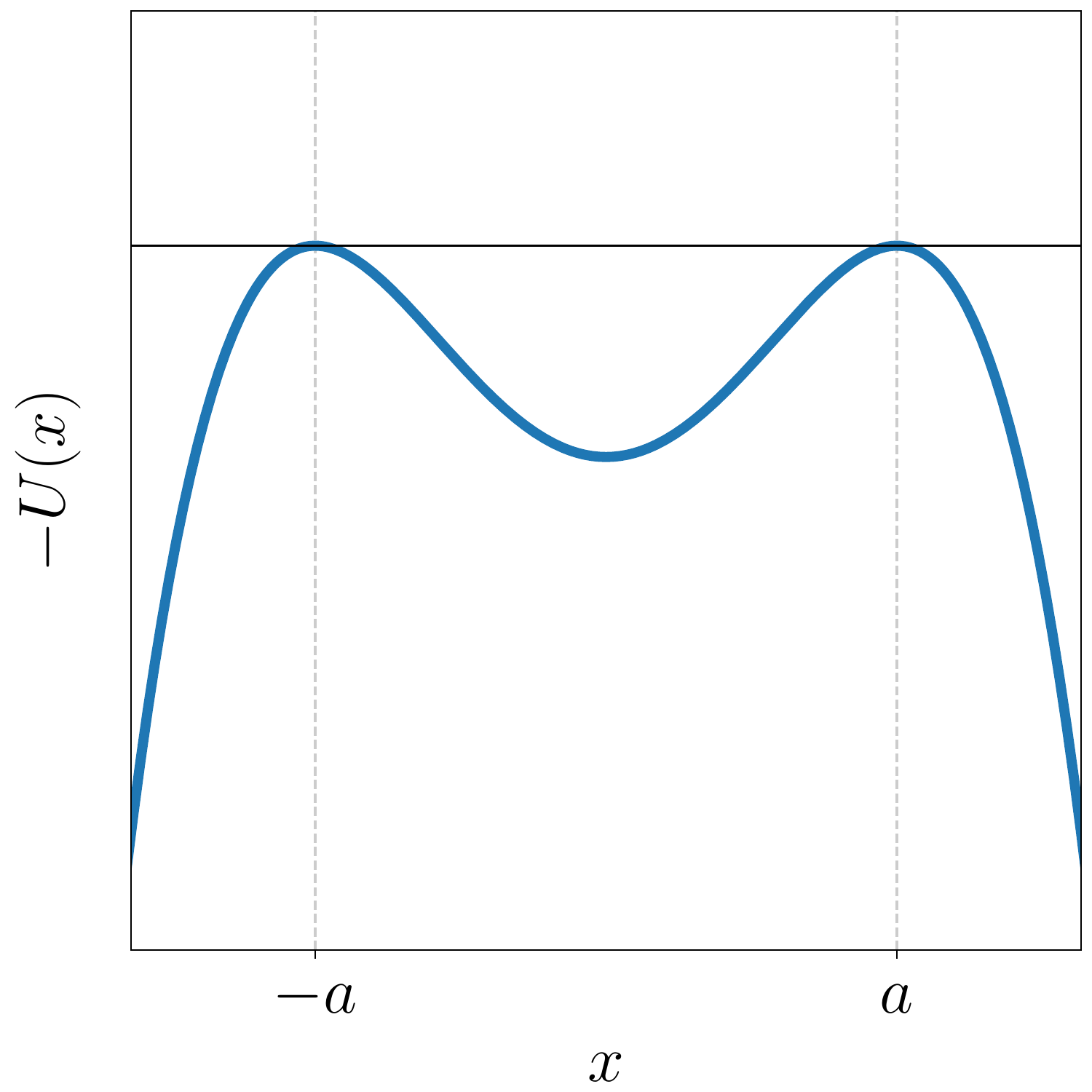}
\caption{
\label{qm_double_well}
Left panel: double-well potential with two symmetric minima in $\pm a$. Right panel: the same potential after the change of sign.}
\end{figure}

A typical application of the semi-classical approximation method is the
calculation of the splitting between the two lowest energy eigenvalues
of a one-dimensional hamiltonian whose potential energy has two degenerate minima, such as the one sketched in the left panel of \fig{qm_double_well}. The standard
calculation can be found for example in Ref.~\cite{landau2013quantum}, and it is
reviewed in~\app{app:doublewell}. 

For our present purposes, it will prove useful to obtain the
result of~\app{app:doublewell} in the context of the path integral formulation of quantum
mechanics. In the case of one-dimensional ordinary quantum
mechanics the path integral calculation turns out to be much more
lengthy and complicated, but it has the advantage that it can be
generalized to quantum field theory beyond the leading semi-classical approximation.

We will assume that
the potential energy $U(x)$ has two degenerate minima at
$x=\pm a$, and that $U''(\pm a)=m\omega^2$, where $m$ is the mass of the particle
and $\omega$ a real constant.
An explicit example of such a potential is 
\beq
U(x)=\frac{m\omega^2}{8a^2}(x^2-a^2)^2.
\label{doublewellpot}
\eeq
It will often be useful to refer to the specific example in \eq{doublewellpot},
for which analytic calculations are possible. Our final result, however, applies
to a more general class of one-dimensional
potential energies with a symmetric double well. In particular, we will keep
$U_0=U(\pm a)$ different from zero.

The first step is switching to imaginary time, $t=i\tau$. The Lagrange function reads
\beq
L(x,\dot x)=-\frac{1}{2}m\dot x^2-U(x),
\eeq
where $x$ is the position of the particle,
\begin{equation}
\dot{x}=\frac{dx(\tau)}{d\tau},
\end{equation}
and the conjugate momentum to the coordinate $x$ is
\beq
p=\frac{\partial L(x,\dot x)}{\partial \dot x}=-m\dot x,
\eeq
so that the Hamiltonian reads
\begin{equation}
H(x,p)=p\dot x-L(x,\dot x)=-\frac{1}{2}m\dot{x}^2+U(x).
\end{equation}
This is called the Euclidean Hamiltonian, because the Minkowski space with
imaginary time has a Euclidean metric (this is a slight abuse of language: the Minkowski
space-time is not relevant here, since we are in a non-relativistic context.)

We consider the probability amplitudes for the particle to start at $x=-a$ at
$\tau=-\frac{T}{2}$
and end up at either $x=+a$ or $x=-a$ at $\tau=\frac{T}{2}$:
\begin{equation}
Z_\pm(T)=\langle \pm a|\e^{-\frac{HT}{\hbar}}|-a\rangle,
\end{equation}
where $|\pm a\rangle$ are position eigenstates with eigenvalues
$\pm a$. Inserting a complete
set $|n\rangle$ of energy eigenstates
\begin{equation}
H|n\rangle=E_n|n\rangle
\end{equation}
we get
\begin{equation}
Z_\pm(T)=
\sum_n\e^{-\frac{E_n T}{\hbar}}\langle\pm a|n\rangle\langle n|-a\rangle.
\label{ampl}
\end{equation}
If the potential barrier between the two minima
is sufficiently high and broad, the two lowest levels $E_1,E_2$ differ from
the lowest level in the absence of tunnelling,
$E_0=\frac{\hbar\omega}{2}+U_0$, by an amount which
is much smaller than the separation of $E_0$ from higher levels, of order
$\hbar\omega$.
Hence, in the large-$T$ limit,
the sum in \eq{ampl} is dominated by the two lowest eigenstates:
\begin{align}
Z_-(T)&=
\e^{-\frac{E_1 T}{\hbar}}\langle -a|1\rangle\langle 1|-a\rangle
+\e^{-\frac{E_2 T}{\hbar}}\langle-a|2\rangle\langle 2|-a\rangle
\\
Z_+(T)&=
\e^{-\frac{E_1 T}{\hbar}}\langle +a|1\rangle\langle 1|-a\rangle
+\e^{-\frac{E_2 T}{\hbar}}\langle +a|2\rangle\langle 2|-a\rangle,
\end{align}
up to terms that vanish more rapidly as $T\to\infty$.
Furthermore, the wave functions of states $|1\rangle$ and $|2\rangle$
in the position representation are even
and odd, respectively; thus
\begin{equation}
\langle+a|1\rangle=\langle -a|1\rangle;\quad
\langle+a|2\rangle=-\langle -a|2\rangle.
\end{equation}
Finally,
\begin{equation}
|\langle 1|+a\rangle|^2=|\langle 2|+a\rangle|^2.
\end{equation}
As a consequence,
\begin{align}
Z_-(T)&=|\langle 1|a\rangle|^2
\left(\e^{-\frac{E_1 T}{\hbar}}+\e^{-\frac{E_2 T}{\hbar}}\right)
\\
Z_+(T)&=|\langle 1|a\rangle|^2
\left(\e^{-\frac{E_1 T}{\hbar}}-\e^{-\frac{E_2 T}{\hbar}}\right)
\end{align}
and therefore, for large $T$,
\begin{equation}
\frac{Z_-(T)-Z_+(T)}{Z_-(T)+Z_+(T)}=\e^{-\frac{(E_2-E_1)T}{\hbar}},
\label{splitting0}
\end{equation}
or
\begin{equation}
E_2-E_1=-\hbar\lim_{T\to+\infty}\frac{1}{T}\log\frac{Z_-(T)-Z_+(T)}{Z_-(T)+Z_+(T)}.
\label{splitting}
\end{equation}
Eq.~(\ref{splitting}) allows us to compute the energy splitting between the two lowest energy
levels, provided the amplitudes $Z_\pm(T)$ can be computed in an independent way.

The path integral formalism provides such an independent calculation.
Indeed, by standard path integral arguments it can be shown that
\begin{equation}
Z_\pm(T)=\int[dx]\,\e^{-\frac{S[x]}{\hbar}},
\label{pathintegral}
\end{equation}
where
$[dx]$ is a functional measure, to be defined accurately below,
over all paths
$x(\tau)$ with
\begin{equation}
x\left(-\frac{T}{2}\right)=-a;\qquad
x\left(\frac{T}{2}\right)=\pm a,
\label{boundary}
\end{equation}
and $S[x]$ is the Euclidean action of the path:
\begin{equation}
S[x]=-\int_{-\frac{T}{2}}^{\frac{T}{2}}d\tau\,L(x,\dot x)
=\int_{-\frac{T}{2}}^{\frac{T}{2}}d\tau\,\left[\frac{1}{2}m\dot{x}^2+U(x)\right].
\end{equation}
Let us now assume that $S[x]$ has a stationary point, i.e.\ a path
$\bar x(\tau)$ such that $S[\bar x]$ is finite and stationary
upon deformations around $\bar x(\tau)$. In this case, the path integral
in \eq{pathintegral} can be computed by
means of a generalization of the saddle-point approximation technique. This is motivated
by the observation that
the exponent in the integrand of \eq{pathintegral} is weighted
by a factor of $\frac{1}{\hbar}$, which is large in the semi-classical
limit $\hbar\to 0$; the contributions to the integral from field configurations
far from the minimum configurations are therefore exponentially suppressed
in this limit.

In the leading saddle-point approximation, the functional
integral is simply given by the value of the integrand computed
in correspondence of the path which minimizes $S[x]$, that is, a
solution of the classical equations of motions. The next-to-leading
correction involves a calculation of the first non-trivial
fluctuations of the exponent around the stationary point;
for this reason we need the idea of a functional Taylor expansion,
and therefore of functional differentiation.
Functional derivatives are defined by
\begin{equation}
\frac{\delta x(\tau)}{\delta x(\tau')}=\delta(\tau-\tau'),
\end{equation}
together with the usual rules of differentiations of sums and products of ordinary functions.
We find
\begin{align}
\frac{\delta S[x]}{\delta x(\tau)}&=
\int_{-\frac{T}{2}}^{\frac{T}{2}}d\tau'\,\left[
m\frac{dx}{d\tau'}\frac{d}{d\tau'}\delta(\tau-\tau')+U'(x)
\delta(\tau-\tau')\right]
\nonumber\\
&=
-m\ddot x(\tau)+U'(x(\tau)),
\end{align}
where we have used
\begin{equation}
\int_{-\frac{T}{2}}^{\frac{T}{2}}d\tau\,f(\tau)\frac{d^n}{d\tau^n}\delta(\tau)
=(-1)^n\frac{d^nf(0)}{d\tau^n}.
\end{equation}
The second functional derivative of the action is given by
\begin{align}
\frac{\delta^2 S[x]}{\delta x(\tau')\delta x(\tau)}
&=-m\frac{d^2}{d\tau^2}\delta(\tau-\tau')+U''(x)\delta(\tau-\tau')
\nonumber\\
&=\left[-m\frac{d^2}{d\tau^2}+U''(x)\right]\delta(\tau-\tau')
\equiv S''[x]\delta(\tau-\tau').
\end{align}
We now take the functional Taylor expansion of $S[x]$ in the vicinity
of a solution $\bar x(\tau)$ of the classical equation of motion:
\begin{equation}
\label{eom}
\left.\frac{\delta S[x]}{\delta x(\tau)}\right|_{x=\bar x}=
-m\ddot{\bar x}+U'(\bar x)=0
\end{equation}
with the boundary conditions \eq{boundary}.
We find
\begin{align}
S[x]&=S[\bar x]+\frac{1}{2}\int_{-\frac{T}{2}}^{\frac{T}{2}}d\tau\,
\int_{-\frac{T}{2}}^{\frac{T}{2}}d\tau'\,\delta x(\tau)S''[\bar x]\delta(\tau-\tau')
\delta x(\tau')
+O((\delta x)^3)
\nonumber\\
&=
S[\bar x]+\frac{1}{2}\int_{-\frac{T}{2}}^{\frac{T}{2}}d\tau\,
\delta x(\tau)S''[\bar x]\delta x(\tau)
+O((\delta x)^3)
\label{actionexpanded}
\end{align}
where $\delta x(\tau)=x(\tau)-\bar x(\tau)$.

As in the case of a finite-dimensional gaussian integral,
it is convenient to introduce a set of eigenfunctions of $S''[\bar x]$:
\begin{equation}
S''[\bar x]\psi_n(\tau)=
\left[-m\frac{d^2}{d\tau^2}+U''(\bar x)\right]\psi_n(\tau)
=\lambda_n\psi_n(\tau),
\label{s2eigev}
\end{equation}
which can be taken to be orthogonal and normalized:
\begin{equation}
\label{eq:normalizpsin}
\int_{-\frac{T}{2}}^{\frac{T}{2}}d\tau\,\psi_n(\tau)\psi_m(\tau)=\delta_{nm};
\qquad \psi_n\left(\pm\frac{T}{2}\right)=0.
\end{equation}
A generic path can be parametrized as
\begin{equation}
x(\tau)=\bar x(\tau)+\sum_n c_n\psi_n(\tau).
\end{equation}
The integration over all possible paths is therefore identified with
an integration over all possible values of the coefficients $c_n$. This suggests that
the functional measure $[dx]$ be defined by
\begin{equation}
[dx]=N\prod_n\frac{dc_n}{\sqrt{2\pi\hbar}}.
\end{equation}
The factors of $\frac{1}{\sqrt{2\pi\hbar}}$ and the overall factor
$N$ have been introduced for later convenience.
The expansion of $S[x]$ around $\bar x$ to second order, \eq{actionexpanded},
takes the form
\begin{equation}
S[x]=S[\bar x]+\frac{1}{2}\sum_n\lambda_n c_n^2+O(c_n^3),
\end{equation}
and the path integral \eq{pathintegral} can be computed as an infinite product of gaussian integrals:
\begin{align}
Z_\pm(T)&=N \e^{-\frac{S[\bar x]}{\hbar}}\prod_n\frac{1}{\sqrt{2\pi\hbar}}
\int_{-\infty}^{+\infty} dc_n\,\e^{-\frac{1}{2\hbar}\lambda_nc_n^2}
\nonumber\\
&=N\e^{-\frac{S[\bar x]}{\hbar}}\prod_n\frac{1}{\sqrt{\lambda_n}}
\label{Zpm}
\end{align}
(the dependence of the r.h.s.~of Eq.~(\ref{Zpm}) on the $\pm$ sign in the l.h.s.~is hidden in the boundary values of the stationary path $\bar x$ and in the eigenvalues $\lambda_n$.)
We now formally define the determinant of an operator on a space
of functions, denoted by the symbol Det, as the product of its eigenvalues:
\begin{equation}
\Det S''[\bar x]=\prod_n\lambda_n,
\end{equation}
so that, at least formally,
\begin{equation}
Z_\pm(T)
=\frac{N}{\sqrt{\Det S''[\bar x]}}\e^{-\frac{S[\bar x]}{\hbar}}.
\label{gauss}
\end{equation}
A few comments are in order.
\begin{itemize}
\item
The result \eq{gauss} holds provided all eigenvalues $\lambda_n$ are strictly
positive. We will see that this is not the case in some relevant
situations, including the present one. We will show later in this section how to deal with this difficulty.
\item
The determinant in \eq{gauss} is divergent, since it is the
infinite product of positive and growing numbers. This infinity 
is usually absorbed by a suitable definition of the constant $N$ in the functional integration
measure. However, we shall not need an explicit expression of $N$,
because it cancels
in the calculation of the splitting between the two lowest energy eigenvalues.
\item
If there are more paths which minimize the Euclidean action, $x_n(\tau),n=0,1,\ldots$,
the path integral receives one contribution \eq{gauss} from each of them:
\begin{equation}
Z_\pm(T)
=N\sum_n\frac{1}{\sqrt{\Det S''[\bar x_n]}}\e^{-\frac{S[\bar x_n]}{\hbar}}.
\label{gaussn}
\end{equation}

\end{itemize}
In order to complete the calculation of the energy splitting $E_2-E_1$
from \eq{splitting}, we must therefore {\it i)} find all the
minimum paths for the Euclidean action, and {\it ii)}~learn how to
compute determinants of operators on spaces of functions. Both
steps are not trivial, and we will carefully go through the details.

\subsubsection{Stationary paths}
We first address the problem of finding all paths $\bar x(\tau)$ which
minimize the Euclidean action. In order to help intuition, it is
useful to view \eq{eom} as the equation of 
motion of a particle under the effect of the potential $-U(x)$
in real time (see the right panel of \fig{qm_double_well}). 

As far as $Z_-(T)$ is concerned, for which $x(T/2)=-a$, one solution is obviously
\beq
\bar x(\tau)=x_0(\tau)=-a
\eeq
for all times $\tau$. We have
\begin{equation}
S[x_0]=U_0T;\qquad S''[x_0]=-m\frac{d^2}{d\tau^2}+m\omega^2,
\end{equation}
and the corresponding contribution to the path integral is
\begin{equation}
I[x_0]=\frac{N}{\sqrt{\Det S''[x_0]}}e^{-\frac{S[x_0]}{\hbar}}.
\label{I0}
\end{equation}
The action $S[x_0]$ vanishes if we set the zero of the potential at
$x=\pm a$, as in the case of \eq{doublewellpot}. This is the choice adopted in \fig{qm_double_well} and in \app{app:doublewell}.
For the time being we keep however $U_0\ne 0$ for greater generality. Note that in this case
$S[x_0]=U_0T$ is only finite for finite $T$.

In the case $x(T/2)=+a$, the particle starts from $x=-a$ at time
$\tau=-T/2$ with zero velocity,
and reaches $x=a$ at time $\tau=T/2$. This solution,
which we will denote by $x_1(\tau)$, is usually called  an {\it instanton}.
The explicit form of $x_1(\tau)$ depends on the shape
of the potential energy $U(x)$, but its main features follow from
general considerations.
\begin{enumerate}
\item Because $U(x)$ is an even function, and the equation of motion is time-reversal invariant,
we have $x_1(-\tau)=-x_1(\tau)$, and therefore $x_1(0)=0$.

\item
From energy conservation,
\begin{equation}
-\frac{1}{2}m\dot x_1^2+U(x_1)=U_0
\label{instenergy}
\end{equation}
we get
\begin{equation}
\frac{dx_1(\tau)}{d\tau}=\sqrt{\frac{2[U(x_1)-U_0]}{m}},
\label{dx1dtau}
\end{equation}
where we have selected the positive sign because $x_1(\tau)$ is an increasing function of $\tau$.
\eq{dx1dtau} implies that the particle spends most of the time close to either
the initial position $x=-a$
or the final position $x=a$; the transition between these two values
occurs within a short time interval of order $1/\omega$ (the {\it instanton size})
around $\tau=0$ (the {\it instanton center}).

\item
When $T\to+\infty$, the instanton center can be located
anywhere; hence, in this limit we must consider all solutions $x_1(\tau-\tau_1)$
with $-\frac{T}{2}<\tau_1<\frac{T}{2}$. The one-instanton
contribution is obtained by summing over all possible locations of the instanton center, i.e.\ by integrating over $\tau_1$ in this range:
\begin{equation}
I[x_1]
=\int_{-T/2}^{T/2}d\tau_1\,\frac{N}{\sqrt{\Det S''[x_1]}}
\e^{-\frac{S[x_1]}{\hbar}}
=\frac{N}{\sqrt{\Det S''[x_1]}}T\e^{-\frac{S[x_1]}{\hbar}}.
\end{equation}
\item Using again \eq{instenergy}, we find that the instanton action
$S[x_1]$ can be expressed either in terms of the potential energy:
\begin{align}
S[x_1]&=\int_{-T/2}^{T/2}d\tau\,2U(x_1)-U_0T
\notag\\
&=\int_{-T/2}^{T/2}d\tau\,2[U(x_1)-U_0]+U_0T
\nonumber\\
&=\int_{-a}^a dx\,\sqrt{2m[U(x)-U_0]}+S[x_0],
\label{instaction}
\end{align}
or in terms of the kinetic energy:
\beq
S[x_1]=\int_{-T/2}^{T/2}d\tau\,m\dot x_1^2+S[x_0].
\label{instactionkin}
\eeq
Both expressions will prove useful. Note that $S[x_0]$ (and therefore $S[x_1]$) is only
finite for
finite values of $T$, while the difference $S[x_1]-S[x_0]$ is finite even in
the limit $T\to+\infty$.
\end{enumerate}
We can check the above features of the instanton solution 
in the case of the potential \eq{doublewellpot}. In this case, \eq{dx1dtau} can be integrated analytically:
\beq
x_1(\tau)=a\frac{e^{\omega\tau}-1}{e^{\omega\tau}+1}.
\eeq
In this explicit example, 
$x_1(\tau)$ is close $\pm a$ for $|\tau|$ larger than a few times $\frac{1}{\omega}$;
indeed $x_1(\pm 4/\omega)\simeq \pm 0.96\,a$.
Hence, as anticipated, $x_1(\tau)$ differs sizeably from $\pm a$ only in an interval $-\delta<\tau<\delta$,
with $\delta$ of order a few times $\omega^{-1}$.
Furthermore,
the instanton action can be computed analytically in this case. Using \eq{instaction}
and $S[x_0]=0$ we find
\beq
S[x_1]=\frac{m\omega}{2a}\int_{-a}^a dx\,(a^2-x^2)=\frac{2ma^2\omega}{3}.
\label{instact0}
\eeq

As anticipated, in the limit $T\to \infty$ the operator $S''[x_1]$
has a {\it zero mode}, that is, an eigenfunction with zero eigenvalue:
\begin{equation}
\label{norm}
\psi_0(\tau)=C\frac{dx_1(\tau)}{d\tau},
\end{equation}
where
\begin{equation}
C=\left[\int_{-\frac{T}{2}}^{\frac{T}{2}}d\tau\,\left|\frac{dx_1(\tau)}{d\tau}\right|^2\right]^{-\frac{1}{2}}=\sqrt{\frac{m}{S[x_1]-S[x_0]}}
\end{equation}
by \eq{instactionkin}. Indeed,
\begin{equation}
\left[-m\frac{d^2}{d\tau^2}+U''(x_1)\right]\frac{dx_1}{d\tau}=
\frac{d}{d\tau}\left[-m\ddot x_1+U'(x_1)\right]=0
\end{equation}
by the equation of motion. The presence of a zero mode implies that the integration over the corresponding coefficient $c_0$ leads to an infinity. Luckily, we do not need to perform this integration,
because an infinitesimal change in $c_0$
is equivalent to an infinitesimal shift in the instanton center $\tau_1$:
\begin{align}
\frac{dx(\tau)}{dc_0}&=\psi_0(\tau)=\sqrt{\frac{m}{S[x_1]-S[x_0]}}
\frac{dx_1}{d\tau}
\\
\frac{dx(\tau)}{d\tau_1}&=\frac{dx_1}{d\tau_1}=-\frac{dx_1}{d\tau},
\end{align}
or equivalently
\begin{equation}
\frac{dc_0}{\sqrt{2\pi\hbar}}
=-\sqrt{\frac{S[x_1]-S[x_0]}{2\pi\hbar m}}d\tau_1.
\eeq
Since we have already performed an integration
over the instanton center $\tau_1$, we may simply omit the integration over
$c_0$, which amounts to omitting the zero eigenvalue in the determinant,
and multiply the result with the jacobain factor $\sqrt{\frac{S[x_1]-S[x_0]}{2\pi\hbar m}}$:
\begin{equation}
I[x_1]
=\frac{N}{\sqrt{\Det'S''[x_1]}}
\sqrt{\frac{S[x_1]-S[x_0]}{2\pi\hbar m}}T\e^{-\frac{S[x_1]}{\hbar}}
=I[x_0] KT\e^{-\frac{S[x_1]-S[x_0]}{\hbar}}
\label{I1}
\end{equation}
where $I[x_0]$ is given in \eq{I0}, we have defined
\begin{equation}
K=\sqrt{\frac{S[x_1]-S[x_0]}{2\pi\hbar m}}
\sqrt{\frac{\Det S''[x_0]}{\Det' S''[x_1]}}
\label{kdef}
\end{equation}
and $\Det'$ stands for a determinant with the zero eigenvalue removed.

More approximate solutions of the equation of motion can be built
out of the following observation. Given an instanton solution $x_1(\tau)$,
then, by time reversal invariance,
\beq
x_1(-\tau)=-x_1(\tau)
\eeq
is also a solution, with boundary conditions
interchanged, and the same action as the instanton.
This is called an {\it anti-instanton}.
It follows that any sequence of an even [odd] number
$n$ of alternate instanton and anti-instanton solutions contributes
to the saddle-point estimate of $Z_-(T)$
[$Z_+(T)$]. We denote these solutions by
$x_n(\tau)$.

In order to compute $S[x_n]$ we observe that,
as in the single-instanton case, we may use energy conservation
to obtain
\beq
S[x_n]=\int_{-T/2}^{T/2}d\tau\,m\dot x_n^2+S[x_0].
\eeq
The function $x_n(\tau)$ is approximately
equal to $\pm a$ almost everywhere in time, except for $n$ small
intervals, of order $1/\omega$
in size, where it behaves as the single instanton or anti-instanton solutions.
We may therefore approximate it by
\beq
x_n(\tau)\simeq x_0+\sum_{k=1}^n(-1)^{k+1}\left[x_1(\tau-\tau_k)-x_0\right],
\label{multiinst}
\eeq
which gives
\beq
\dot x_n^2=\left[\sum_{k=1}^n(-1)^{k+1}\dot x_1(\tau-\tau_k)\right]^2
\simeq
\sum_{k=1}^n \dot x^2_1(\tau-\tau_k),
\eeq
where in the last step we have used the fact that $x_1(\tau-\tau_k)$ is approximately
constant for $\tau$ far from $\tau_k$.
Hence, in the large-$T$ limit,
\begin{equation}
S[x_n]=n\left(S[x_1]-S[x_0]\right)+S[x_0],
\end{equation}
and
\beq
I[x_n]=e^{-\frac{S[x_0]}{\hbar}}
\left(e^{-\frac{S[x_1]-S[x_0]}{\hbar}}\right)^n
\int [dx]\,\exp\left[-\frac{1}{2}\int_{-T/2}^{T/2}d\tau\,
x(\tau)S''[x_n]x(\tau)\right].
\eeq

In order to compute the functional integral, we observe that $S''[x_n]$ is approximately given by
\begin{equation}
S''[x_n]\simeq\left\{\begin{array}{ll}
S''[x_1(\tau-\tau_k)] & \tau_k-\delta < \tau < \tau_k+\delta
\\
S''[x_0] & {\rm elsewhere}
\end{array}\right.
\label{S2approx}
\end{equation}
where $k=1,\ldots,n$ and $\delta\sim\omega^{-1}$.
This is a consequence of the fact that $x_n(\tau)$ is a multi-instanton
solution in which the different instantons and anti-instantons are
well separated from each other (this is usually called a dilute instanton gas; the validity of this assumption
must be verified, which we will do in a moment.)

Let us now denote by $[dx]_k$ the functional integration measure over paths $x(\tau)$
that are sizably different from zero only in the range
$\tau_k-\delta < \tau < \tau_k+\delta$, where
$\tau_k, k=1,\ldots, n$ are the centers of the $n$ instantons and anti-instantons.
Using the approximation \eq{S2approx} we obtain 
\begin{align}
&\int [dx]\,\exp\left[-\frac{1}{2}\int_{-T/2}^{T/2} d\tau\,x(\tau)S''[x_n]x(\tau)\right]
\notag\\
&\qquad\qquad
\simeq
\int [dx]\,\exp\left[-\frac{1}{2}\int_{-T/2}^{T/2} d\tau\,x(\tau)S''[x_0]x(\tau)\right]
\notag\\
&\qquad\qquad
\times
\prod_{k=1}^n \frac{
\int [dx]_k\,\exp\left[-\frac{1}{2}\int_{-T/2}^{T/2}d\tau\,
x(\tau)S''[x_1(\tau-\tau_k)]x(\tau)\right]}
{ \int [dx]_k\,\exp\left[-\frac{1}{2}\int_{-T/2}^{T/2}d\tau\,
x(\tau)S''[x_0])]x(\tau)\right]}.
\end{align}
The ratio under the product sign is in fact independent of $k$. Hence
\begin{align}
&\prod_{k=1}^n
\frac{
\int [dx]_k\,\exp\left[-\frac{1}{2}\int_{-T/2}^{T/2}d\tau\,
x(\tau)S''[x_1(\tau-\tau_k)]x(\tau)\right]}
{\int [dx]_k\,\exp\left[-\frac{1}{2}\int_{-T/2}^{T/2}d\tau\,
x(\tau)S''[x_0]x(\tau)\right]}
\nonumber\\
&=\left[
\frac{
\int [dx]_k\,\exp\left[-\frac{1}{2}\int_{-T/2}^{T/2}d\tau\,
x(\tau)S''[x_1(\tau-\tau_k)]x(\tau)\right]
}{
\int [dx]_k\,\exp\left[-\frac{1}{2}\int_{-T/2}^{T/2}d\tau\,
x(\tau)S''[x_0]x(\tau)\right]}
\right]^n
\nonumber\\
&=\left[
\frac{
\int [dx]\,\exp\left[-\frac{1}{2}\int_{-T/2}^{T/2}d\tau\,
x(\tau)S''[x_1(\tau-\tau_k)]x(\tau)\right]
}{
\int [dx]\,\exp\left[-\frac{1}{2}\int_{-T/2}^{T/2}d\tau\,
x(\tau)S''[x_0]x(\tau)\right]}
\right]^n
\label{ratiodet}
\end{align}
In the last step the functional integrals
have been extended to all paths, not restricted by the condition of being different from zero only around $\tau_k$. This is allowed in the ratio,
because $S''[x_1(\tau-\tau_k)]$ and $S''[x_0]$ essentially coincide
for $\tau$ not too close to $\tau_k$.
Hence, leaving aside for the moment the problem
of the zero mode of $S''[x_1]$, we find
\beq
\int [dx]\,\exp\left[-\frac{1}{2}\int_{-T/2}^{T/2} d\tau\,
x(\tau)S''[x_n]x(\tau)\right]
\simeq
\frac{N}{\sqrt{\Det S''[x_0]}}
\left[
\sqrt{\frac{\Det S''[x_0]}{\Det S''[x_1]}}\right]^n.
\eeq

To complete the calculation of the $n$-instanton contribution,
we must remove the zero eigenvalue from
each factor of ${\rm det\,}S''[x_1]$, include $n$ jacobian factors
\beq
\sqrt{\frac{S[x_1]-S[x_0]}{2\pi\hbar m}}
\eeq
and integrate over all  possible locations $\tau_1,\ldots,\tau_n$
of the instanton centers. This yields a factor of
\begin{equation}
\left[\sqrt{\frac{S[x_1]-S[x_0]}{2\pi\hbar m}}\right]^n
\int_{-T/2}^{T/2}d\tau_1\int_{\tau_1}^{T/2}d\tau_2\ldots
\int_{\tau_{n-1}}^{T/2}d\tau_n
=
\left[\sqrt{\frac{S[x_1]-S[x_0]}{2\pi\hbar m}}\right]^n\frac{T^n}{n!}.
\end{equation}
So finally
\begin{align}
I[x_n]&=\frac{N}{\sqrt{\Det S''[x_0]}}\frac{K^nT^n}{n!}
\e^{-\frac{n(S[x_1]-S[x_0])}{\hbar}}\e^{-\frac{S[x_0]}{\hbar}}
=I[x_0]\frac{1}{n!}
\left[KT\e^{-\frac{S[x_1]-S[x_0]}{\hbar}}\right]^n
\end{align}
with $K$ defined in \eq{kdef}. Note that our previous results Eqs.~(\ref{I0},\ref{I1}) are recovered for $n=0,1$ respectively.

We are now in a position to check the validity of the dilute instanton gas assumption.
When all contributions $I[x_n]$ are summed up, the dominant contribution
to the sum comes from values of $n$ for which
\beq
\frac{1}{n!}
\left[KT\e^{-\frac{S[x_1]-S[x_0]}{\hbar}}\right]^n
\sim
\frac{1}{(n-1)!}
\left[KT\e^{-\frac{S[x_1]-S[x_0]}{\hbar}}\right]^{n-1}
\eeq
or
\beq
\frac{n}{T}\sim K\e^{-\frac{S[x_1]-S[x_0]}{\hbar}}.
\label{instdensity}
\eeq
We see that the instanton density $\frac{n}{T}$
for the relevant contributions is exponentially suppressed,
as long as $S[x_1]-S[x_0]\gg\hbar$, which is the condition of validity
of the semi-classical approximation (we will be able to check this
statement in the explicit example of \eq{doublewellpot} after
computing $K$.) This is an {\it a posteriori} confirmation of the reliability
of our assumption that instantons and anti-instantons are well separated from each other
in the relevant multi-instanton configurations.

\subsubsection{Functional determinants}
\label{sec:funcdet}
We now turn to the calculation of the ratio
\begin{equation}
\frac{\Det S''[x_0]}{\Det'S''[x_1]}
\label{detratio}
\end{equation}
which appears in the definition of $K$, \eq{kdef}.
The procedure is illustrated in Ref.~\cite{Coleman:1985rnk}; we will reproduce here the
argument presented there, including some details which are omitted in Ref.~\cite{Coleman:1985rnk}.
We first show how to compute the ratio
\begin{equation}
\label{ratio}
\frac{\Det O_{1}}{\Det O_{2}},
\end{equation}
where
\begin{equation}
O_i=-m\frac{d^2}{d\tau^2}+W_i(\tau);\qquad i=1,2
\label{Oi}
\end{equation}
and $W_i(\tau)$ are arbitrary functions of $\tau$
in the range $-T/2\leq\tau\leq T/2$.
Then, we remove the contribution of the minimum eigenvalue,
which is different
from zero as long as $T$ is kept finite, from the denominator of
\eq{ratio}. Finally, we take the limit $T\to+\infty$.

Let us consider solutions $\phi^{(i)}_\lambda(\tau)$ of the
Cauchy problem
\begin{align}
&O_{i}\phi^{(i)}_\lambda(\tau)=\lambda\phi^{(i)}_\lambda(\tau)
\label{equation}
\\
&\phi^{(i)}_\lambda\left(-\frac{T}{2}\right)=0
\label{insol}
\\
&\dot{\phi}^{(i)}_\lambda\left(-\frac{T}{2}\right)=1.
\label{insol1}
\end{align}
The important point here is that the function $\phi^{(i)}_\lambda(\tau)$, which certainly exists, because it is the unique solution to the Cauchy problem
Eqs.~(\ref{equation}, \ref{insol}, \ref{insol1}), is not necessarily an eigenvector of $O_i$,
because $\phi^{(i)}_\lambda(T/2)$ is in general different from zero.
Also note that $\phi^{(i)}_\lambda(\tau)$ is not, in general,
normalized to one.

We now define
\begin{equation}
F(\lambda)=\frac{\Det (O_{1}-\lambda)}{\Det (O_{2}-\lambda)};
\qquad
G(\lambda)=\frac{\phi^{(1)}_\lambda\left(\frac{T}{2}\right)}
{\phi^{(2)}_\lambda\left(\frac{T}{2}\right)}.
\end{equation}
The function $F(\lambda)$ has a simple zero whenever $\lambda$ is an eigenvalue
$\lambda_n^{(1)}$ of $O_{1}$,
and a simple pole whenever $\lambda$ is an eigenvalue $\lambda_n^{(2)}$ of $O_{2}$.
The function $G(\lambda)$ has exactly the same poles and zeros.
By Liouville's theorem, the ratio $\frac{F(\lambda)}{G(\lambda)}$ is therefore
a constant. Since both $F(\lambda)$ and $G(\lambda)$ tend to 1
as $\lambda\to \infty$ in any direction except the real axis, where both functions have simple poles,
the constant is equal to one. Hence
\begin{equation}
F(\lambda)=G(\lambda)
\end{equation}
in the whole complex plane $\lambda$. In particular, for $\lambda\to 0$,
\begin{equation}
\label{ratio0}
\frac{\Det O_{1}}{\Det O_{2}}=
\frac{\phi^{(1)}_0\left(\frac{T}{2}\right)}
{\phi^{(2)}_0\left(\frac{T}{2}\right)},
\end{equation}
which is the desired result. 

We are interested in the case
\begin{equation}
O_1=S''[x_0]=-m\frac{d^2}{d\tau^2}+m\omega^2\qquad O_2=S''[x_1]=-m\frac{d^2}{d\tau^2}+U(x_1).
\end{equation}
We need the solutions of the Cauchy problems
\begin{align}
&\left(-m\frac{d^2}{d\tau^2}+m\omega^2\right)\phi_0(\tau)=0;\qquad
\phi_0\left(-\frac{T}{2}\right)=0;\qquad\dot{\phi}_0\left(-\frac{T}{2}\right)=1
\label{S20}
\\
&\left(-m\frac{d^2}{d\tau^2}+U''(x_1)\right)\psi_0(\tau)=0;\qquad
\psi_0\left(-\frac{T}{2}\right)=0;\qquad\dot{\psi}_0\left(-\frac{T}{2}\right)=1.
\label{S21}
\end{align}
The solution of \eq{S20} is easily found:
\begin{equation}
\label{phi0}
\phi_0(\tau)=\frac{\e^{\omega(\tau+T/2)}-\e^{-\omega(\tau+T/2)}}{2\omega}
=\frac{1}{\omega}\sinh\omega\left(\tau+\frac{T}{2}\right).
\end{equation}
\eq{S21} requires some work. We have
\begin{equation}
\psi_0(\tau)=c_1f(\tau)+c_2g(\tau),
\end{equation}
where $f(\tau)$ and $g(\tau)$ are two independent solutions with constant Wronskian  determinant $w(\tau)$.
We may normalize them so that
\begin{equation}
w(\tau)=f(\tau)\dot g(\tau)-\dot f(\tau)g(\tau)=1.
\label{Wronskian}
\end{equation}
The initial conditions
\begin{align}
&\psi_0\left(-\frac{T}{2}\right)=
c_1f\left(-\frac{T}{2}\right)+c_2g\left(-\frac{T}{2}\right)=0
\\
&\dot{\psi}_0\left(-\frac{T}{2}\right)=
c_1\dot f\left(-\frac{T}{2}\right)+c_2\dot g\left(-\frac{T}{2}\right)=1
\end{align}
give
\begin{equation}
c_1=-g\left(-\frac{T}{2}\right)\qquad
c_2=f\left(-\frac{T}{2}\right)
\end{equation}
with the choice \eq{Wronskian}.
Since we are interested in $\psi_0(T/2)$ for large $T$,
we only need the asymptotic behaviours of $f(\tau)$ and $g(\tau)$.
We already know that $f(\tau)=\dot x_1(\tau)$.
The asymptotic behaviour of $f(\tau)$ can be obtained from \eq{dx1dtau}
and from the observation that $x_1\to a$ for $\tau\to+\infty$.
In this limit
\begin{equation}
\frac{dx_1(\tau)}{d\tau}=\sqrt{\frac{2}{m}[U(x_1)-U_0]}
\simeq 
-\omega(x_1-a)
\qquad {\rm for}\;\tau\to+\infty
\end{equation}
which gives, in the same limit,
\begin{equation}
a-x_1(\tau)\simeq\hat a\e^{-\omega\tau};\qquad
f(\tau)=\dot x_1(\tau)\simeq \omega \hat a\e^{-\omega\tau},
\label{largetauinst}
\end{equation}
where $\hat a$ is a positive constant, which is determined as follows. We have
\begin{equation}
\tau=\sqrt{\frac{m}{2}}\int_0^{x_1}\frac{dx}{\sqrt{U(x)-U_0}}
\label{tau1}
\end{equation}
from \eq{dx1dtau}, and
\begin{equation}
\tau=-\frac{1}{\omega}\log\frac{\epsilon}{\hat a}+
O(\epsilon^2)
\label{tau2}
\end{equation}
from \eq{largetauinst}, where $\epsilon=a-x_1$. Equating \eq{tau1} and \eq{tau2} we get
\begin{equation}
\hat a=\lim_{\epsilon\to 0}\epsilon
\exp\left[m\omega\int_0^{a-\epsilon}\frac{dx}{\sqrt{2m[U(x)-U_0]}}\right].
\label{Adef}
\end{equation}
The integral in the exponent in \eq{Adef} is logarithmically divergent as $\epsilon\to 0$, due to the behaviour of the integrand at the upper integration bound:
\beq
U(x)-U_0=\frac{1}{2}m\omega^2(x-a)^2+O((x-a)^3)).
\eeq
We have
\begin{align}
m\omega\int_0^{a-\epsilon}\frac{dx}{\sqrt{2m[U(x)-U_0]}}&=
\int_0^{a-\epsilon}dx\,\left[\frac{m\omega}{\sqrt{2m[U(x)-U_0]}}-\frac{1}{a-x}\right]
+\int_0^{a-\epsilon}\frac{dx}{a-x}
\nonumber\\
&=
\int_0^{a-\epsilon}dx\,\left[\frac{m\omega}{\sqrt{2m[U(x)-U_0]}}-\frac{1}{a-x}\right]
+\log\frac{a}{\epsilon},
\end{align}
and the integral is now convergent as $\epsilon\to 0$. Therefore
\begin{equation}
\hat a=a\exp\int_0^a dx\,\left[\frac{m\omega}{\sqrt{2m[U(x)-U_0]}}-\frac{1}{a-x}\right].
\label{Adw}
\end{equation}
In the case of the potential in \eq{doublewellpot} we find
\beq
\sqrt{2m[U(x)-U_0]}=\frac{m\omega}{2a}(a^2-x^2);\qquad \hat a=2a.
\eeq

The asymptotic behaviour of $g(\tau)$ can now be obtained from
\eq{Wronskian}, which can be rewritten
\begin{equation}
\frac{d}{d\tau}\frac{g(\tau)}{f(\tau)}=\frac{1}{f^2(\tau)}.
\end{equation}
The solution for $\tau\to+\infty$ is
\begin{equation}
g(\tau)=\frac{1}{2\hat a\omega^2}\e^{\omega\tau}
\qquad {\rm for}\;\tau\to+\infty.
\end{equation}
Hence, in the large-$T$ limit,
\begin{align}
\psi_0\left(\frac{T}{2}\right)&=
-f\left(\frac{T}{2}\right)g\left(-\frac{T}{2}\right)
+f\left(-\frac{T}{2}\right)g\left(\frac{T}{2}\right)
\nonumber\\
&=
\frac{1}{2\omega}\left[\e^{-\frac{\omega T}{2}}\e^{\frac{\omega T}{2}}+
\e^{\frac{\omega T}{2}}\e^{-\frac{\omega T}{2}}\right]
=\frac{1}{\omega}.
\label{psi0T2}
\end{align}
Using \eq{phi0} we finally obtain
\begin{equation}
\frac{\Det S''[x_0]}{\Det S''[x_1]}
=\frac{\phi_0\left(\frac{T}{2}\right)}{\psi_0\left(\frac{T}{2}\right)}
=\sinh\omega T.
\label{detratio2}
\end{equation}

The next step is the removal of the smallest eigenvalue from the denominator
of \eq{detratio2}. To this purpose, we consider the differential equation
\begin{equation}
-m\ddot\psi_\lambda(\tau)+U''(\bar
x_1)\psi_\lambda(\tau)=\lambda\psi_\lambda(\tau); \quad
\psi_\lambda\left(-\frac{T}{2}\right)=0;\quad
\dot\psi_\lambda\left(-\frac{T}{2}\right)=1
\label{mineig}
\end{equation}
and we take $\lambda$ to be small, because we
know that the smallest eigenvalue is zero for large $T$:
\begin{equation}
\psi_\lambda(\tau)=\psi_0(\tau)+\lambda\eta(\tau)+O(\lambda^2).
\end{equation}
Replacing in \eq{mineig} and expanding to first order in $\lambda$
we get
\begin{equation}
-m\ddot\eta(\tau)+U''(x_1)\eta(\tau)=\psi_0(\tau);
\qquad \eta\left(-\frac{T}{2}\right)=\dot\eta\left(-\frac{T}{2}\right)
=0.
\label{deltaeq}
\end{equation}
This differential equation can be turned into an integral equation
\begin{align}
\eta(\tau)&=-\frac{1}{m}\int_{\frac{T}{2}}^\tau d\tau''\int_{-\frac{T}{2}}^{\tau''}d\tau'\,
\left[\psi_0(\tau')-U''(x_1)\eta(\tau')\right]
\notag\\
&=-\frac{1}{m}\int_{-\frac{T}{2}}^\tau d\tau'(\tau-\tau')
\left[\psi_0(\tau')-U''(x_1)\eta(\tau')\right]
\end{align}
and solved by iteration. The result has the form
\begin{equation}
\label{delta}
\eta(\tau)=\int_{-\frac{T}{2}}^\tau d\tau'\,k(\tau,\tau')\psi_0(\tau')
\end{equation}
where the kernel $k(\tau,\tau')$ is a solution of
\begin{equation}
\left(-m\frac{d^2}{d\tau^2}+U''(x_1)\right)k(\tau,\tau')=0;
\quad k(\tau,\tau)=0;
\quad \left[\frac{d}{d\tau}k(\tau,\tau')\right]_{\tau'=\tau}=-\frac{1}{m},
\end{equation}
as one can check directly by replacing \eq{delta} in \eq{deltaeq}.
Hence, $k(\tau,\tau')$ is a linear combination of
$f(\tau)$ and $g(\tau)$ with $\tau'$-dependent coefficients.
We find
\begin{equation}
k(\tau,\tau')=
\frac{1}{m}\left[f(\tau)g(\tau')-g(\tau)f(\tau')\right],
\end{equation}
which obeys the relevant initial conditions thanks to the choice
\eq{Wronskian}. So finally
\begin{equation}
\psi_\lambda(\tau)=\psi_0(\tau)
+\frac{\lambda}{m}\int_{-\frac{T}{2}}^\tau d\tau'\,
\left[f(\tau)g(\tau')-g(\tau)f(\tau')\right]\psi_0(\tau')
+O(\lambda^2).
\end{equation}
The smallest eigenvalue is determined by the condition
$\psi_\lambda(T/2)=0$, which gives
\begin{equation}
\label{smallest}
\frac{1}{\omega}
+\frac{\lambda}{m}\int_{-\frac{T}{2}}^{\frac{T}{2}} d\tau'\,
\left[
f\left(\frac{T}{2}\right)g(\tau')-g\left(\frac{T}{2}\right)f(\tau')
\right]\psi_0(\tau')=0,
\end{equation}
where we have used \eq{psi0T2}. For large $T$ the integrand reads
\begin{align}
&\left[
f\left(\frac{T}{2}\right)g(\tau')-g\left(\frac{T}{2}\right)f(\tau')\right]
\left[-g\left(-\frac{T}{2}\right)f(\tau')+f\left(-\frac{T}{2}\right)g(\tau')\right]
\nonumber\\
&\simeq
\left[\omega \hat a\e^{-\frac{\omega T}{2}}g(\tau')
     -\frac{1}{2\hat a\omega^2}\e^{\frac{\omega T}{2}}f(\tau')\right]
\left[\frac{1}{2\hat a\omega^2}\e^{\frac{\omega T}{2}}f(\tau')
      +\omega \hat a \e^{-\frac{\omega T}{2}}g(\tau')\right]
\nonumber\\
&\simeq
-\frac{1}{4\omega^4 \hat a^2}f^2(\tau')\e^{\omega T},
\end{align}
and therefore in this limit \eq{smallest} becomes
\begin{equation}
\label{smallest1}
1-\frac{\lambda}{m\omega^3}
\frac{1}{4\hat a^2}\e^{\omega T}\int_{-\frac{T}{2}}^{\frac{T}{2}} d\tau_1\,
f^2(\tau_1)
=1-\frac{\lambda}{m\omega^3}
\frac{1}{4\hat a^2}\e^{\omega T}\frac{S[x_1]-S[x_0]}{m}
=0,
\end{equation}
where we have used \eq{norm}. Thus
\begin{equation}
\label{smallest3}
\lambda=\frac{4\hat a^2m^2\omega^3}{S[x_1]-S[x_0]}\e^{-\omega T}.
\end{equation}
Observe that the smallest eigenvalue tends to 0 as
$T\to\infty$, as expected.
We conclude that
\begin{equation}
\frac{\Det S''[x_0]}{\Det'S''[x_1]}
=\lambda\frac{\Det S''[x_0]}{\Det S''[x_1]}
=
\frac{4\hat a^2m^2\omega^3}{S[x_1]-S[x_0]}\e^{-\omega T}
\sinh\omega T,
\end{equation}
which is finite and $T$ independent as $T\to+\infty$:
\begin{equation}
\lim_{T\to+\infty}\frac{\Det S''[x_0]}{\Det'S''[x_1]}
=\frac{2\hat a^2m^2\omega^3}{S[x_1]-S[x_0]}.
\end{equation}

\subsubsection{Energy splitting of the lowest energy levels}
Collecting all our results, we find
\begin{align}
Z_-(T)=&\sum_{n\,{\rm even}}I[x_n]
=I[x_0]\sum_{n\,{\rm even}}
\frac{1}{n!}\left(KT\e^{-\frac{S[x_1]-S[x_0]}{\hbar}}\right)^n
\nonumber\\
=&I[x_0]
\cosh\left(KT\e^{-\frac{S[x_1]-S[x_0]}{\hbar}}\right)
\\
Z_+(T)
=&\sum_{n\,{\rm odd}}I[x_n]
=I[x_0]\sum_{n\,{\rm odd}}
\frac{1}{n!}\left(KT\e^{-\frac{S[x_1]-S[x_0]}{\hbar}}\right)^n
\nonumber\\
=&I[x_0]\sinh\left(KT\e^{-\frac{S[x_1]-S[x_0]}{\hbar}}\right)
\end{align}
where
\beq
K=\sqrt{\frac{S[x_1]-S[x_0]}{2\pi\hbar m}}
\sqrt{\frac{\Det S''[x_0]}{\Det'S''[x_1]}}
=\hat a\omega\sqrt{\frac{m\omega}{\pi\hbar}}.
\label{kdef2}
\eeq
The energy splitting $E_2-E_1$ is given by \eq{splitting}, which reads
\begin{equation}
E_2-E_1=\hbar\omega \left[2\hat a\sqrt{\frac{m\omega}{\pi\hbar}}\right] \e^{-\frac{S[x_1]-S[x_0]}{\hbar}}
\label{splitPI}
\end{equation}
where
\begin{equation}
S[x_1]-S[x_0]=\int_{-a}^adx\,\sqrt{2m[U(x)-U_0]}.
\label{SPI}
\end{equation}
Note that the result is independent of the constant $N$, which appears in the
functional integration measure, and of $\Det S''[x_0]$, because the factors
of $I[x_0]$, cancel in the ratio. 

Recalling the expression \eq{instdensity} for the instanton density in the dominant
multi-instanton solutions, and the expression \eq{instact0} for the
instanton action, we obtain, in the case of the potential \eq{doublewellpot},
\beq
\frac{n}{T}=2\omega\sqrt{\frac{m a^2\omega}{\pi\hbar}}
\exp\left(-\frac{2ma^2\omega}{3\hbar}\right),
\eeq
which is exponentially small for $ma^2\omega\gg\hbar$, thus justifying the use of
the dilute instanton gas approximation.

\subsection{Decay of a metastable state}
\label{sect:decmet}
Although admittedly cumbersome,
the procedure described in the previous subsection lends itself to an
extension to the case of the decay probability per unit time of an unstable state. Let us consider a
one-dimensional potential energy $U(x)$ with a narrow local minimum in
$x=a$, with  $U(a)=U_0$ and $U''(a)=m\omega^2$, separated from a lower, much broader minimum
by a potential barrier. An example of such a potential is sketched in the left panel of \fig{qm_double_well3}. As in the case of the
double potential well, if the barrier were infinitely broad and high,
there would be two distinct sets of energy eigenstates. When tunnelling
is taken into account, we expect the eigenstates of the potential well
with a local minimum in $x=a$ to become unstable, or equivalently, the energy eigenvalues
to acquire an imaginary part. We shall see that the path integral formulation of the problem
confirms this expectation, and also allows us to compute the imaginary
part of the lowest eigenvalue, which is related to
the decay probability per unit time of the metastable ground state.

\begin{figure}[t]
\centering
\includegraphics[width=0.45\textwidth]{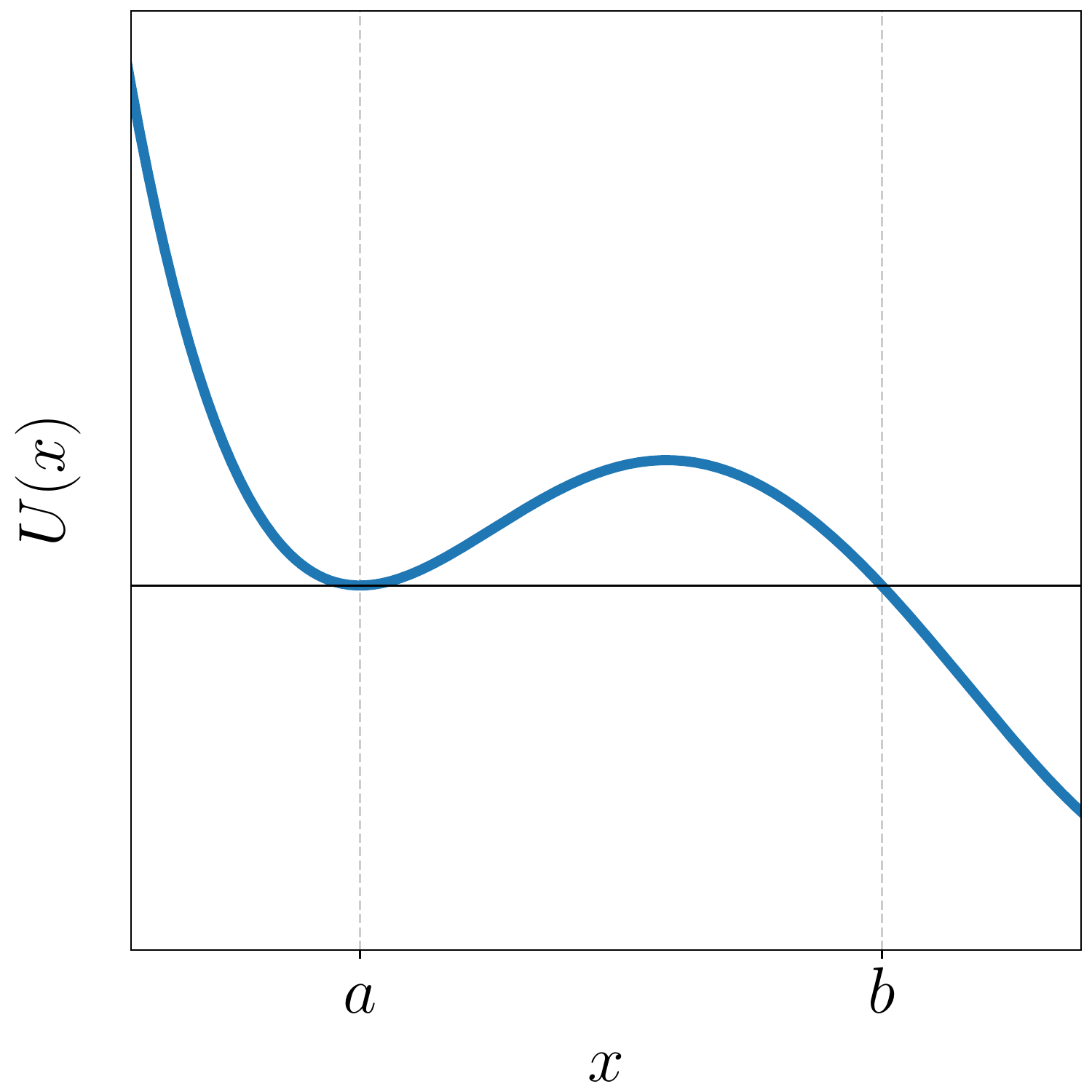} \qquad
\includegraphics[width=0.45\textwidth]{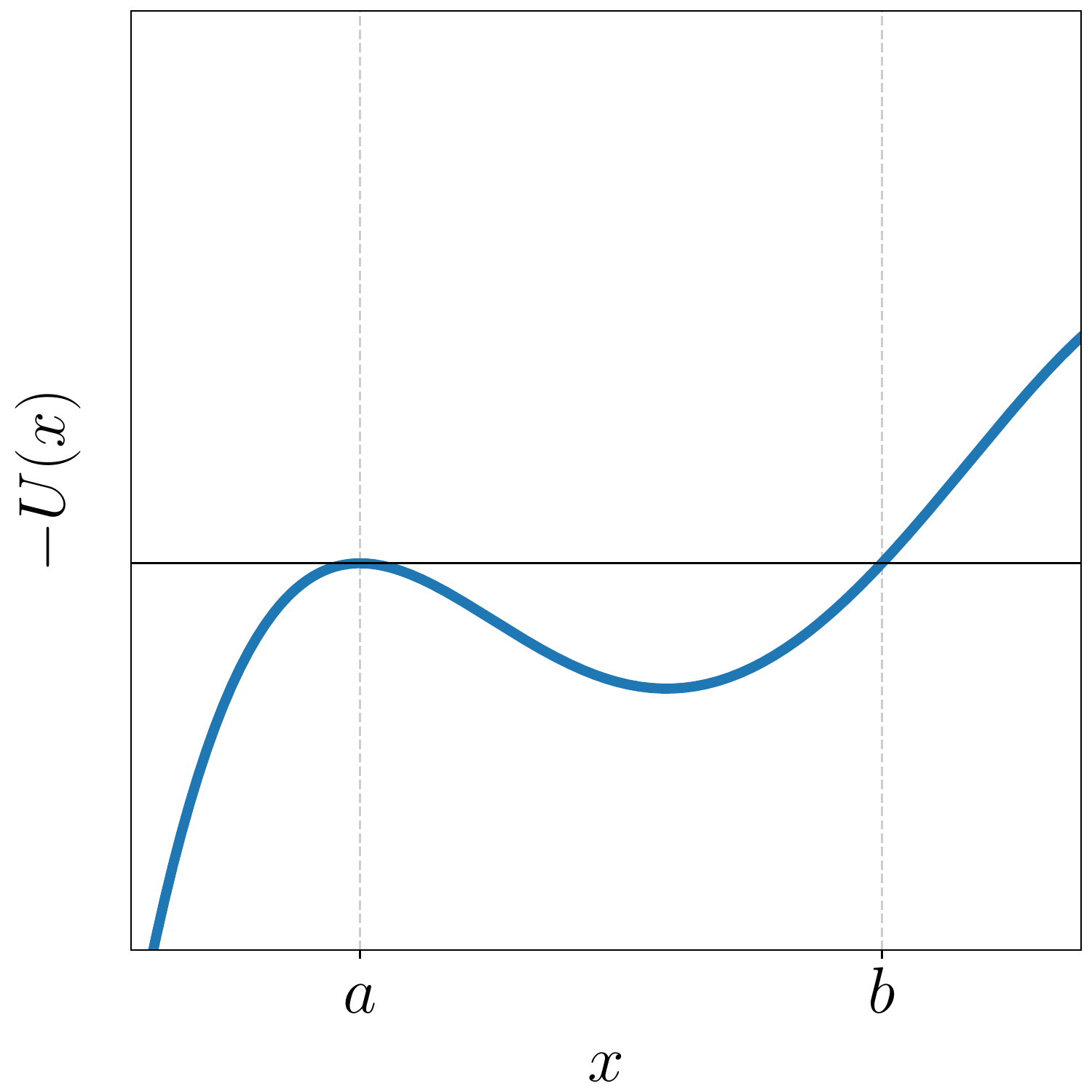}
\caption{
\label{qm_double_well3}
Left panel: potential of a metastable state. Right panel: the same potential after the change of sign.}
\end{figure}

The calculation is essentially the same as in the case of the double well,
with a few modifications. We consider the amplitude
\begin{equation}
\label{metastab}
Z(T)=\langle a|\e^{-\frac{HT}{\hbar}}|a\rangle
=\sum_n\e^{-\frac{E_nT}{\hbar}}|\langle a|n\rangle|^2,
\end{equation}
where $|n\rangle$ is a complete set of eigenstates of the Hamiltonian
with tunnelling neglected. We expect
$E_0$, the energy of the lowest-lying metastable state,
to acquire a negative imaginary part when tunnelling is turned on:
\begin{equation}
{\rm Im\,}E_0=-\frac{\Gamma}{2}
\end{equation}
related to the decay probability $\gamma$
of the metastable ground state per unit time
by $\Gamma=\hbar\gamma$.

The contribution to the amplitude
\eq{metastab} of the state
with minimum real part of $E_n$ is isolated by taking
the limit $T\to+\infty$. We find
\begin{equation}
E_0=-\hbar\lim_{T\to+\infty}\frac{1}{T}\log Z(T);
\qquad
\gamma=2{\rm Im\,}\lim_{T\to+\infty}\frac{1}{T}\log Z(T).
\label{decprob}
\end{equation}
As in the previous case, the amplitude $Z(T)$ is given by
\begin{equation}
\label{metastapi}
Z(T)=\int[dx]\,\e^{-\frac{S[x]}{\hbar}},
\end{equation}
and the functional integral can be computed by the saddle-point technique.

We therefore look for
stationary points of the action, i.e.\ solutions $\bar x(\tau)$ of the classical equation of motion,
with $\bar x(-T/2)=\bar x(T/2)=a$. One of them is the constant solution $x_0(\tau)=a$.
Other solutions with the same boundary conditions are sequences
of the so-called
{\it bounce} solution, namely, a solution $x_b(\tau)$ which starts at $x=a$
in the remote Euclidean past $\tau=-T/2$ with zero velocity,
reaches the point $x=b$ beyond the potential barrier with $U(b)=U(a)$
at time $\tau=0$, and bounces off to reach again $x=a$ at $\tau=T/2$.
The path integral is given by the sum of saddle-point contributions around all
multibounce solutions,
sequences of $n$ bounces located at different instants,
this time with no constraints on the parity of $n$, integrated
over all possible locations of the $n$ bounces, in analogy to the case of the double potential well.

The details of the bounce solution obviously depend on the specific form of
the potential energy
$U(x)$; however, as in the case of the instanton, its general features are independent
of the details. As in the case of the double-well,
the behaviour of the bounce solution is better understood if the 
Euclidean equation of motion
\begin{equation}
\frac{d^2 x}{d\tau^2}=U'(x)
\end{equation}
is viewed as the real-time equation of motion n a potential $-U(x)$ (see right panel of \fig{qm_double_well3}),
which has therefore a maximum at $x=a$ and a well for
$a<x<b$.
\begin{enumerate}
\item
The bounce $x_b(\tau)$ actually equals $a$
only in the limit $\tau\to -\infty$. Indeed, since the bounce energy is $U(a)$,
in the vicinity of $x=a$ we have
\begin{equation}
\frac{dx_b}{d\tau}= \sqrt{\frac{2[U(x_b)-U(a)]}{m}}\simeq\omega(x_b-a)
\end{equation}
with the solution
\begin{equation}
x_b(\tau)=a+C\e^{\omega\tau},
\end{equation}
which approaches $a$ at $\tau\to-\infty$.
\item For the same reason, the bounce has zero velocity at time $\tau=0$, when
it reaches the point $x=b$, where the potential energy has another zero:
\begin{equation}
\left.\frac{dx_b(\tau)}{d\tau}\right|_{\tau=0}=\sqrt{\frac{2[U(b)-U(a)]}{m}}=0.
\end{equation}
\item In the limit $T\to+\infty$, the bounce can reach $x=b$ at any time $\tau_1$.
An integration over all possible values of $\tau_1$ is therefore necessary,
in analogy with the integration over all possible instanton centers in the previous case.
\end{enumerate}
Using the results of the previous section, we get
\begin{align}
Z(T)&=\frac{N}{\sqrt{\Det S''[x_0]}}
e^{-\frac{S[x_0]}{\hbar}}\sum_{n=0}^\infty \frac{K^nT^n}{n!}
\e^{-\frac{n(S[x_b]-S[x_0])}{\hbar}}
\nonumber\\
&=\frac{N}{\sqrt{\Det S''[x_0]}}e^{-\frac{S[x_0]}{\hbar}}
\exp \left[KT\e^{-\frac{S[x_b]-S[x_0]}{\hbar}}\right].
\end{align}

We now turn to a careful evaluation of $K$. A naive expectation would be
an expression like \eq{kdef2}, with the instanton solution $x_1(\tau)$
replaced by the bounce $x_b(\tau)$. There is however an important difference:
because the bounce has zero velocity at $\tau=0$, the zero mode $\dot x_b(\tau)$ has a node, and therefore cannot be the eigenvector of $S''[x_b]$ with smallest eigenvalue (the eigenvalue equation for $S''[x_b]$ is formally the same as a time-independent Schr\"odinger equation in one dimension, and we can rely on well known results in that context.)
This in turn implies that $S''[x_b]$ has an eigenvector with negative eigenvalue,
and therefore its determinant (with the zero eigenvalue removed) is negative. This is expected, because as a consequence
$K$ is purely imaginary, as appropriate for a metastable state.
The contribution of the negative mode to the gaussian integral, however, requires some care.

Let us investigate in some more detail the origin of the negative eigenvalue of
$S''[x_b]$. The presence of a negative eigenvalue means that
$x_b$ is not a minimum of the Euclidean action in the space of paths, but rather
a saddle point: there is a direction in the space of configurations along which the action $S[x]$ 
has a maximum at $x(\tau)=x_b(\tau)$. It is not difficult to identify this direction. Let us
consider the value of the Euclidean action in correspondence of the sequence of
paths $x(\tau)$ with $x(\pm\infty)=a$ and a maximum at $\tau=0$, and let us
parametrise them by their
value at the maximum, $c=x(0)$. Members of this family are the constant solution,
$x_0(\tau)=a$,  for which $c=a$, and the bounce, for which $c=b$ (we assume $b>a$ for definiteness). Other values of
$c$ correspond to paths that either go back to $x=a$ without reaching the  classically
allowed region beyond the barrier (those with $c<b$) or go beyond the turning point,
$c>b$. The Eulidean action, restricted to this family of paths,
is an ordinary function $S(c)$ of $c$.
We have
\beq
S(a)=S[x_0]=U(a)T,
\eeq
which is a local minimum, because increasing $c$ with respect to
$a$ increases both terms in the action. The action increases monotonically until $c=b$
(the bounce) is reached; in $c=b$ the action has a stationary point,
\beq
\left.\frac{dS(c)}{dc}\right|_{c=b}=0,
\eeq
because the bounce is another solution of the equation of motion. If we now increase $c$ further, the action
starts decreasing down to $-\infty$, because the path spends more and more time in the region where $U(x)<U(a)$. Hence, the path with $c=b$ (the bounce) must be a maximum,
because there is no other stationary point. We conclude that the bounce is a minimum configuration
among those paths which turn back at $x=b$, but a maximum in the direction of paths
which turn back at different values (either larger or smaller than $b$). This clarifies the origin
of the negative mode.

Since, however, the action goes to $-\infty$ as $c$ increases beyond $c=b$, the contribution
of this direction to the gaussian integral is divergent. It must be defined by an analytic continuation from the case when $x=a$ is a global minimum, to the case of interest.
Such an analytic continuation can be performed restricting ourselves to
the direction in the configuration space which corresponds to a local maximum.
Following Ref.~\cite{Weinberg:1996kr}, we may cast this contribution in the form of an integral over the parameter $c$:
\begin{equation}
J=\frac{1}{\sqrt{2\pi\hbar}}\int_{-\infty}^{+\infty}dc\,\e^{-\frac{S(c)}{\hbar}}.
\end{equation}
As long as the potential energy has an absolute minimum in $x=a$, $J$ is convergent and real. If we now continuously deform
the potential energy so that a new, lower minimum appears for $x>b$, the integral
is finite and real between $-\infty$ and $b$, but diverges between $b$ and $+\infty$
because $S(c)\to-\infty$. We may regularize the integral by deforming the integration
path (the real $c$ axis) away from the real axis for ${\rm Re\,} c>b$. This gives
$J$ an imaginary part, which can be evaluated by the steepest descent method:
\begin{align}
{\rm Im\,}J&={\rm Im\,}\frac{1}{\sqrt{2\pi\hbar}}\int_b^{b+i\infty}dc\,\e^{-\frac{S(c)}{\hbar}}
=\frac{1}{\sqrt{2\pi\hbar}}\int_0^{+\infty}dc\,\e^{-\frac{S(b+ic)}{\hbar}}
\notag\\
&\simeq
\frac{\e^{-\frac{S(b)}{\hbar}}}{\sqrt{2\pi\hbar}}\int_0^{+\infty}dc\,\e^{\frac{S''(b)}{2\hbar}c^2}
=\frac{1}{2}\e^{-\frac{S(b)}{\hbar}} \frac{1}{\sqrt{|S''(b)|}}
\end{align}
where $S''(b)<0$. Note the factor of $\frac{1}{2}$, arising from the gaussian integration
over one half of the gaussian peak. Taking this into account, the value of $K$
in the case of an unstable minimum is given by
\begin{equation}
K=\frac{i}{2}\sqrt{\frac{S[x_b]-S[x_0]}{2\pi\hbar m}}
\frac{\sqrt{\Det S''[x_0]}}{\sqrt{|\Det'S''[x_b]|}}.
\label{kdeftunn}
\end{equation}
We are now in a position to use Eq.~(\ref{decprob}) to compute the decay probability
per unit time of the metastable ground state:
\begin{align}
\gamma&=2{\rm Im\,}\lim_{T\to+\infty}\frac{1}{T}\log Z(T)
\notag\\
&=2{\rm Im\,}\lim_{T\to+\infty}\left[
\frac{1}{T}
\log\left(\frac{N}{\sqrt{\det S''[x_0]}}e^{-\frac{S[x_0]}{\hbar}}\right)
+K\e^{-\frac{S[x_b]-S[x_0]}{\hbar}}\right]
\notag\\
&=
\sqrt{\frac{S[x_b]-S[x_0]}{2\pi\hbar m}}
\frac{\sqrt{\Det S''[x_0]}}{\sqrt{|\Det'S''[x_b]|}}\e^{-\frac{S[x_b]-S[x_0]}{\hbar}}
\notag\\
&\equiv Ae^{-B}.
\label{gammaqm}
\end{align}
Note that the prefactor $A$ has the correct dimension of an inverse time:
\begin{equation}
[A]={\rm mass}^{-\frac{1}{2}}\times \left(\frac{\rm mass}{{\rm time}^2}\right)^{\frac{1}{2}}
=\frac{1}{\rm time}.
\end{equation}
The exponent
\begin{equation}
B=\frac{S[x_b]-S[x_0]}{\hbar}
\end{equation}
in Eq.~\eqref{gammaqm} can be computed in terms of the potential energy
as in the case of the previous section, using
\begin{equation}
-\frac{1}{2}m\left(\frac{dx_b}{d\tau}\right)^2+U(x_b)=U(a).
\label{bounceenergy}
\end{equation}
We find
\begin{align}
S[x_0]&=\int_{-T/2}^{T/2}d\tau\,\left[\frac{1}{2}m\left(\frac{dx_0}{d\tau}\right)^2+U(x_0)\right]
=U(a)T
\\
S[x_b]-S[x_0]
&=\int_{-T/2}^{T/2}d\tau\,\left[\frac{1}{2}m\left(\frac{dx_b}{d\tau}\right)^2+U(x_b)-U(a)\right]
\notag\\
&=\int_{-T/2}^{T/2}d\tau\,2[U(x_b)-U(a)].
\label{bounceaction}
\end{align}
The integral in the last line of Eq.~(\ref{bounceaction}) can be turned to a space integral,
using Eq.~(\ref{bounceenergy}) again, to obtain
\begin{equation}
B
=\frac{2}{\hbar}\int_a^b dx\,\sqrt{2m[U(x)-U(a)]}.
\label{bounceaction2}
\end{equation}
Equations~(\ref{gammaqm},\ref{bounceaction2}) are the main result of this section.

As a useful consistency check, we may compute the real part of $E_0$ by the same technique; we expect
it to be $U(a)+\frac{\hbar\omega}{2}$, as appropriate for a potential well
with a minimum in $x=a$ and $U''(a)=m\omega^2$.
We find
\begin{align}
{\rm Re\,}E_0&=-\hbar{\rm Re\,}\lim_{T\to\infty}\frac{1}{T}\log Z(T)
\notag\\
&=-\hbar\lim_{T\to+\infty}
\frac{1}{T}\left[\log N-\frac{1}{2}\log\Det S''[x_0]-\frac{S[x_0]}{\hbar}\right].
\end{align}
The constant $N$ is $T$-independent, and disappears from this
expression in the large-$T$ limit, while $S[x_0]=U(a)T$. Finally,
$\Det S''[x_0]$ is proportional to $\phi_0(T)$, given in Eq.~(\ref{phi0}). Hence
\begin{align}
{\rm Re\,}E_0&=\U(a)+\frac{\hbar}{2}\lim_{T\to+\infty}\frac{1}{T}
\log\frac{\sinh\omega T}{\omega}
\notag\\
&=U(a)+\frac{\hbar\omega}{2}
\end{align}
as expected.

\subsection{Tunnelling with many degrees of freedom}
\label{subsec:Tunnelling in many dimensions}
The extension of the above results to quantum systems with more than one
degree of freedom was developed in Refs.~\cite{Banks:1973ps,Banks:1974ij}.
We consider a generic quantum system with $N$ degrees of freedom.
The relevant generalized coordinates are collected in an $N$-dimensional vector
$\v x$. The Lagrangian of the system is
\begin{equation}
L(\v x,\dot{\v x})
=\frac{1}{2}\left|\frac{d\v x(t)}{dt}\right|^2-U(\v x)
=\frac{1}{2}\left|\dot{\v x}(t)\right|^2-U(\v x)
\end{equation}
(the generalized coordinates $\v x$ are not necessarily cartesian
coordinates, hence the absence of a factor of $m$ in the kinetic
term.)

We assume that the potential energy $U(\v x)$ has a local minimum in
$\v x=\v a$, and that the system is initially in the ground state
of the potential well in $\v x=\v a$, whose energy $E$ differs
from $U(\v a)$ by an amount of order $\hbar$. By tunnell effect, the
system has a finite probability per unit time $\gamma$ to
penetrate the potential barrier, escape from the local minimum, and
emerge at some point $\v x=\v b$ on a surface $\Sigma$ in the
configuration space, defined implicitly by $U(\v x)=U(\v a)$ for all $\v x$ on
$\Sigma$. It is shown in Ref.~\cite{Banks:1973ps} that
\begin{equation}
\gamma=A\e^{-B},
\end{equation}
where the prefactor $A$ is not fixed in the leading semi-classical
approximation, while the exponent
$B$ can be written in analogy with the one-dimensional case:
\begin{equation}
B=\frac{2}{\hbar}\int_0^{s_f}ds\,\sqrt{2[U(\v x(s))-E]}
\label{Bmany}
\end{equation}
where $\v x(s)$ is a path in configuration space
which starts at the local minimum $\v x=\v a$ and ends at a point
$\v b$ on the surface $\Sigma$:
\begin{equation}
\v x(0)=\v a;\qquad \v x(s_f)=\v b,
\end{equation}
normalized so that
\begin{equation}
\left|\frac{d\v x(s)}{ds}\right|^2=
\sum_i\frac{dx_i(s)}{ds}\frac{dx_i(s)}{ds}=1.
\label{paramtext}
\end{equation}
The path $\v x(s)$, including its endpoint $\v b$, must be such
that the exponent $B$ is minimum, that is, tunnelling takes place
along the path (or paths) with maximum decay probability. The problem is
therefore reduced to the variational problem of finding those
particular paths.

An analogous problem arises in classical mechanics, when one is
interested in finding the trajectory of motion, with no reference to
the time dependence of the generalized coordinates. The argument is reviewed
in~\app{app:maup}; it is based on a modified version of Hamilton's
principle, sometimes called the Maupertuis principle. Here, we recall
the relevant result. The action of the system is considered as a function of the configuration
$\v x(t)$ at time $t$, and $t$ itself:
\begin{equation}
S(\v x(t),t)=\int_{t_0}^t L(\v x(t'),\dot{\v x}(t'))dt',
\end{equation}
where $\v x(t')$ is a solution of the equations of motion
with $\v x(t_0)=\v a$. If the energy $E$ of the system is conserved,
one finds
\begin{equation}
S(\v x(t),t)=S_0(\v x(t))-E(t-t_0),
\end{equation}
where
\begin{equation}
S_0(\v x)=\sum_i\int_{a_i}^{x_i}p_i(\v x')dx'_i
\end{equation}
and
\begin{equation}
p_i(\v x)=\frac{\partial L}{\partial x_i}
\end{equation}
is the conjugate momentum of $x_i$. It can be shown that
$S_0(\v x(t))$, sometimes called the reduced action,  is stationary upon
variations of $\v x(t')$ around a solution of the equations of motion
with $\v a$ and $\v x(t)$ kept fixed, but $t$ allowed to vary.
Furthermore, it can be shown that
\begin{equation}
S_0(\v x(t))=\int_0^{s_f}\sqrt{2[E-U(\v x(s))]}\,ds,
\label{S0mauptext}
\end{equation}
where $\v x(s)$ is
a parametrization of the physical trajectory normalized as in \eq{paramtext},
with $\v b=\v x(s_f)=\v x(t)$.

This variational problem differs from the one we would like to solve,
namely the minimization of the exponent $B$, \eq{Bmany}, in two
respects.  First, the quantities under square root in
Eqs.~(\ref{Bmany}) and (\ref{S0mauptext}) have opposite signs. This is
expected, since the physical solution of the equations of motion
refers to the classically allowed region $E\ge U(\v x)$, while the
exponent $B$ depends on a path which is defined in the classically
forbidden region $E\le U(\v x)$. Second, the minimization of the
exponent $B$ in \eq{Bmany} should be performed not only with respect
to the path $\v x(s)$, but also with respect to its endpoint $\v
b$, while the endpoint $\v x(t)$ is kept fixed in the minimization
of $S_0(\v x(t))$, \eq{S0mauptext}.

Leaving aside for a while the problem of the endpoint,
the two variational problems become exactly the same switching to
imaginary (or Euclidean) time
\begin{equation}
t=i\tau.
\end{equation}
The Lagrangian is transformed into
\begin{equation}
L(\v x,\dot{\v x})
=-\frac{1}{2}\sum_i\frac{dx_i}{d\tau}\frac{dx_i}{d\tau}-U(\v x)
=-\frac{1}{2}\left|\dot{\v x}(\tau)\right|^2-U(\v x).
\end{equation}
Correspondingly,
\begin{equation}
p_i=\frac{\partial L}{\partial\dot x_i}=-\dot x_i
\end{equation}
and therefore
\begin{equation}
H(\v p,\v x)=\sum_i p_i \dot x_i-L(\v x,\dot{\v x})
=-\frac{1}{2}\sum_i\frac{dx_i}{d\tau}\frac{dx_i}{d\tau}+U(\v x)=E.
\label{EuclideanE}
\end{equation}
The Euclidean action is defined by
\begin{equation}
S(\v x(\tau),\tau)=-\int_{\tau_0}^\tau L(\v x(\tau'),\dot{\v x}(\tau')).
d\tau'
\end{equation}

By the same procedure described in \app{app:maup} one finds
\begin{equation}
S(\v x(\tau),\tau)=S_0(\v x(\tau))+E(\tau-\tau_0)
\end{equation}
where $S_0(\v x(\tau))$ is the reduced action:
\begin{equation}
S_0(\v x(\tau))=-\sum_i\int_{a_i}^{x_i(\tau)}p_idx_i
=\int_0^{s_f}\sqrt{2[U(\v x(s))-E]}\,ds
=\frac{\hbar}{2}B.
\label{reducedaction}
\end{equation}
Hence, the paths which minimize the exponent $B$ are the solutions
of the equation of motion in Euclidean time~\cite{Coleman:1977py}
\begin{equation}
\frac{d^2x_i}{d\tau^2}=+\frac{\partial U(\v x)}{\partial x_i}
\label{fma}
\end{equation}
with $\v x(\tau_0)=\v a$ and $\v x(\tau)=\v b$, where $\v b$ is some point
such that $U(\v b)=U(\v a)$.

We now turn to the problem of minimizing the path with respect to its endpoint.
We note that the reduced action, and hence the exponent $B$,
is automatically minimized with respect to variations of the endpoint
$\v b=\v x(\tau)$
of the path $\v x(s)$. Indeed,
\begin{equation}
\frac{\partial}{\partial b_i}S_0(\v b)=
-\sqrt{2[U(\v b)-E]}\left.\frac{dx_i(s)}{ds}\right|_{s=s_f}
\end{equation}
which is zero for all points $\v b$ on the surface $\Sigma$.

We now observe that \eq{fma} can be interpreted as an ordinary equation of
motion (that is, in ordinary time) in a potential $\tilde U(\v x)=-U(\v x)$
which has a local maximum in $\v x=\v a$.
Of course, the explicit form of the solution of the equation of motion depends on the potential
$U(\v x)$. Some of its features can however be obtained
on general grounds:
\begin{enumerate}
\item The solution $\v x(\tau)$ can only start at $\v a$ at
  $\tau_0=-\frac{T}{2}\to-\infty$. Indeed, the assumption that $U(\v x)$ has
  a minimum in $\v x=\v a$ implies that
\begin{equation}
U(\v x)=E+\frac{1}{2}\sum_{i,j}U_{ij}(\v a)(x_i-a_i)(x_j-a_j)
+O(|\v x-\v a|^3),
\end{equation}
where the matrix $U_{ij}(\v a)$ has positive eigenvalues
$\omega^2_i$.  Hence, after a suitable rotation in the configuration
space, the equations of motion for $\v x$ close to $\v a$ read
\begin{equation}
\frac{d^2}{d\tau^2}(x_i-a_i)=\omega^2_i(x_i-a_i),
\end{equation}
which has the general solution
\begin{equation}
x_i(\tau)-a_i=A_i\e^{\omega_i\tau}+B_i\e^{-\omega_i\tau}.
\end{equation}
Hence, $x_i(\tau)\to 0$ for either $A_i=0,\tau\to+\infty$ or $B_i=0,\tau\to-\infty$.
Since $\tau_0$ is the initial time, the only possibility is
$\tau_0\to-\infty$.

\item Since $\tau_0=-\infty$, we may perform a time translation such
  that $\v x=\v b$ at $\tau=0$.
\item The solution reaches $\v x=\v b$ with zero velocity.
Indeed, from \eq{EuclideanE},
\begin{equation}
\left|\frac{d\v x(\tau)}{d\tau}\right|^2=2[U(\v x)-E],
\label{velocity}
\end{equation}
and $U(\v x)\to E$ for $\v x\to\v b$.
\end{enumerate}
The exponent $B$ can be related to the action of a solution of the
equation of motion computed over the whole Euclidean time range
$-\infty<\tau<+\infty$.  Indeed, by time reversal invariance, the
time-reversed configuration $\v x(-\tau)$ is also a solution of the
EoM, with the same action. Hence,
\begin{equation}
\v x_b(\tau)=\v x(\tau)+\v x(-\tau)
\end{equation}
is a solution with
\beq
\v x_b(\pm\infty)=\v a;\qquad \v x_b(0)=\v b;\qquad \dot{\v x}_b(0)=0
\eeq
and Euclidean action
\begin{equation}
S[x_b]
=2S[x]=2S_0[x]+ET.
\end{equation}
In analogy with the one-dimensional case, such a solution is usually called a bounce,
because it starts in the remote past $\tau=-\frac{T}{2}$ at $\v x_b=\v a$, reaches
$\v x_b=\v b$ at $\tau=0$, and bounces back to $\v x_b=\v a$ at
$\tau=+\frac{T}{2}$.

The constant solution
\begin{equation}
\v x_0(\tau)=\v a
\end{equation}
is also a solution of the equations of motion, with the same boundary conditions.
The corresponding Euclidean action is
\begin{equation}
S[x_0]=ET
\end{equation}
Hence
\begin{equation}
\lim_{T\to +\infty}\left(S[x_B]-S[x_0]\right)=\lim_{T\to+\infty}2S_0[x]=\hbar B.
\label{Sbounce}
\end{equation}
Alternatively, one may choose the zero of the potential energy such
that $U(\v a)=E=0$; in this case, $S=S_0$ for all solutions with
zero energy, and $S[x_B]=\hbar B$, which is consistent with \eq{Sbounce}
since $S[x_0]=0$ in this case.

\section{Tunnelling in quantum field theory}
\label{tunnellingQFT}

After reviewing barrier penetration in ordinary quantum mechanics, we now consider its generalization to
quantum field theory.
We will follow closely the approach of Ref.~\cite{Coleman:1977py}. We will start by simply translating the results obtained in the previous section in the language of quantum field theory, and we will then focus on the physical interpretation of tunnelling processes. Specifically, we will pay special attention to the generalization of the concept of a potential barrier in quantum field theory. Unless explicitly stated, we will adopt the natural unit convention $\hbar=c=1$ from now on.

\subsection{From quantum mechanics to quantum field theory}
\label{sec:QFT}

We consider the quantum theory of a real scalar field $\phi$,
characterized by a scalar potential $V(\phi)$ with a local minimum in $\phi=\phi_{\rm FV}$
(the false vacuum) and a deeper, absolute minimum in $\phi=\phi_{\rm TV}$ (the true vacuum)
as shown in the left panel of~\fig{barrier_qft_1}. In the example of \fig{barrier_qft_1} the scalar potential is defined so that $V(\phi_{\rm FV})=0$; this is a convenient choice in many respects, but we will keep $V(\phi_{\rm FV})\ne 0$ in our discussion for greater generality.
\begin{figure}[ht]
  \centering
\includegraphics[width=0.45\textwidth]{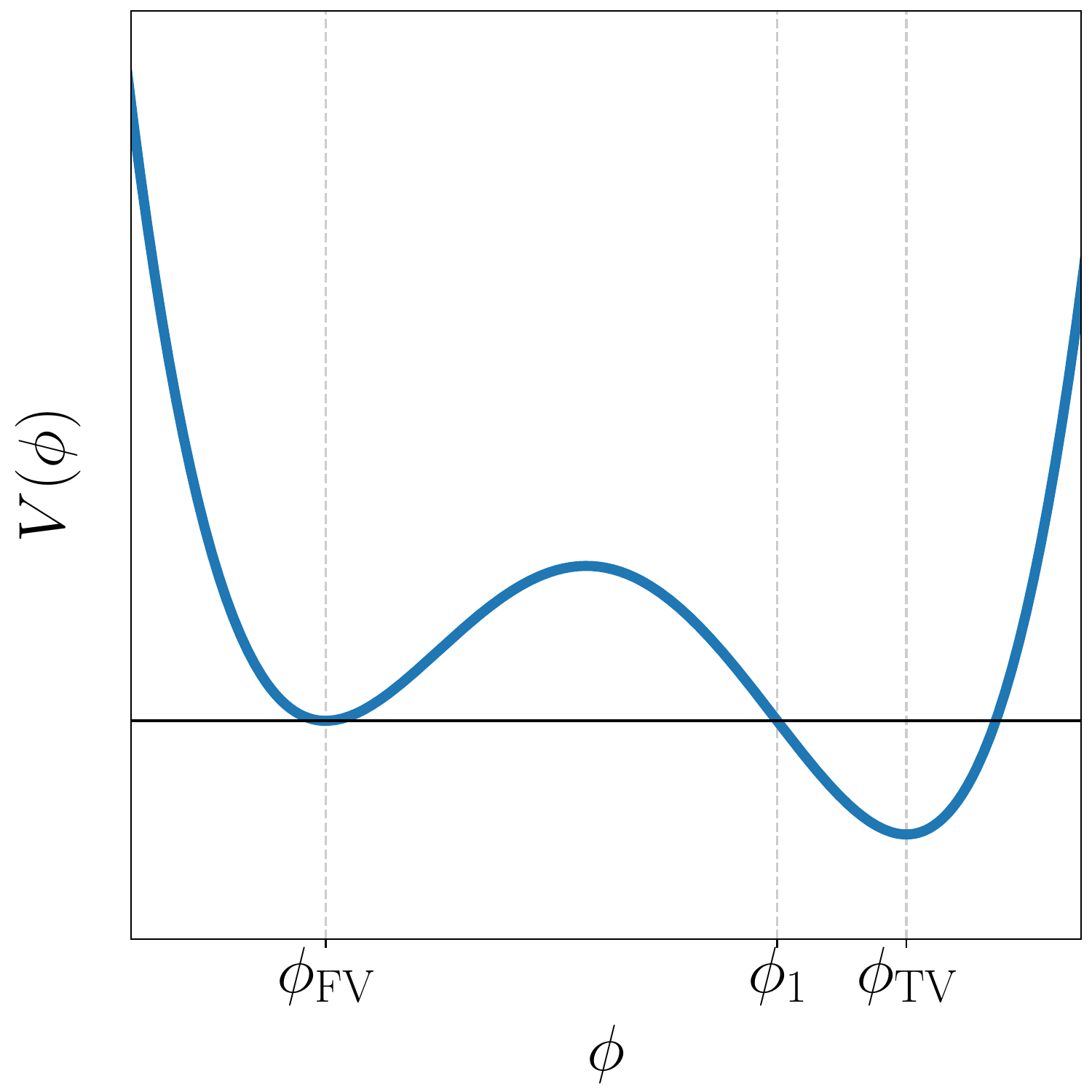} \qquad
\includegraphics[width=0.45\textwidth]{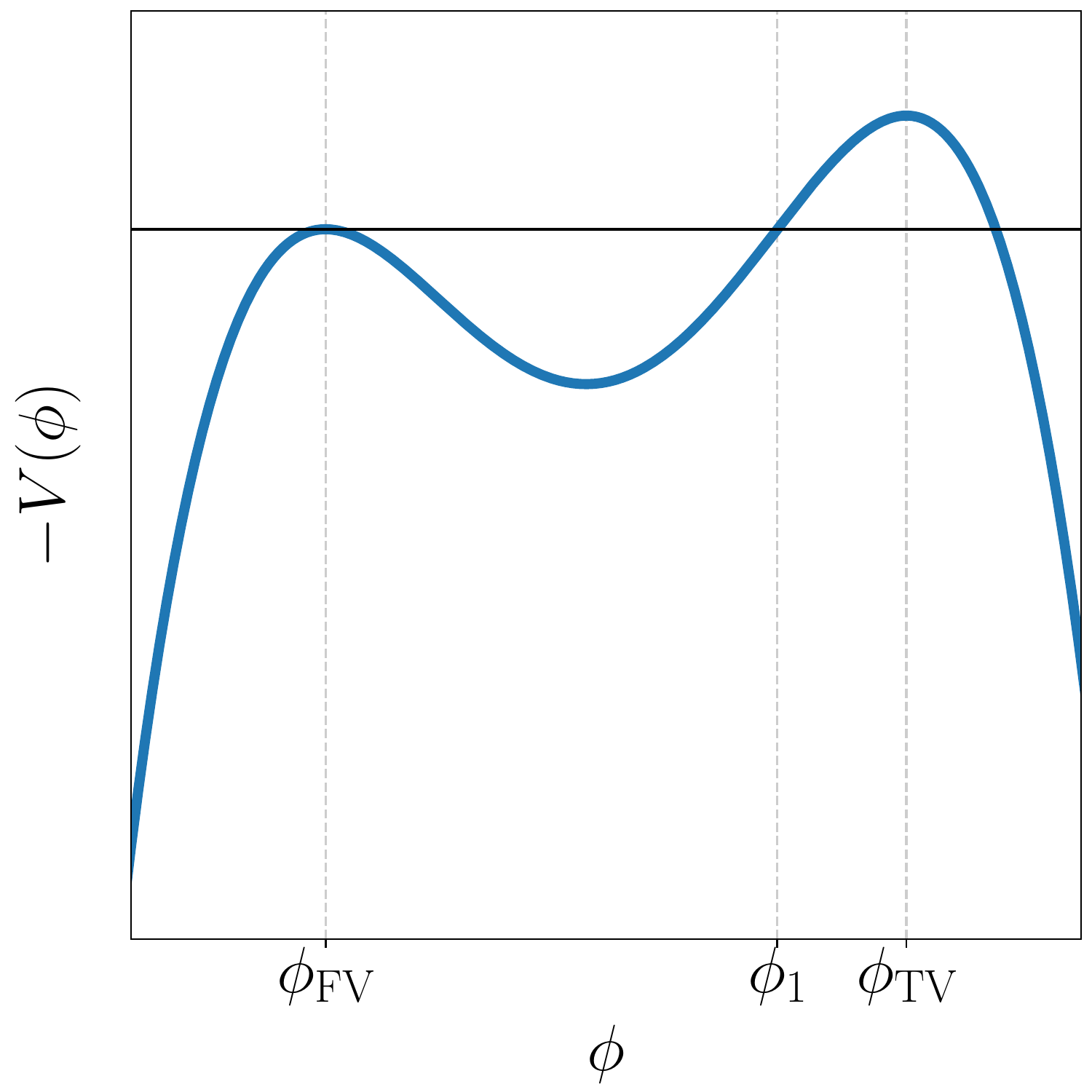}
\caption{
\label{barrier_qft_1}
Left panel: asymmetric double-well potential with a local minimum in $\phi_{\rm FV}$ and a global one in $\phi_{\rm TV}$. Right panel: the same potential after the change of sign. $\phi_1$ labels the turning point.}
\end{figure}

Our goal is the computation of the probability
per unit time $\gamma$ that the system, initially in the false vacuum configuration $\phi_{\rm FV}$,
decays to the true vacuum configuration $\phi_{\rm TV}$. To this purpose, we consider our field theory as an ordinary quantum theory with an infinite number of degrees of freedom, labelled
by points in three-dimensional space: $x_i(\tau)\rightarrow\phi_{\v x}(\tau)=\phi(\tau,{\v x})$. The formalism of \sect{subsec:Tunnelling in many dimensions} is readily generalized,
to conclude that
\begin{equation}
  \label{eq: gamma}
\gamma=A\e^{-B},
\end{equation}
where the overall factor $A$ is undetermined in the leading semi-classical approximation, while the exponent $B$ can be computed by a generalization of the argument presented in the previous section. In particular, we need bounce solutions $\phi_b(\tau,\v x)$ to the Euclidean
field equation
\begin{equation}
  \left(\frac {\partial^2}{\partial \tau^2}+\bm \nabla^2\right)\phi_b(\tau,\v x)=V'(\phi_b),
  \label{eq:qft eom}
\end{equation}
with the boundary conditions
\begin{align}
  \label{eq:qft boundary1}
  &\lim_{\tau\to\pm\infty}\phi_b(\tau,\v x)=\phi_{\rm FV}
  \\
  \label{eq:qft boundary2}
  &\left.\frac{\partial\phi_b(\tau,\v x)}{\partial \tau}\right|_{\tau=0}=0.
\end{align}
The exponent $B$ is then given by
\begin{equation}
B=S[\phi_b]-S[\phi_{\rm FV}],
\label{eq:QFTB}
\end{equation}
where $S[\phi]$ is the Euclidean action of the theory,
\beq
S[\phi]=\int d^4x\,
\left[\frac{1}{2}(\partial_\mu\phi)^2+V(\phi)\right].
\eeq

By construction, the bounce solution $\phi_b$ approaches $\phi_{\rm FV}$, and hence $V(\phi_b)\to V(\phi_{\rm FV})$,
at infinite Euclidean time, see \eq{eq:qft boundary1}.
By inspection of \eq{eq:QFTB} we conclude that $\phi_b$ must approach $\phi_{\rm FV}$
also at space infinity, in order to keep the exponent $B$ finite:
\begin{equation}
  \lim_{|\vec x|\to\infty}\phi_b(\tau,\v x)=\phi_{\rm FV}.
  \label{eq:qft boundary3}
\end{equation}

The exponent $B$ can be written in a more compact form by using the virial theorem.
Since the action, and hence $B$, is stationary in correspondence of solutions of the field equations, it will be stationary, in particular, upon variations of the following type:
\begin{equation}
  \phi_b(x)\rightarrow \phi_b(ax).
  \label{virialtr}
\end{equation}
After such transformations,
\begin{align}
B\rightarrow B_a&=
  \int d^4x\left[\frac 12 \frac{\partial \phi_b(ax)}{\partial x_\mu}\frac{\partial \phi_b(ax)}{\partial x_\mu}+V(\phi_b(ax))-V(\phi_{\rm FV})\right]
  \notag\\
  &=\frac 1{a^4}\int d^4y\,\left[\frac {a^2}2 \frac{\partial \phi_b(y)}{\partial y_\mu}\frac{\partial \phi_b(y)}{\partial y_\mu}+V(\phi_b(y))-V(\phi_{\rm FV})\right],
\end{align}
with the change of integration variable $y=ax$. The stationarity condition
\begin{equation}
\left.  \frac {dB_a}{da}\right|_{a=1}=0
\end{equation}
provides a relationship between the potential and the derivative terms in $B$,
when computed at stationary points:
\begin{equation}
  \int d^4 x \,\frac 12  \left(\partial_\mu \phi_b\right)^2=-2\int d^4 x\,
\left[V(\phi_b)-V(\phi_{\rm FV})\right]
\label{viriale}
\end{equation}
and therefore
\begin{equation}
  \label{eq:flatvirial}
B=-\int d^4 x \,\left[V(\phi_b)-V(\phi_{\rm FV})\right].
\end{equation}

Before looking for an explicit bounce solution of the Euclidean field equation, it will be useful to illustrate its physical meaning. In ordinary quantum mechanics, the system initially occupies the false ground state; after an average time $T=1/\gamma$, the system simply makes a quantum jump and appears at the escape point beyond the barrier with zero kinetic energy (that is, the bounce configuration at time $\tau=0$). Afterwards, it evolves classically.

A similar picture holds in quantum field theory: at $\tau=0$ the field makes a quantum jump from the state in which it is
uniformly equal to its false vacuum configuration, to the one described by the bounce solution, then it evolves classically. Therefore, the tunnelling process does not take place
between two spatially uniform configurations, but rather from a spatially uniform one
(the false vacuum) to a space-dependent one (the bounce), which, as we will see, has the
shape of a bubble of true vacuum $\phi_b=\phi_{\rm TV}$ of finite radius, in a false vacuum
$\phi_b=\phi_ {\rm FV}$ background. Indeed, as we will show in more detail in
\sect{barrierQFT}, the tunnelling between two spatially homogeneous configurations would require  penetration through an infinitely high potential energy barrier; the amplitude for such a process is zero. Instead we have a non-zero amplitude for the tunnelling between a spatially homogeneous configuration and one with a region of approximate true vacuum surrounded by a false vacuum background. This configuration then evolves classically, and eventually converts the false vacuum in the true vacuum everywhere in space.
Since the bubble of true vacuum can appear anywhere, we expect the decay rate to be proportional to the volume of three-dimensional space $\mathcal V_3$.

The result, as described in Ref.~\cite{Coleman:1977py}, is closely similar to the boiling
processes in a superheated fluid: the false vacuum corresponds to the superheated
fluid state, while the true vacuum to the vapor state. Due to thermodynamical fluctuations,
bubbles of vapor state appear at different points: when a bubble is large enough that the vapor pressure inside the bubble exceeds the external pressure plus the contribution of the surface tension, the bubble grows until it converts all the fluid into vapor, otherwise it shrinks back.
By substituting thermodynamical fluctuations with quantum ones, this picture describes the decay of the false vacuum: once a bubble of true vacuum appears, large enough that its growth is energetically favourable, then it expands throughout the Universe, thereby achieving the transition to the true vacuum.

Let us go back to the problem of searching for a bounce solution of \eq{eq:qft eom}.
It was shown in Ref.~\cite{Coleman:1977th} that, if the field equation has bounce solutions,
there always exists one bounce which is $\O(4)$ invariant, i.e.\ a function of
\beq
r=\sqrt{\tau^2+\left|\v x\right|^2}
\eeq
alone, whose Euclidean action is smaller than that of any $\O(4)$ non-invariant bounce. We may therefore restrict our search to
bounce solutions with $\O(4)$ invariance, $\phi_b(r)$.
This simplifies considerably our task. Indeed, the field equation (\ref{eq:qft eom})  becomes
an ordinary differential equation:
\begin{equation}
  \frac{d^2 \phi_b}{dr^2}+\frac 3r \frac{d\phi_b}{dr}=V'(\phi_b),
  \label{eq:qfteomrho}
\end{equation}
while the two boundary conditions \eq{eq:qft boundary1} and \eq{eq:qft boundary3}
combine into a single one:
\begin{equation}
  \lim_{r\to\infty} \phi_b(r)=\phi_{\rm FV}.
  \label{eq:qft boundary1 rho}
\end{equation}
Requiring that the solution is regular at $\v x=0$ we get the further boundary condition
\begin{equation}
\left.  \frac{d\phi_b(r)}{dr}\right|_{r=0}=0.
  \label{eq:qft boundary2 rho}
\end{equation}

\subsection{Existence of a bounce solution}
\label{overunder}
We now prove that a bounce solution with the correct boundary conditions actually exists in the case of an asymmetric double-well scalar potential $V(\phi)$ with a local minimum in $\phi=\phi_{\rm FV}$ and a global minimum in $\phi=\phi_{\rm TV}$. The proof was originally presented in Ref.~\cite{Coleman:1977py}; here we reproduce the same argument almost word by word. 

It is useful to view the field $\phi(r)$ as the position of a classical particle of unit mass in one dimension, as a function of the time $r$.
Then the field equation,
\eq{eq:qfteomrho}, can be regarded as the classical equation of motion of such a particle
in a potential $-V$, which is further subject to a damping force with a time-dependent coefficient which decreases as $1/r$. The potential $-V$ is shown
in the right panel of \fig{barrier_qft_1}.
We are looking for a solution  $\phi_b(r)$ that starts at rest at
$r=0$ from some initial value 
$\phi_b(0)$ and reaches $\phi_b=\phi_{\rm FV}$ with zero velocity at $r\to \infty$.
We will demonstrate that such a solution exists by showing that if the initial position is too close to the left of $\phi_{\rm TV}$ then the particle passes through $\phi_{\rm FV}$ with non-zero velocity (the particle {overshoots}); if instead the initial position is too far to the left of from $\phi_{\rm TV}$, the particle does not reach $\phi_{\rm FV}$ (the particle {undershoots}). As a consequence, there must be an intermediate initial position such that the particle comes at rest exactly at $\phi_{\rm FV}$.

It is easy to demonstrate the undershoot: if the initial condition $\phi(0)$ is smaller than $\phi_1$, then the particle's energy is smaller than the potential energy at $\phi=\phi_{\rm FV}$, which therefore cannot be reached. The presence of
the viscous term just makes things worse by further lowering the energy; indeed, the energy $E$ is not conserved because of the viscous force, and
\beq
\frac{dE}{dr}=\frac{d}{dr}\left[\frac{1}{2}\left(\frac{d\phi}{dr}\right)^2-V(\phi)\right]=
-\frac{3}{r}\left(\frac{d\phi}{dr}\right)^2<0
\eeq
by the field equation.

The overshoot would also be obvious if we could neglect the damping force:
the energy of a particle starting at rest with $\phi_1<\phi(0)<\phi_{\rm TV}$ would be larger than the potential
energy at $\phi_{\rm FV}$, and would therefore reach $\phi_{\rm FV}$ with non-zero
kinetic energy. The effect of the viscous term can be taken into account as follows. Let us take $\phi(0)$ to be close to $\phi_{\rm TV}$.
For $\phi$ close to $\phi_{\rm TV}$ we may expand the potential in powers
of $\phi-\phi_{\rm TV}$ up to second order:
\beq
V(\phi)=V(\phi_{\rm TV})+\frac{1}{2}\mu^2(\phi-\phi_{\rm TV})^2+O((\phi-\phi_{\rm TV})^3),
\label{eq:noviscous eom}
\eeq
where $\mu^2=V''(\phi_{\rm TV})>0$.
In this limit, \eq{eq:qfteomrho} takes the form
\begin{equation}
\left(\frac{d^2}{dr^2}+\frac{3}{r}\frac{d}{dr}-\mu^2\right)\left[\phi(r) -\phi_{\rm TV}\right]=0.
\end{equation}
It is easy to check that the solution of this equation is
\begin{equation}
\phi(r)-\phi_{\rm TV}=2\left[\phi(0) -\phi_{\rm TV}\right]\frac{I_1(\mu r)}{\mu r}
\end{equation}
and $I_1(x)=-iJ_1(ix)$, where $J_1(x)$ is the Bessel function of order 1,
solution of the differential equation
\beq
J_1''(x)+\frac{J_1'(x)}{x}+\left(1-\frac{1}{x^2}\right)J_1(x)=0.
\eeq
We see that, provided $\phi(0)$ is sufficiently close to $\phi_{\rm TV}$, the particle will spend
an arbitrarily large amount of time in the vicinity of
$\phi_{\rm TV}$.\footnote{Note that the same conclusion holds even in the absence of the damping force; in such case we would have found
$\phi(r)-\phi_{\rm TV}=\left[\phi(0)-\phi_{\rm TV}\right]\cosh \mu r$.}
But after an arbitrarily large time $r$ the viscous term, which decreases as $r^{-1}$,
becomes negligible, and in the absence of a damping force the particle overshoots.
The announced result is therefore proved.

As an example, we have employed the overshoot-undershoot technique to find the bounce solution in the case of the scalar potential
\begin{align}
&V(\phi)=g\left[\frac{1}{4}\phi^4
-\frac{1}{3}\phi^3(M+\phi_{\rm TV})
+\frac{1}{2}\phi^2M\phi_{\rm TV}
\right]
\label{V432add}
\\
&g>0;\qquad M=0.3\, \phi_{\rm TV},
\end{align}
which has a local minimum for $\phi=0$ and an absolute minimum for $\phi=\phi_{\rm TV}$. In fig.~\ref{fig:ou} we plot $h(x)=\frac{\phi(r)}{\phi_{\rm TV}}$
as a function of $x=\sqrt{g}r\phi_{\rm TV}$ for three different choices of $\phi(0)$. The first case is an undershoot: $\phi(0)$ is too much smaller than $\phi_{\rm TV}$,
the particle gets close the false vacuum but then turns back and oscillates around $\phi=M$. Overshoot is shown in the second panel: the particle has too much energy to come at rest at $\phi=0$; it reaches the false vacuum with positive velocity, and goes on to negative values of the coordinate. In the third panel, $\phi(0)$ has the right value to reach the false-vacuum value $\phi=0$ at large $r$ with zero velocity.
\begin{figure}[h]
\centering
	{\includegraphics[width=.45\textwidth]{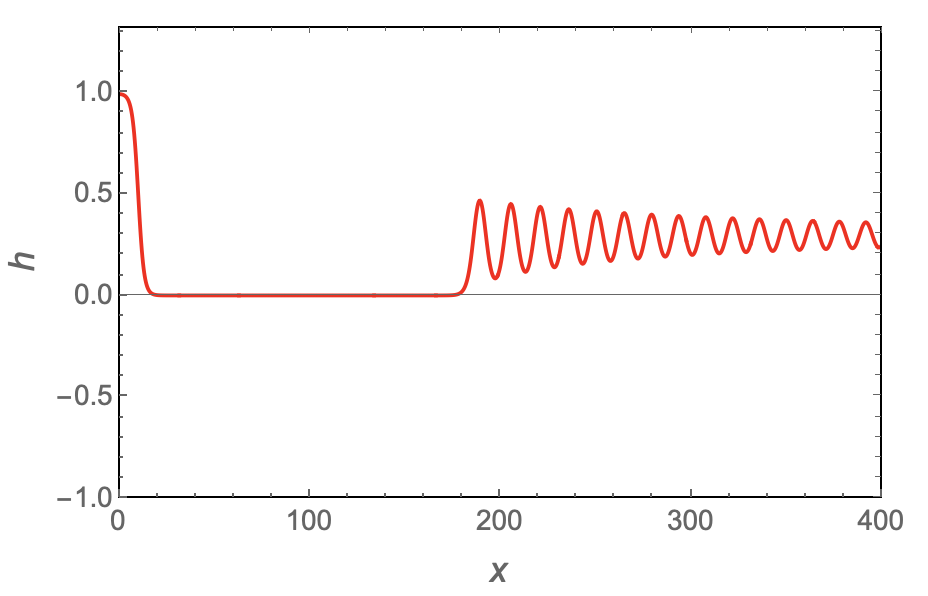}} \quad 
		{\includegraphics[width=.45\textwidth]{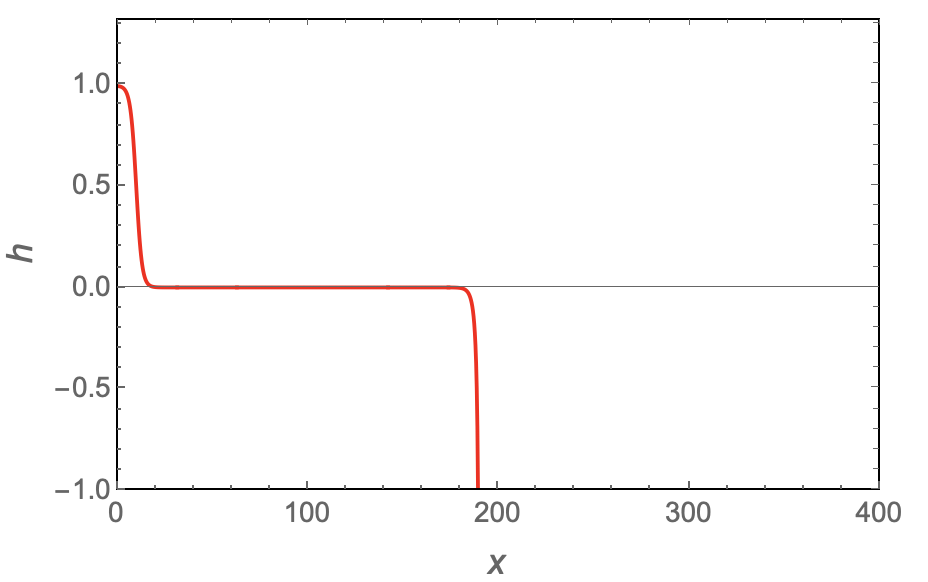}}\quad
		{\includegraphics[width=.45\textwidth]{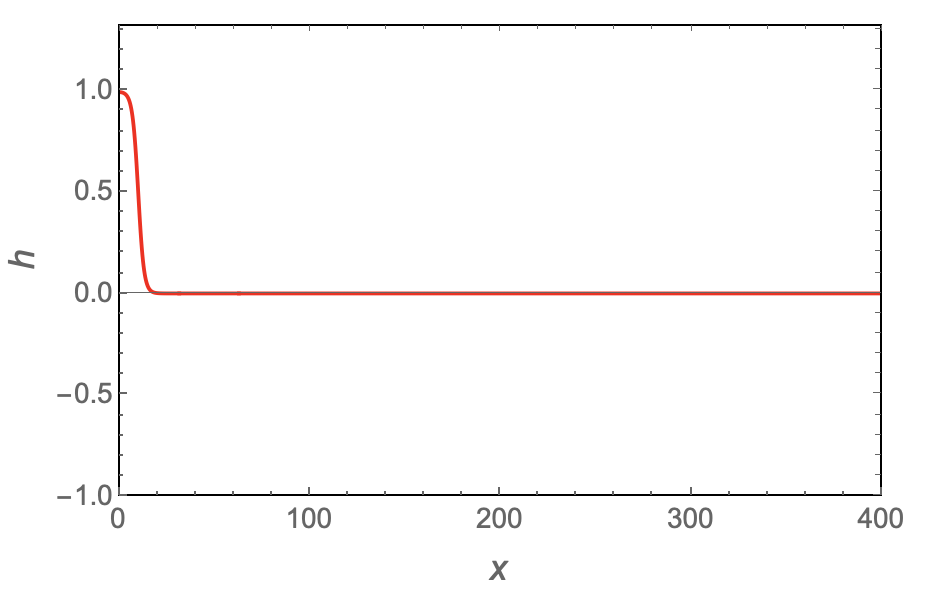}}
\caption{An example of the overshoot-undershoot mechanism.}
\label{fig:ou}
\end{figure}

The case when the two minima are nearly degenerate (in the above example, the case when $M$ is close to $\phi_{\rm TV}/2$) is especially interesting. This case corresponds to the so-called thin-wall approximation: the bounce is essentially a constant, equal to the true vacuum  field value, for $r$ smaller than some value $R$, which is a function of the parameters in the theory. Then, within a small region around $r=R$ the bounce drops rapidly to the false vacuum field value. To an excellent approximation, the bounce is therefore a four-dimensional sphere of radius $R$ where the field takes the true-vacuum value, while $\phi=\phi_{\rm FV}$ in the rest of Euclidean space-time. In the thin-wall approximation, the calculation of the tunnelling rate can be carried on analytically; the leading semi-classical tunnelling rate (that is, the exponent $B$) was computed in the original Coleman's paper~\cite{Coleman:1977py}, while a closed-form calculation at one loop in this limit has been presented recently~\cite{Ivanov:2022osf}.

To summarize, the decay probability
per unit time of a metastable state in the leading semi-classical approximation
is given by
\begin{equation}
\gamma=Ae^{-B}
  \label{eq:gamma qft}
\end{equation}
where
\begin{align}
B&=S[\phi_b]-S[\phi_{\rm FV}]
\notag\\
&=\int d^4x\,\left[\frac 12\partial _\mu\phi_b\partial_\mu \phi_b+V(\phi_b)-V(\phi_{\rm FV})\right]
\nonumber\\
&=-\int d^4x\,[V(\phi_b)-V(\phi_{\rm FV})]
\label{SB}
\end{align}
where $\phi_b(r)$ is the $O(4)$-invariant bounce.
The overall factor $A$ is undetermined in the leading semi-classical limit; we
expect it to be proportional to the three-dimensional volume ${\mathcal V}_3$
as a consequence of integration over the position of the true-vacuum bubble.
This expectation will be confirmed by the inclusion of the first quantum corrections.

\subsection{Time evolution of the true-vacuum bubble}
\label{timeev}

The decay of the false vacuum takes place as illustrated in \sect{sec:QFT}:
at some Euclidean time, say $\tau=0$, and at some point $\v x$ of three-dimensional space, an $O(4)$-symmetric bubble of true vacuum is formed in the false-vacuum background by quantum fluctuations:
\begin{equation}
\phi_b(|\v x|)=\phi_{\rm TV};\qquad
\left.\frac{\partial}{\partial\tau}\phi_b\left(\sqrt{|\v x|^2+\tau^2}\right)\right|_{\tau=0}=0;\qquad
\lim_{\tau\to\infty}\phi_b\left(\sqrt{|\v x|^2+\tau^2}\right)=\phi_{\rm FV}.
\end{equation}
Then, the bubble evolves in real (i.e.\ Minkowski space) time according to the classical equations of motion. The qualitative behavior of the bubble is easy to obtain by analytical continuation of the bounce to Minkowski space:
\beq
\phi_b(\sqrt{|\v x|^2+\tau^2}) \ \to \ \phi_b(\sqrt{|\v x|^2-t^2}).
\eeq
To visualize the time evolution of the bubble, we may study the time dependence of
the three-dimensional surface in space-time separating the true vacuum region from the false vacuum background, which in the thin-wall limit is the hyperbole
\beq
|\v x|^2-t^2=R^2.
\eeq
We find that the bubble expands in three-dimensional space with velocity
\beq
v=\frac{1}{\sqrt{1+\frac{R^2}{t^2}}},
\eeq
which approches the velocity of light as $t\to\infty$ quite rapidly: for example, we shall see that
$R\sim 10^{-16}\,{\rm GeV}^{-1}\sim 10^{-40}\,{\rm s}$ in the case of the standard model. 

\subsection{Potential barriers in quantum field theory}
\label{barrierQFT}

This section is devoted to a discussion of a feature of tunnelling phenomena in quantum field theory which may appear as counter-intuitive, and deserves some comments. In some cases, tunnelling from a metastable ground state to a lower minimum configuration takes place
even if the scalar potential of the theory, considered as an ordinary function of the scalar fields,
has no barrier in the usual sense.

This apparent paradox can be explained by observing that the tunnelling process in quantum field theory does not take place between two
spatially homogeneous field configurations, but rather through the formation of a bubble of true vacuum in a false-vacuum background.
Such field configuration (the bounce) is therefore not spatially homogenous, and the gradient term in the potential energy gives a non-zero contribution to the full potential energy.

In order to identify a quantity with the same properties as a potential barrier
in ordinary quantum mechanics, it will be useful to cast the exponent in the semi-classical decay rate
\begin{equation}
B=S[\phi_b]=\int d^4x\,\left[\frac 12 \left(\partial_\mu\phi_b\right)^2+V(\phi_b)\right]
\label{eq:compare qft}
\end{equation}
(we are now taking $V(\phi_{\rm FV})=0$)
in a form similar to the WKB formula for the exponential factor in quantum mechanics, \eq{reducedaction},
\begin{equation}
  B_{\rm QM}=2\int_0^{s_f} ds\,\sqrt{2U(\v x(s))}
  \label{eq:compare wkb}
\end{equation}
where $U(\v x)$ is the potential energy and the integral is taken over a
a suitable path $\v x=\v x(s)$ in configuration space,
which
connects the local minimum to a point beyond the potential barrier. This section is devoted
to the identification of such a quantity, following an argument originally presented by 
K.M.~Lee and E.J.~Weinberg in Ref.~\cite{Lee:1985uv}.

Our starting point is the Lagrangian density of a scalar theory in Eucidean time
\beq
 {\mathcal L}=-\frac 12 \left(\frac{\partial\phi}{\partial\tau}\right)^2-\frac 12 \left(\bm\nabla\phi\right)^2-V(\phi)
\eeq
and the corresponding Hamiltonian density
\beq
 {\mathcal H}=-\frac 12 \left(\frac {\partial\phi}{\partial\tau}\right)^2
 + \frac 12 \left(\bm\nabla\phi\right)^2+V(\phi).
    \label{eq:hamiltonian}
\eeq
We assume that the total energy
\beq
E[\phi]=\int d^3x\,{\mathcal H}
\eeq
has a local minimum
(the false vacuum) for a constant and uniform configuration
of the field $\phi=\phi_{\rm FV}$.
We define the potential energy $U[\phi]$ as the term in the total energy that does not contain
time derivatives of the field:
\begin{equation}
  U[\phi]=\int d^3x\,\left[\frac 12 \left(\bm\nabla\phi\right)^2+V(\phi)\right],
  \label{eq:U(phi)}
\end{equation}
which is a functional of the field $\phi$ and an ordinary function of the Euclidean time $\tau$. Then the Euclidean action can be written as
\beq
    S[\phi]=-\int d^3x\,d\tau\,\mathcal L
   =\int d^3x\,\left\{\int_{-\infty}^\infty d\tau\,\frac 12 \left(\frac{\partial\phi}{\partial\tau}\right)^2+\int_{-\infty}^\infty d\tau\,U[\phi]\right\}.
\eeq
For $\phi=\phi_b$, this formula can be simplified by energy conservation: the total energy of the system in the false vacuum configuration must be the same as the energy in the bounce configuration:
\begin{equation}
  E[\phi_{\rm FV}]=E[\phi_b],
\end{equation}
which is zero by our choice of the additive constant in the scalar potential. Hence
\begin{equation}
  E[\phi_b]=-\frac 12\int d^3x\, \left(\frac{\partial\phi_b}{\partial\tau}\right)^2+U[\phi_b]=0
\end{equation}
which gives
\begin{equation}
  U[\phi_b]=\frac 12\int d^3x\, \left(\frac{\partial\phi_b}{\partial\tau}\right)^2.
\label{eq:energyqft}
\end{equation}
Thus, 
\begin{equation}
S[\phi_b]=\int_{-\infty}^\infty d\tau\, 2U[\phi_b].
  \label{eq:actiontau}
\end{equation}
To complete the analogy with \eq{eq:compare wkb}, 
we perform a change of integration variable in \eq{eq:actiontau}
from $\tau$ to $s$, defined by
\begin{equation}
  ds^2=\int d^3x\, \left[d\phi_b(x)\right]^2,
  \label{eq:dsqft}
\end{equation}
in analogy with the infinitesimal path  length defined
in the case of quantum mechanics with many degrees of freedom,
\begin{equation}
\sum_i\frac{dx_i}{ds}\frac{dx_i}{ds}=1;\qquad  ds^2=\sum_i dx_i^2.
\end{equation}
We have therefore
\begin{equation}
   \left(\frac{ds}{d\tau}\right)^2=\int d^3x\, \left(\frac{\partial \phi_b}{\partial\tau}\right)^2=2U[\phi_b],
\end{equation}
where in the second step we have used \eq{eq:energyqft}.
This allows us to write \eq{eq:actiontau} as
\begin{equation}
   S[\phi_b]=\int_{-\infty}^\infty d\tau\, 2U[\phi_b]
   =2\int_{0}^\infty d\tau\, 2U[\phi_b]
   =2\int_\gamma ds \,\frac{d\tau}{ds} 2U[\phi_b]
   =2\int_\gamma ds\, \sqrt{2U[\phi_b]},
   \label{eq:stau}
\end{equation}
where $\gamma$ is the path in configuration space.
We have therefore achieved our goal: we have written the Euclidean action as a line integral
over a path in configuration space. Correspondingly, $U[\phi_b]$ can be considered as the quantum field theory
generalization of the potential barrier of ordinary quantum mechanics, as one can check by comparison with \eq{eq:compare wkb}.
It is important to emphasize that the integral is independent of the parametrization chosen for the path, which is the one which minimizes the Euclidean action.

The potential energy $U[\phi_b]$ 
is the sum of a potential term $U_V[\phi_b]$ and a gradient term $U_\nabla[\phi_b]$:
\begin{align}
\label{eq:Uphi}
    &U[\phi_b]=U_\nabla[\phi_b]+U_V[\phi_b]\\
\label{eq:Tphi}
    &U_\nabla[\phi_b]=\int d^3x\,\frac 12 \left(\bm\nabla\phi_b\right)^2\\
\label{eq:Vphi}
    &U_V[\phi_b]=\int d^3x\,V(\phi_b).
\end{align}
The identification of $U[\phi_b]$ as the barrier in quantum field theory allows us to draw some interesting consequences.
\begin{itemize}
  \item It is now clear why tunnelling in quantum field theory requires bounce solutions and cannot directly take place as a transition between spatially homogeneous configurations: the barrier associated with such a tunnelling process would be infinitely high. Specifically, in such cases we would find $U_\nabla[\phi_b]=0$ and $U_V[\phi_b]$ proportional to the space volume
 ${\mathcal V}_3$.
  \item The gradient and the potential terms may have opposite signs: even if the scalar potential has no barrier separating different minima, a potential energy barrier may appear because of the gradient term. An important examples of this class is the standard model,
 as we shall see in~\sect{sec:UVandNP}.
  \item The barrier penetration integral $B = S[\phi_b]$ is calculated as an integral over the three-dimensional space. In particular, the potential term is obtained by integrating the scalar potential, computed at the bounce configuration. Hence, the value of $S[\phi_b]$ is affected by all energy scales between $0$ and $\phi_b(0)$. We will come back to this issue in \sect{sec:UVandNP}, when discussing the impact of non-standard physics on the metastability of the electroweak ground state.
 \end{itemize}

\subsection{Quantum corrections}
\label{quantumcorr}

The first quantum corrections to the leading semi-classical approximation
of the decay rate of an unstable ground state are computed~\cite{Callan:1977pt}
by a generalization
to quantum field theory of the analogous calculation, presented in \sect{sect:decmet}, of the tunnelling rate in the case of an ordinary quantum theory with many degrees of freedom.
This generalization is rather straightforward, except for a few differences, which must be taken into proper account.

\subsubsection{Zero modes}
\label{sec:zeromodes}

In the case of ordinary quantum mechanics, a factor of
$\sqrt{\frac{S[x_b]-S[x_0]}{2\pi\hbar m}}$ appears as a consequence of the presence of
a zero mode in the spectrum of the second derivative of the action due time translation invariance. Here
we have four such factors (with $m=1$), corresponding to one zero mode for translations in each space-time direction. Correspondingly, the integration over all possible bounce centers, which yields a factor of $T$ in the case of ordinary quantum mechanics in one space dimension, is now replaced by an integration over all possible
bounce centers in space-time, yielding a factor of $T{\mathcal V}_3$, where ${\mathcal V}_3$ is the three-dimensional volume.

This is a good place for a more general discussion of zero modes of the second derivative of the action, which will turn out to play an important role in the application of the formalism to the standard model. In general, zero modes of $S''[\phi_b]$ appear whenever the theory has an invariance property
under a class of transformations of the field which we define in infinitesimal form as
\beq
\phi(x)\to\phi'(x)=\phi(x)+\epsilon\Delta\phi(x).
\label{transf}
\eeq
In such case, 
\beq
\Delta\phi_b(x)= \frac{\partial\phi'_b(x)}{\partial\epsilon}
\eeq
is a zero mode of $S''[\phi_b]$. Here is the proof. As a consequence of the invariance of the theory upon the transformations \eq{transf}, $\phi'_b(x)$ is also a bounce solution, and $S[\phi_b']=S[\phi_b]$. Therefore
\beq
S''[\phi_b]\Delta\phi_b=
S''[\phi'_b]\frac{\partial\phi'_b}{\partial\epsilon}=(-\partial^2+V''(\phi'_b))\frac{\partial\phi'_b}{\partial\epsilon}
=\frac{\partial}{\partial\epsilon}(-\partial^2+V'(\phi'_b))\phi'_b=0
\eeq
by the field equations.
For example, under an infinitesimal translation
\beq
x_\mu\to x_\mu+\epsilon_\mu
\eeq
the bounce transforms as
\beq
\phi_b(x)\to\phi'_b(x)=\phi_b(x+\epsilon)=\phi_b(x)+\epsilon_\mu\partial_\mu\phi_b(x).
\eeq
If the theory is invariant under translations, then
\beq
\phi_\mu(x)=N\partial_\mu\phi_b(x)
\eeq
are four zero modes of $S''[\phi_b]$. The normalization factor $N$ is finite: 
Eqs.~(\ref{eq:QFTB}, \ref{viriale}, \ref{eq:flatvirial}) give
\beq
\int d^4x\,(\partial_\mu\phi_b)^2=4\left(S[\phi_b]-S[\phi_{\rm FV}]\right),
\eeq
so
\beq
N=\frac{1}{\sqrt{S[\phi_b]-S[\phi_{\rm FV}]}}.
\eeq
From our experience in quantum mechanics, we know how to deal with these zero modes in the computation of the path integral:
the calculation of the path integral is performed by integrating over all possible values of the coefficients $c_i$ of the expansion of the scalar field in the basis of orthonormal eigenfunctions of $S''[\phi]$, which includes normalized zero-modes:
\beq
\phi(x)=\phi_b(x)+\sum_i c_i \phi_i(x)=\phi_b(x)+c_\mu N \partial_\mu\phi_b(x)+\ldots.
\eeq
As a consequence, integrating over $c_\mu$ is equivalent to integrating over all possible locations of the bounce center:
\beq
\frac{dc_\mu}{\sqrt{2\pi}} = \frac{1}{\sqrt{2\pi}}\frac{d\epsilon_\mu}{N}=\sqrt{\frac{S[\phi_b]-S[\phi_{\rm FV}]}{2\pi}}d\epsilon_\mu.
\eeq
These integrations contribute to the path integral a factor of 
\beq
\left[\sqrt{\frac{S[\phi_b]-S[\phi_{\rm FV}]}{2\pi}}\right]^4{\mathcal V}_3 T
\eeq
where ${\mathcal V}_3 T$ is  the space-time volume; note however that the integration in the directions of zero-modes
can be converted into an integration over a collective coordinate (the bounce center in this case) only provided that the zero-modes are normalizable.

One might wonder whether analogous zero modes arise from Lorentz transformations, which in Eulidean space-time correspond to 4-dimensional rotations. The six functions
\beq
\phi_{\mu\nu}(x)=\(x_\mu\partial_\nu-x_\nu\partial_\mu\)\phi_b(x)
\eeq
are in fact eigenfunction of $S''[\phi_b]$ with zero eigenvalues, as one can check directly.  However, they vanish because of the above-mentioned $\O(4)$ invariance of the bounce.

To conclude this section we mention two more possible origins of zero modes; both will be discussed in more detail in \sect{sec:SMstability}.
The first one is the presence of an internal symmetry. The simplest example is a complex scalar theory, invariant under $U(1)$ transformations
\beq
\phi(x)\to\phi'(x)=\phi(x)+i\epsilon\phi(x).
\eeq
In this case and $\Delta\phi_b(x)=i\phi_b(x)$ is a zero mode. 
The corresponding contribution to the path integral is obtained integrating over all possible $U(1)$ transformations:
\beq
\frac{dc}{\sqrt{2\pi}} = \frac{1}{\sqrt{2\pi}}\frac{d\epsilon}{N}=\frac{d\epsilon}{\sqrt{2\pi}}
\left[\int d^4x\,|\phi_b(x)|^2\right]^{-\frac{1}{2}}.
\eeq
The integration contributes to the path integral a factor of 
\beq
\frac{{\mathcal V}}{\sqrt{2\pi}}
\left[\int d^4x\,|\phi_b(x)|^2\right]^{-\frac{1}{2}}
\label{zmint}
\eeq
where ${\mathcal V}=2\pi$ is the volume of the symmetry group. In this case, the bounce itself must be normalizable. A full discussion of 
the impact on the calculation of the path integral of internal symmetry zero modes is presented in Ref.~\cite{Kusenko:1996bv};
quite obviously, they play a role in the case of the standard model.

As a final example, let us assume that the theory is invariant upon scale transformations, defined by
\beq
\phi(x)\to \phi'(x)=\e^{\epsilon d}\phi(\e^\epsilon x),
\eeq
where $d$ is the mass dimension of the field ($d=1$ for boson fields in four space-time dimensions). 
In infinitesimal form
\beq
\phi(x)\to \phi'(x)=\phi(x)+\epsilon(1+x_\mu\partial_\mu)\phi(x).
\eeq
Then, by the argument given above,
\beq
\phi_s(x)=(1+x_\mu\partial_\mu)\phi_b(x)
\eeq
is a zero mode of $S''[\phi_b]$. This example is especially interesting, because the only scale-invariant theories of one scalar fields in four dimensions are specified by the scalar potential
\beq
V(\phi)=\frac{1}{4}\lambda\phi^4,
\eeq
with $\lambda$ a constant, which is an excellent approximation of the standard model scalar potential at large field values. Furthermore, the field equation for an $\O(4)$-invariant bounce,
\beq
\frac{d^2\phi_b(r)}{dr^2}+\frac{3}{r}\frac{d\phi_b(r)}{dr}=\lambda\phi_b^3(r)
\eeq
has an analytical solution for $\lambda<0$:
\beq
\phi_b(r)=\frac{\phi_b(0)}{1+\frac{|\lambda|}{8}\phi_b^2(0)r^2},
\label{FLbounce}
\eeq
usually referred to as the Fubini-Lipatov bounce, originally presented in Refs.~\cite{Fubini:1976jm,Lipatov:1976ny}
(alternative derivations of \eq{FLbounce} can be found in Ref.~\cite{Lee:1985uv} and in Ref.~\cite{DiLuzio:2015iua}; the latter is reviewed in~\app{app:FLinstanton}.)

Unfortunately, the zero mode of scale transformations
\beq
\phi_s(r)=(1+x_\mu\partial_\mu)\phi_b(x)=
\phi_b(0)\frac{1-\frac{|\lambda|}{8}\phi_b^2(0)r^2}{\left(1+\frac{|\lambda|}{8}\phi_b^2(0)r^2\right)^2}
\eeq
is not normalizable: $\phi^2_s(r)\sim r^{-4}$ for $r\to\infty$. It follows that the corresponding contribution to the path integral cannot be replaced by an integration over  a collective coordinate, as in the case of translation zero modes.

\subsubsection{Renormalization}

A second important difference between quantum field theory and ordinary quantum mechanics is the appearance of ultraviolet 
divergences in the computation of the action beyond the semi-classical
approximation. Ultraviolet divergences require the usual renormalization procedure;
specifically, the classical action must be expressed in terms of renormalized fields
and parameters, and supplemented by appropriate counterterms:
\beq
S[\phi]\to S[\phi]+\hbar S_{\rm ct}[\phi]
\label{oneloopS}
\eeq
where $S_{\rm ct}[\phi]$ is the action induced by
one-loop counterterms, and we have (temporarily) restored the dependence on $\hbar$ for later convenience. 
In general, the presence of the counterterms will modify the bounce solution:
\beq
\phi_b\to\phi_b+\hbar\phi_{\rm ct}
\eeq
where $\phi_b$ is a bounce solution for $S[\phi]$:
\beq
\left.\frac{\delta S[\phi]}{\delta\phi}\right|_{\phi=\phi_b}=0.
\eeq
However, to one-loop order, the explicit expression of the correction $\phi_{\rm ct}$
is not needed. Indeed,
\begin{align}
&S[\phi_b+\hbar\phi_{\rm ct}]
+\hbar S_{\rm ct}[\phi_b+\hbar\phi_{\rm ct}]
\notag\\
&=S[\phi_b]+\hbar\int d^4x\,
\phi_{\rm ct}\left.\frac{\delta S[\phi]}{\delta\phi}\right|_{\phi=\phi_b}
+\hbar S_{\rm ct}[\phi_b]+O(\hbar^2)
\notag\\
&=S[\phi_b]+\hbar S_{\rm ct}[\phi_b]+O(\hbar^2).
\end{align}
Taking these differences into proper account, and setting back $\hbar=1$, we obtain
the following result for the decay probability per unit time and per unit volume
$\gamma/{\mathcal V}_3$ of a metastable state:
\begin{equation}
\frac{\gamma}{{\mathcal V}_3}=
 \frac{\left(S[\phi_b]-S[\phi_{\rm FV}]\right)^2}{(2\pi)^2}
\sqrt{
    \frac{  \Det S''[\phi_{\rm FV}]   }
           { |\Det'S''[\phi_b] |}
           }\e^{-\left(S[\phi_b]-S[\phi_{\rm FV}]\right)}
           \e^{-\left(S_{\rm ct}[\phi_b]-S_{\rm ct}[\phi_{\rm FV}]\right)},
\label{QFTdecayrate}
\end{equation}
\eq{QFTdecayrate} is dimensionally correct:
since there are four zero modes, the pre-exponential factor has the dimension of the
square root of $S''[\phi]$ (an inverse length) to the fourth power.

\section{The effective potential}
\label{effatceffpot}

The appropriate tool to determine the ground state of a quantum field theory beyond tree level
is the effective potential \cite{JonaLasinio:1964cw,Coleman:1973jx},
since it automatically encodes radiative corrections while retaining the advantage
of the semi-classical approximation, namely the possibility of simultaneously surveying all the vacua of the theory
by minimizing a function in analogy to the tree-level potential.
Furthermore, we have seen in the previous sections that functional methods
are extremely useful in the study of the lifetime of metastable states
in quantum field theory.

In this section we briefly review the
effective action formalism and apply it to the case of the standard model
of electroweak interactions, in order to study the possible instability of the ground state at large field values.

\subsection{Effective action and effective potential}
\label{sec:effpot}

Let us consider the theory of a single real scalar field $\Phi$ described by the Lagrangian density
$\mathcal{L} (\Phi, \partial_\mu \Phi)$ with a linear coupling of $\Phi(x)$ to an external source $J(x)$:
\begin{equation}
\mathcal{L} \to \mathcal{L} + J(x) \Phi(x).
\end{equation}
The vacuum-to-vacuum transition amplitude $Z[J]$  is usually written in terms
of a functional $W[J]$ as
\beq
\label{defWJ}
Z[J]=\braket{0|0}_J\equiv e^{iW[J]} = \int \mathcal{D} \Phi \, e^{i \int d^4 x \,\left[ \mathcal{L} (\Phi, \partial_\mu \Phi) + J \Phi \right] },
\eeq
where in the last step we have employed the standard path integral representation.
Functional derivatives of $W[J]$ with respect to $iJ$ at $J=0$ give the connected Green's functions of the theory;
for this reason, $W[J]$ is called the connected generating functional.
$W[J]$ admits a functional Taylor expansion
\beq
\label{Wjexp}
W[J]=\sum_n \frac {i^n}{n!} \int d^4x_1 \ \dots \ d^4x_n G^{(n)}(x_1,\dots,x_n)J(x_1)\dots J(x_n),
\eeq
where $G^{(n)}$ denotes the $n$-point connected Green's functions. These, in turn,
are computed in perturbation theory as the sum of all connected Feynman diagrams
with $n$ external lines.

Next, one then defines the \emph{classical field} $\phi_c(x)$ as
\begin{equation}
  \phi_c (x) \equiv \frac{\delta W[J]}{\delta J(x)} = \frac{\bra{0}\Phi(x)\ket{0}_J}{\braket{0|0}_J},
  \label{phic}
\end{equation}
where the last expression follows from the path integral representation in \eq{defWJ}.
Note that $\phi_c(x)$ is by definition a functional of $J$; we are implicitly assuming that
this functional is single-valued. This is related to the convexity issue of the effective potential,
to be discussed in~\sect{convex}.
We now introduce the \emph{effective action}
\index{Effective action}$\Gamma[\phi_c]$, defined by the functional Legendre transformation
\begin{equation}
  \Gamma[\phi_c]=W[J]-\int d^4x \ J(x) \phi_c(x).
  \label{effectiveaction}
\end{equation}
The effective action can be expanded in powers of the classical fields, in a way similar to
\eq{Wjexp}, yielding
\beq
\label{expEP}
\Gamma[\phi_c]=\sum_{n=0}^\infty \frac 1{n!} \int dx_1 \dots dx_n \Gamma^{(n)}(x_1,\dots, x_n) \phi_c(x_1)\dots\phi_c(x_n).
\eeq
The coefficients $\Gamma^{(n)}$ can be shown to correspond to the one-particle-irreducible (1PI) Green's functions of the theory
(see for example Ref.~\cite{Weinberg:1996kr}).

The functional $\Gamma[\phi_c]$ is the appropriate tool to study spontaneous symmetry breaking.
Indeed, the condition for spontaneous symmetry breaking is that $\phi_c$ is different from zero
even when the source $J$ is equal to zero, as can be read off \eq{phic}. On the other hand, one has
\begin{equation}
    \label{dg}
    \frac{\delta\Gamma[\phi_c]}{\delta\phi_c(x)}
    = \int d^4y \ \frac{\delta W[J]}{\delta J(y)}\frac{\delta J(y)}{\delta \phi_c(x)}
    -J(x) -\int d^4y \ \frac{\delta J(y)}{\delta \phi_c(x)}\phi_c(y)
    = -J(x),
\end{equation}
or in other words
\beq
\label{mincond}
\left.\frac{\delta\Gamma[\phi_c]}{\delta\phi_c(x)}\right|_{J=0}= 0.
\eeq
We conclude that spontaneous symmetry breaking takes place when the classical field that minimizes the effective action is different from zero.

The effective action can be computed in perturbation theory
by the usual technique of Feynman diagram. In particular, it can be shown that
in the tree-level approximation the effective action coincides
with the classical action:
\beq
\Gamma[\phi_c]=\int d^4x\,\left[{\mathcal L}(\phi_c,\partial_\mu\phi_c)+J\phi_c\right]
+\text{loop corrections}.
\eeq
In some cases,
however, an expansion in powers of external momenta is more useful.
To define such an expansion, we consider the Fourier transforms of the functions
$\Gamma^{(n)}(x_1,\dots, x_n)$:
\begin{align}
\Gamma^{(n)}(x_1,\dots, x_n) &= \int \frac{d^4 p_1}{(2\pi)^4}\ldots\frac{d^4 p_n}{(2\pi)^4}
e^{i(p_1x_1+\dots+p_nx_n)} \nonumber \\
&\times (2\pi)^4 \delta(p_1+\ldots+p_n) \tilde\Gamma^{(n)} (p_1,\ldots,p_n) ,
\end{align}
and expand $\tilde\Gamma^{(n)}$ in powers of momenta around $p_i = 0$:
\beq
\tilde\Gamma^{(n)} (p_1,\ldots,p_n) = \tilde\Gamma^{(n)} (0) + \ldots .
\eeq
The effective action becomes
\begin{align}
\label{eaexp}
\Gamma[\phi_c] &= \sum_{n=0}^\infty \frac 1{n!} \int dx_1 \dots dx_n \, \phi_c(x_1)\dots\phi_c(x_n) \nonumber \\
&\times \int \frac{d^4 p_1}{(2\pi)^4}\ldots\frac{d^4 p_n}{(2\pi)^4}
e^{i(p_1x_1+\dots+p_nx_n)} \int d^4x \, e^{-ix(p_1+\ldots+p_n)} \left[ \tilde\Gamma^{(n)} (0) + \ldots \right] \nonumber \\
&= \int d^4x \, \sum_{n=0}^\infty \frac 1{n!} \tilde\Gamma^{(n)} (0) \phi^n_c(x) + \ldots .
\end{align}
The first term in this expansion is usually written as
\beq
- \int d^4x \, V_{\rm eff}(\phi_c) ,
\eeq
where
\beq
\label{defep}
V_{\rm eff}(\phi_c) = - \sum_{n=0}^\infty \frac 1{n!} \tilde\Gamma^{(n)} (0) \phi^n_c
\eeq
is an ordinary function of $\phi_c$, usually called the \emph{effective potential} of the theory, since it does not contain derivatives of the classical field.
The following terms, originating from higher powers of momenta in the expansion of
$\tilde\Gamma^{(n)}$, contain instead two or more derivatives of $\phi_c$.

If we require translational invariance of the vacuum state, then $\partial_\mu \phi_c=0$
for $J=0$, and the minimum condition \eq{mincond} reduces to
\beq
 \label{ssb2}
\frac{dV_{\rm eff}(\phi_c)}{d\phi_c} = 0 .
\eeq
The advantage of \eq{ssb2} is that the ground state of the theory
can be determined, including quantum corrections,
via the minimization of a single ordinary function, the effective potential,
in analogy with the tree-level case.

\subsection{The background field method}
\label{sec:bkgd}

The discussion of the previous section shows that efficient techniques for the computation
of the effective potential are needed.
A compact way to compute
the loop-expanded effective potential\footnote{The loop expansion corresponds to
an expansion in $\hbar$ which
does not affect the shift of the fields, since $\hbar$ multiplies
the total Lagrangian density.
In order to show this let us restore $\hbar$ in the definition of the Lagrangian,
$\mathcal L(\phi,\partial_\mu\phi)\to \frac 1\hbar \mathcal L(\phi,\partial_\mu\phi)$,
and denote with $P$ the power of $\hbar$ associated with a given Feynman diagram. Each vertex carries a factor $\hbar^{-1}$, while each propagator (being the inverse of the quadratic part of $\mathcal L$) carries a factor $\hbar$.
Denoting by $I$ the number of propagators and by $V$ the number of vertices of a graph, we have $P=I-V$.
On the other hand, the number of loops $L$ in a diagram corrresponds to the number of independent momenta:
every internal line contribute to one integration momentum,
while every vertex to a $\delta$ function which reduces
by one unit the number of integration momenta (except for an overall $\delta$ which is left for energy-momentum conservation),
which yields $L=I-V+1$. We can hence establish an equivalence between the $\hbar$ and loop expansions: $P=L-1$.
}
is given by the \emph{background field method} \cite{Jackiw:1974cv}, 
which we review here for the one-loop case.
Let us consider the Lagrangian density for a real scalar field $\Phi$:
\begin{equation}
\mathcal{L} (\Phi,\partial_\mu\Phi) = \frac{1}{2} \partial_\mu \Phi \partial^\mu \Phi  - V(\Phi),
\end{equation}
with $V(\Phi)$ denoting the tree-level scalar potential. We perform the following change of variable in the functional integral \eq{defWJ}:
\begin{equation}
\Phi(x) = \phi_c(x) + \phi(x)\,
\end{equation}
where $\phi_c(x)$ is a
background field which we identify with the classical field of the previous section,
while $\phi(x)$ is a dynamical (quantum) field.
Organized in powers of $\phi$ (up to quadratic terms), the exponent of the r.h.s.~of \eq{defWJ} takes the form
\begin{align}
&\int d^4x \, \left[\mathcal{L} (\phi_c + \phi ) + J (\phi_c + \phi) \right]
\notag\\
&\qquad =\int d^4x \, \left[ \mathcal{L} (\phi_c) + J \phi_c \right]
+ \int d^4x \, \phi(x)
\left[ \left. \frac{\delta \mathcal{L}}{\delta \Phi} \right|_{\Phi=\phi_c} + J \right]
\nonumber \\
&\qquad+ \frac{1}{2} \int d^4x \, d^4y \, \phi(x) i \mathscr{D}^{-1}(\phi_c,x-y) \phi (y)
+ \mathcal{O} (\phi^3),
\end{align}
where
\begin{align}
i \mathscr{D}^{-1}(\phi_c,x-y) &=
\left. \frac{\delta}{\delta\Phi(x)}\frac{\delta} {\delta\Phi(y)}
\int d^4z\,\left[\mathcal{L}(\Phi)+J(z)\Phi(z)\right] \right|_{\Phi=\phi_c}
\notag\\
&=\left[-\partial^2-V''(\phi_c(x))\right]\delta^{(4)}(x-y)
\end{align}
is the inverse propagator in configuration space evaluated at $\Phi=\phi_c$.
We can now explicitly perform the path integral in the gaussian approximation:
\begin{align}
\label{pathintgauss}
e^{iW[J]} &\simeq e^{i \int d^4x \, \left[\mathcal{L} (\phi_c) + J \phi_c \right]}
\int \mathcal{D} \phi \,
\exp{\left( \frac{i}{2} \int d^4x \, d^4y \, \phi(x) i \mathscr{D}^{-1}(\phi_c,x-y) \phi (y) \right)} \nonumber \\
&= \mathcal{N} e^{i \int d^4x \, \left[ \mathcal{L} (\phi_c) + J \phi_c \right]}
\left[ \det i \mathscr{D}^{-1}(\phi_c,x-y) \right]^{-\frac{1}{2}},
\end{align}
up to an irrelevant normalization constant $\mathcal{N}$. The result \eq{pathintgauss} follows from the
invariance of the path integral measure, $\mathcal{D} \Phi = \mathcal{D} \phi$,
and the classical field equation~$\left.\frac{\delta\mathcal{L}}{\delta \Phi} \right|_{\phi_c} + J = 0$.
The latter approximation is sufficient as long as we are interested in the one-loop effective
action.
After replacing \eq{pathintgauss} in the definition of the effective action, \eq{effectiveaction}, we obtain
\beq
\label{effact}
\Gamma[\phi_c] =  S[\phi_c]+ \frac{i}{2} \log \Det i \mathscr{D}^{-1}(\phi_c,x-y),
\eeq
where
\beq
S[\phi_c]=\int d^4x \, \mathcal{L} (\phi_c,\partial_\mu\phi_c)
\eeq
is the classical action. The calculation of one-loop corrections to the effective
action involves the usual renormalization procedure: although not explicitly shown
in \eq{effact}, it is understood that the final result involves the choice of a regularization
procedure and of a renormalization scheme, and the introduction of suitable counterterms.
Typically, this results in the dependence of the renormalized parameters on an arbitrary
energy scale, to compensate the explicit dependence of the correction term
on the same scale.

The effective potential is now immediately obtained by evaluating \eq{effact}
for a space-time-independent classical field configuration $\phi_c$ (see \eq{eaexp}).
We get
\beq
- \mathcal{V}_4 \ V_{\rm eff} (\phi_c) = - \mathcal{V}_4 \ V_{0} (\phi_c) + \frac{i}{2} \log \Det i \mathscr{D}^{-1}(\phi_c,x-y),
\eeq
where $\mathcal{V}_4 \equiv \int d^4x$ denotes the space-time volume, originating from the fact that $\phi_c$ is space-time independent.
The homogeneity of $\phi_c$ also allows for a simple evaluation of the functional determinant, through the identity
\beq
\label{functdet}
\log \Det i \mathscr{D}^{-1}(\phi_c,x-y) = \tr \log i \mathscr{D}^{-1}(\phi_c,x-y).
\eeq
The functional trace, denoted by $\tr$,  is taken by setting $x=y$ and integrating over
the space-time. This gives
\beq
\log \Det i \mathscr{D}^{-1}(\phi_c,x-y)
=\mathcal{V}_4 \log i \mathscr{D}^{-1}(\phi_c,0)
=\mathcal{V}_4 \int \frac{d^4 k}{(2\pi)^4} \log i \tilde{\mathscr{D}}^{-1}(\phi_c,k),
\eeq
where $\tilde{\mathscr{D}}^{-1}(\phi_c,k)$ denotes the four-dimensional Fourier transform of $\mathscr{D}^{-1}(\phi_c,x-y)$.

An explicit example is provided by a renormalizable scalar theory, with
\beq
V_0(\Phi)=\frac{1}{2}m^2\Phi^2+\frac{1}{4}\lambda\Phi^4.
\eeq
In this case
\beq
i \mathscr{D}^{-1}(\phi_c,x-y) =\left(-\partial_\mu\partial^\mu-m^2-3\lambda
\phi_c^2\right)\delta^{(4)}(x-y)
\eeq
and
\beq
i \tilde{\mathscr{D}}^{-1}(\phi_c,k) =k^2-m^2-3\lambda\phi_c^2
\eeq
The background field method can be consistently extended to include higher orders in the loop
expansion (see e.g.~Ref.~\cite{Martin:2017lqn} for a state-of-the-art calculation of the three-loop effective potential).
In this review we will limit ourselves to the explicit calculation of the effective potential
at one loop.

\subsection{The one-loop effective potential of the standard model}

The results of the previous section 
can be readily generalized to the case of a generic quantum field interacting with $\Phi$,
thereby obtaining the following closed form for the one-loop effective potential:
\beq
\label{Veff1loop}
V_{\rm eff} (\phi) = V_{0} (\phi) + i \sum_n \eta_n\int \frac{d^4 k}{(2 \pi)^4} {\rm Tr} \log i \tilde{\mathscr{D}}_n^{-1}(\phi,k) ,
\eeq
where the sum is extended to all fields in the theory, labelled by the index $n$,
and $i \tilde{\mathscr{D}}_n^{-1}(\phi_c,k)$ denotes the corresponding
inverse propagators of the dynamical fields after the
background field $\phi_c$ shift in $\Phi$. The trace acts on all internal indices (e.g.~Lorentz or gauge); it is therefore the conventional trace in finite-dimension linear spaces, and we denote it by the usual symbol $\rm Tr$ to distinguish it from the functional trace,
denoted by $\tr$. The factor $\eta_n$
is the power of the functional determinant due to the gaussian path integral; it takes
the values $-\frac{1}{2}$ for bosonic fields, and $+1$
for fermionic fields (matter fermions or ghosts). Finally, we have denoted the classical field
$\phi_c$ by $\phi$ to simplify notations.

We are now ready to apply the background field method to the calculation of the one-loop effective potential of the standard model of electroweak interactions.
In order to set the notation, we split the classical Lagrangian density of
the electroweak sector into a gauge term, a Higgs term  and Yukawa term,
the latter including only the leading contribution from the top quark:
\begin{equation}
\label{Lclassical}
\mathcal{L}_{\rm{C}} = \mathcal{L}_{\rm{YM}} + \mathcal{L}_{\rm{H}} + \mathcal{L}_{\rm{top}},
\end{equation}
with
\begin{align}
\label{LYM}
\mathcal{L}_{\rm{YM}} &=
-\frac{1}{4} \left( \partial_\mu W^a_\nu - \partial_\nu W^a_\mu  + g \epsilon^{abc} W^b_\mu W^c_\nu \right)^2
-\frac{1}{4} \left( \partial_\mu B_\nu - \partial_\nu B_\mu  \right)^2 ,\\
\label{LH}
\mathcal{L}_{\rm{H}} &= \left( D_\mu H \right)^\dag \left( D^\mu H \right) - V(H) ,\\
\label{LF}
\mathcal{L}_{\rm{top}} &= \overline{Q}_L i \slashed{D} Q_L + \overline{t}_R i \slashed{D} t_R +
\left(- y_t \overline{Q}_L \epsilon H^* t_R + \rm{h.c.}\right) ,
\end{align}
where $W^a_\mu$ ($a=1,2,3$) and $B_\mu$ are the $SU(2)_L$ and $U(1)_Y$ gauge fields,
$H$ is the standard model Higgs doublet, $\epsilon = i \sigma^2$ is the $2\times 2$ antisymmetric matrix,
and
\beq
Q_L = \left(\begin{array}{c}t_L\\b_L\end{array}\right)
\eeq
is the left-handed quark doublet of the third generation.
QCD indices are suppressed in the quark sector.
The covariant derivative is defined by
\begin{equation}
\label{defcovder}
D_\mu = \partial_\mu - i g T^a W^a_\mu - i g' Y B_\mu,
\end{equation}
where $T^a$ are the $SU(2)_L$ generators in the relevant representation
($T^a=\sigma^a/2$ with $\sigma^a$ the three Pauli matrices for doublets,
$T^a=0$ for singlets),
and $Y$ is the hypercharge of the standard model fields. We have $Y=Q-T^3$, where
$Q$ is the electric charge in units of the proton charge, and hence $Y(H) = 1/2$, $Y(Q_L) = 1/6$, $Y(t_R) = 2/3$.
The Higgs potential is
\begin{equation}
\label{eq:Higgspot}
V (H) = m^2 H^\dag H + \lambda (H^\dag H)^2,
\end{equation}
with $m^2<0$.
Gauge invariance allows us to perform the shift of the Higgs doublet,
required by the background-field-method calculation of the effective potential,
in a specific direction in the $SU(2)_L \times U(1)_Y$ space:
\begin{equation}
\label{Hshift}
H(x) \rightarrow
\frac{1}{\sqrt{2}}
\left(
\begin{array}{c}
\chi^1(x) + i \chi^2(x) \\
\phi + h(x) + i \chi^3(x)
\end{array}
\right),
\end{equation}
where $\phi$ denotes the classical background field, $h$ the physical Higgs field and
$\chi^a$ ($a=1,2,3$) the Goldstone fields.
At tree-level the effective potential simply reads
\begin{equation}
\label{V0}
V_0 (\phi) = \frac{1}{2} m^2\phi^2 + \frac{1}{4} \lambda\phi^4,
\end{equation}
while in order to compute the one-loop quantum correction, $V_{1} (\phi)$, according to \eq{Veff1loop}
one needs to work out the inverse propagators of the dynamical fields in the shifted standard model Lagrangian, including the gauge fixing term.
Although the calculation is most easily carried out in the Landau gauge, it is instructive to discuss a more
general gauge. A convenient choice is the so-called Fermi gauge 
in which the gauge fixing term is
\begin{equation}
\label{gflagFermi}
\mathcal{L}^{\rm{Fermi}}_{\rm{g.f.}}  = -\frac{1}{2 \xi_W} \left( \partial^\mu W^a_\mu \right)^2
-\frac{1}{2 \xi_B} \left( \partial^\mu B_\mu \right)^2.
\end{equation}
The Landau gauge is obtained for $\xi_W=\xi_B=0$. 
We start from the determination of the quadratic ($\phi$-dependent) term
of the Lagrangian, $\mathcal{L}^{(2)} + \mathcal{L}^{\rm{Fermi}}_{\rm{g.f.}}$, after the shift
in \eq{Hshift}. A straightforward calculation yields
\begin{align}
\label{LYMquadFermi}
\mathcal{L}^{(2)}_{\rm{YM}} &=
\tfrac{1}{2} W^a_\mu \left( \partial^2 \, g^{\mu\nu} - \partial^\mu \partial^\nu \right) \delta^{ab} W^b_\nu
+ \tfrac{1}{2} B_\mu \left( \partial^2 \, g^{\mu\nu} - \partial^\mu \partial^\nu \right) B_\nu,\\
\label{LHquadFermi}
\mathcal{L}^{(2)}_{\rm{H}} &=
\tfrac{1}{2} h \left( - \partial^2 - \bar{m}_h^2 \right) h
+ \tfrac{1}{2} \chi^a \left( - \partial^2 - \bar{m}_\chi^2 \right) \delta^{ab} \chi^b  \nonumber \\
&+ \tfrac{1}{2} \bar{m}_W^2 W^a_\mu W^{a\mu}
+ \tfrac{1}{2} \bar{m}_B^2 B_\mu B^{\mu} - \bar{m}_W \bar{m}_B W^3_\mu B^{\mu}
\nonumber \\
&  - \bar{m}_W \partial_\mu \chi^1 W^{2\mu}
- \bar{m}_W \partial_\mu \chi^2 W^{1\mu}
+ \bar{m}_W \partial_\mu \chi^3 W^{3\mu}
- \bar{m}_B \partial_\mu \chi^3 B^{\mu} ,\\
\label{LFquadFermi}
\mathcal{L}^{(2)}_{\rm{top}} &= \overline{t} \left( i \slashed{\partial} - \bar{m}_t \right) t ,
\end{align}
where we have defined $\phi$-dependent masses
\begin{align}
\label{mhphiFermi}
\bar{m}_h^2 &= m^2 + 3 \lambda \phi^2,\\
\label{mchiphi}
\bar{m}_{\chi}^2 &= m^2 + \lambda \phi^2  ,\\
\label{defmwFermi}
\bar{m}_W &= \tfrac{1}{2} g \phi ,\\
\label{defmbFermi}
\bar{m}_B &= \tfrac{1}{2} g' \phi ,\\
\label{mtphiFermi}
\bar{m}_t &= \frac{y_t}{\sqrt{2}} \phi .
\end{align}
Note that $\mathcal{L}^{\rm{Fermi}}_{\rm{g.f.}}$ is already quadratic in the gauge boson fields,
while bilinear ghost terms are $\phi$-independent.\footnote{Recall that the ghost Lagrangian is given by
$$
\mathcal{L}_{\rm ghost} = \sum_{\alpha\beta} \eta_\alpha^\dag \frac{\delta F^\alpha}{\delta \theta^\beta} \eta_\beta,
$$
where $F^a_W = \partial^\mu W^a_\mu$ and $F_B = \partial^\mu B_\mu$ are gauge-fixing
functionals, $\delta W^a_\mu = g \epsilon^{abc} \theta^b W^c_\mu - \partial_\mu \theta^a$
and $\delta B_\mu = - \partial_\mu \theta^B$ are gauge transformations, $\theta^a$ and $\theta^B$
the corresponding gauge parameters.
}
Hence, in the Fermi gauge the ghost contribution decouples from the one-loop effective potential.

A slight complication in the Fermi gauge arises in the
Goldstone--gauge boson mixing in \eq{LHquadFermi}.
To deal with it we define an extended field vector
\begin{equation}
\label{extX}
X^T = \left(V^T_\mu , \chi^T\right),
\end{equation}
where
\begin{equation}
\label{defVchiFermi}
V^T_\mu = \left( W^1_\mu, W^2_\mu, W^3_\mu, B_\mu \right) ,\qquad
\chi^T = \left( \chi^1, \chi^2, \chi^3 \right) .
\end{equation}
The quadratic part of the Goldstone--gauge sector can be cast into
\begin{equation}
\label{quadgoldgauge}
\frac{1}{2} X^T \left( i \mathscr{D}_X^{-1} \right) X
= \frac{1}{2} \left( V_\mu^T, \chi^T \right)
\left(
\begin{array}{cc}
i \left( \mathscr{D}^{-1}_V \right)^\mu_{\nu} & \bar{m}^T_{\text{mix}} \, \partial^\mu  \\
- \bar{m}_{\text{mix}} \, \partial_\nu & i \mathscr{D}^{-1}_\chi
\end{array}
\right)
\left(
\begin{array}{c}
V^\nu \\
\chi
\end{array}
\right)
,
\end{equation}
with
\begin{equation}
\label{defmmix}
\bar{m}_{\text{mix}} =
\left(
\begin{array}{cccc}
0 & - \bar{m}_W & 0 & 0 \\
- \bar{m}_W & 0 & 0 & 0 \\
0 & 0 & \bar{m}_W & -\bar{m}_B
\end{array}
\right).
\end{equation}
After Fourier transformation ($\partial_\mu \rightarrow i k_\mu$) the inverse propagator matrix becomes
\begin{equation}
\label{invpropX}
i \tilde{\mathscr{D}}_X^{-1} =
\left(
\begin{array}{cc}
i ( \tilde{\mathscr{D}}^{-1}_V )^\mu_{\nu} & i k^\mu \bar{m}^T_{\text{mix}}  \\
- i k_\nu \bar{m}_{\text{mix}} & i \tilde{\mathscr{D}}^{-1}_\chi
\end{array}
\right) ,
\end{equation}
where $(\tilde{\mathscr{D}}^{-1}_V)^\mu_{\nu}$ can be conveniently
split into a transversal and a longitudinal part
\begin{equation}
\label{invpropgaugeFermi}
( \tilde{\mathscr{D}}^{-1}_V )^\mu_\nu  =
i \tilde{\mathscr{D}}^{-1}_{T} \,(\Pi_{T})^\mu_\nu
+ i \tilde{\mathscr{D}}^{-1}_{L} \,(\Pi_{L})^\mu_\nu,
\end{equation}
where
\begin{equation}
\label{defPiTLFermi}
(\Pi_{T})^\mu_\nu = g^\mu_\nu - \frac{k^\mu k_\nu}{k^2},
\qquad
(\Pi_{L})^\mu_\nu = \frac{k^\mu k_\nu}{k^2} ,
\end{equation}
and
\begin{align}
\label{invpropTFermi}
i \tilde{\mathscr{D}}^{-1}_{T} &=
\left(
\begin{array}{cccc}
- k^2 + \bar{m}_W^2 & 0 & 0 & 0 \\
0 & - k^2 + \bar{m}_W^2 & 0 & 0 \\
0 & 0 & - k^2 + \bar{m}_W^2 & -\bar{m}_W \bar{m}_B \\
0 & 0 & -\bar{m}_W \bar{m}_B & - k^2 + \bar{m}_B^2
\end{array}
\right) ,\\
\label{invpropLFermi}
i \tilde{\mathscr{D}}^{-1}_{L} &=
\left(
\begin{array}{cccc}
- \xi_W^{-1} k^2 + \bar{m}_W^2  & 0 & 0 & 0 \\
0 & - \xi_W^{-1} k^2 + \bar{m}_W^2 & 0 & 0 \\
0 & 0 & - \xi_W^{-1} k^2 + \bar{m}_W^2  & -\bar{m}_W \bar{m}_B \\
0 & 0 & -\bar{m}_W \bar{m}_B & -\xi_B^{-1} k^2 + \bar{m}_B^2
\end{array}
\right) .
\end{align}
The Goldstone boson, Higgs and top quark inverse propagators are
\begin{align}
\label{invpropchiFermi}
i \tilde{\mathscr{D}}^{-1}_{\chi} &=
\left(
\begin{array}{ccc}
k^2 - \bar{m}_\chi^2 & 0 & 0 \\
0 & k^2 - \bar{m}_\chi^2  & 0 \\
0 & 0 &k^2 - \bar{m}_\chi^2
\end{array}
\right) ,\\
\label{invprophFermi}
i \tilde{\mathscr{D}}^{-1}_{h} &=
k^2 - \bar{m}_h^2 ,\\
\label{invproptFermi}
i \tilde{\mathscr{D}}^{-1}_{t} &= \slashed{k} - \bar{m}_t .
\end{align}
The next step is the evaluation of $\log\det i \tilde{\mathscr{D}}_n^{-1}$, for $n = X, h ,t$.
Let us express the determinant of the block matrix in \eq{invpropX} as
\begin{align}
\label{evaluationdetX}
\det i \tilde{\mathscr{D}}_X^{-1} &= \det i \tilde{\mathscr{D}}_\chi^{-1} \det
\left(
i ( \tilde{\mathscr{D}}^{-1}_V )^\mu_{\nu}
- k^\mu k_\nu \bar{m}^T_{\text{mix}} \left( i \tilde{\mathscr{D}}^{-1}_{\chi} \right)^{-1} \bar{m}_{\text{mix}}
\right)\\
&= \det i \tilde{\mathscr{D}}_\chi^{-1} \det
\left(
i \tilde{\mathscr{D}}^{-1}_{T} (\Pi_{T})^\mu_\nu
+ \left( i \tilde{\mathscr{D}}^{-1}_{L}  - k^2 \bar{m}^T_{\text{mix}} \left( i \tilde{\mathscr{D}}^{-1}_{\chi} \right)^{-1} \bar{m}_{\text{mix}} \right) (\Pi_{L})^\mu_\nu
\right),
 \nonumber
\end{align}
where in the last step we used the decomposition in \eq{invpropgaugeFermi}.
It is convenient to perform a Lorentz transformation in $d$ space-time
dimensions, $k_\mu \rightarrow (k_0, 0, 0, 0, \ldots)$, such
that $(\Pi_{L})^\mu_\nu \rightarrow {\rm diag}(1,0,0,0,\ldots)$
and $(\Pi_{T})^\mu_\nu \rightarrow \rm{diag}(0,1,1,1,\ldots)$.
Then, using the Lorentz invariance of the determinant, we obtain
\begin{align}
\label{evaluationlogdetX}
\log\det i \tilde{\mathscr{D}}_X^{-1} &=
(d-1) \log \det \tilde{\mathscr{D}}^{-1}_{T}
\nonumber \\
&+ \log \det i \tilde{\mathscr{D}}_\chi^{-1}
\det \left( i \tilde{\mathscr{D}}^{-1}_{L}  - k^2 \bar{m}^T_{\text{mix}} \left( i \tilde{\mathscr{D}}^{-1}_{\chi} \right)^{-1} \bar{m}_{\text{mix}} \right).
\end{align}
The two terms in the r.h.s.~of \eq{evaluationlogdetX} yield respectively
\begin{equation}
\label{logdetTFermi}
\log \det i \tilde{\mathscr{D}}^{-1}_T =
2 \log \left( -k^2 + \bar{m}_W^2 \right)
+ \log \left( -k^2 + \bar{m}_{Z}^2 \right)
+ \ldots
\end{equation}
and
\begin{align}
\label{rhsexpl}
& \log \det i \tilde{\mathscr{D}}_\chi^{-1}
\det \left( i \tilde{\mathscr{D}}^{-1}_{L}
- k^2 \bar{m}^T_{\text{mix}} \left( i \tilde{\mathscr{D}}^{-1}_{\chi} \right)^{-1} \bar{m}_{\text{mix}} \right)  \\
& =
2 \log \left( k^4 -k^2 \bar{m}_\chi^2 + \bar{m}_\chi^2 \xi_W \bar{m}_W^2 \right)
+ \log \left( k^4 -k^2 \bar{m}_\chi^2 + \bar{m}_\chi^2 (\xi_W \bar{m}_W^2 + \xi_B \bar{m}_B^2) \right) + \ldots \nonumber \\
& = 2 \log \left( k^2 - \bar{m}_{A^{+}}^2 \right)
+ 2 \log \left( k^2 - \bar{m}_{A^{-}}^2 \right) + \log \left( k^2 - \bar{m}_{B^{+}}^2 \right) + \log \left( k^2 - \bar{m}_{B^{-}}^2 \right) + \ldots ,\nonumber
\end{align}
where the ellipses stand for $\phi$-independent terms, and we have defined the additional
$\phi$-dependent masses
\begin{align}
\label{defmassZFermi}
\bar{m}_{Z}^2 &= \bar{m}_W^2 + \bar{m}_B^2 ,\\
\label{defmassApm}
\bar{m}_{A^{\pm}}^2 &= \frac{1}{2} \bar{m}_\chi \left(  \bar{m}_\chi \pm \sqrt{ \bar{m}_\chi^2 - 4 \xi_W \bar{m}_W^2} \right) ,\\
\label{defmassBpm}
\bar{m}_{B^{\pm}}^2 &= \frac{1}{2} \bar{m}_\chi \left(  \bar{m}_\chi \pm \sqrt{ \bar{m}_\chi^2
- 4 (\xi_W \bar{m}_W^2 + \xi_B \bar{m}_B^2) } \right) .
\end{align}
For the evaluation of the fermionic determinant of \eq{invproptFermi} 
it is sufficient to employ 
a naive treatment of $\gamma_5$ in dimensional
regularization, i.e.~$\{ \gamma_5, \gamma_\mu \} = 0$ in $d$ dimensions,  
and make the standard choice
$\Tr \mathbf{1}_\text{Dirac} = 4$ in $d$ dimensions.\footnote{Subtleties 
related to the proper definition of $\gamma_5$ in $d$ dimensions only arise at higher orders in 
perturbation theory, while a different choice for the trace, e.g.~$\Tr \mathbf{1}_\text{Dirac} = 2^{d/2}$, would just lead to a different renormalization scheme \cite{Collins:1984xc}.}
Explicitly, one has
\begin{align}
\log \det \left( \slashed{k} - \bar{m}_t \right)
&= \Tr \log \left( \slashed{k} - \bar{m}_t \right)
= \Tr \log \gamma^5 \left( \slashed{k} - \bar{m}_t \right) \gamma^5
= \Tr \log \left( - \slashed{k} - \bar{m}_t \right) \nonumber \\
&= \frac{1}{2} \left[ \Tr \log \left( \slashed{k} - \bar{m}_t \right) + \Tr \log \left( - \slashed{k} - \bar{m}_t \right) \right]
= \frac{1}{2}  \Tr \log \left( - k^2 + \bar{m}_t^2 \right) \nonumber \\
&= \frac{1}{2} 4 \times 3 \log \left( -k^2 + \bar{m}_t^2 \right) ,
\end{align}
where the extra factors in the last step are due to the trace over Dirac and color indices.

Including all the relevant degrees of freedom
and working in dimensional regularization with $d = 4 - 2 \epsilon$,
the one-loop contribution to the effective potential~\eq{Veff1loop}
can be cast in the form
\begin{align}
\label{EP1loopSMdrexplFermi}
V_1(\phi,\epsilon)&= -\frac{i}{2} \mu^{2\epsilon} \int \frac{d^d k}{(2\pi)^d}
\left[
- 12 \log \left( - k^2 + \bar{m}_t^2 \right) + (d-1) \left( 2 \log \left( -k^2 + \bar{m}_W^2 \right)
\right. \right. \nonumber \\
& \left. + \log \left( -k^2 + \bar{m}_Z^2 \right) \right) + \log \left( k^2 - \bar{m}_h^2 \right)
+ 2 \log \left( k^2 - \bar{m}_{A^{+}}^2 \right) + 2 \log \left( k^2 - \bar{m}_{A^{-}}^2 \right)
\nonumber \\
& \left. + \log \left( k^2 - \bar{m}_{B^{+}}^2 \right) + \log \left( k^2 - \bar{m}_{B^{-}}^2 \right)
\right] ,
\end{align}
where we have introduced a mass parameter $\mu$ in order to keep the
correct dimension of the scalar potential.

The loop integral is easily evaluated after Wick rotation, yielding (see for instance Ref.~\cite{Peskin:1995ev})
\begin{equation}
\label{integralFermi}
-\frac{i}{2} \mu^{2\epsilon} \int \frac{d^d k}{(2\pi)^d} \log (-k^2 + m^2) =
\frac{1}{4} \frac{m^4}{(4\pi)^2} \left( \log \frac{m^2}{\mu^2} -\frac{3}{2} -\Delta_\epsilon \right),
\end{equation}
where we have introduced the modified minimal subtraction ($\overline{\rm{MS}}$) term \cite{Bardeen:1978yd}
\begin{equation}
\label{defDeltaepsFermi}
\Delta_\epsilon = \frac{1}{\epsilon} - \gamma_E + \log 4 \pi.
\end{equation}
After the $\epsilon$-expansion
the one-loop contribution to the effective potential is given by
\begin{align}
\label{1loopEPbareFermi}
& V_1(\phi,\epsilon)= V_1^{\rm pole}(\phi,\epsilon) + \frac{1}{4 (4 \pi)^2} \left[
-12 \bar{m}_t^4 \left( \log\frac{\bar{m}_t^2}{\mu^2} - \frac{3}{2} \right)
+6 \bar{m}_W^4 \left( \log\frac{\bar{m}_W^2}{\mu^2} - \frac{5}{6} \right) \right.
\nonumber \\
& \left.+3 \bar{m}_Z^4 \left( \log\frac{\bar{m}_Z^2}{\mu^2} - \frac{5}{6}\right)
+\bar{m}_h^4 \left( \log\frac{\bar{m}_h^2}{\mu^2} - \frac{3}{2}\right)
+2 \bar{m}_{A^+}^4 \left( \log\frac{\bar{m}_{A^+}^2}{\mu^2} - \frac{3}{2}\right)  \right.
\nonumber \\
& \left. +2 \bar{m}_{A^-}^4 \left( \log\frac{\bar{m}_{A^-}^2}{\mu^2} - \frac{3}{2} \right)
+ \bar{m}_{B^+}^4 \left( \log\frac{\bar{m}_{B^+}^2}{\mu^2} - \frac{3}{2} \right)
+ \bar{m}_{B^-}^4 \left( \log\frac{\bar{m}_{B^-}^2}{\mu^2} - \frac{3}{2}\right)
\right],
\end{align}
where all divergent terms as $\epsilon\to 0$ are collected in
$V_1^{\rm pole}(\phi,\epsilon)$, which can be expressed as a function of the standard model couplings
as
\begin{align}
\label{1loopdivFermi}
V_1^{\rm pole}(\phi,\epsilon)&= \frac{\Delta_\epsilon}{(4\pi)^2}
\left[
-m^4
- \left(3 \lambda - \frac{1}{8} \xi _B g'^2 - \frac{3}{8} \xi _W g^2   \right) m^2 \phi^2 \right.  \\
& \left. +
   \left(
   -\frac{3}{64} g'^4
   -\frac{3}{32} g'^2 g^2
   -\frac{9}{64} g^4
   +\frac{3}{4} y_t^4
   -3 \lambda ^2
   +\frac{1}{8} \xi _B g'^2 \lambda
   +\frac{3}{8}  \xi _W g^2  \lambda
   \right) \phi^4
\right]. \nonumber
\end{align}
While the $m^4$-dependent pole in \eq{1loopdivFermi} can be always subtracted by a constant shift in the
effective potential,
the remaining divergences are cancelled by the multiplicative renormalization of the bare field and couplings
appearing in $V^{(0)}_{\rm{eff}}$ (cf.~\eq{V0}):
\begin{equation}
\label{renV0parFermi}
\phi_0 = Z_{\phi}^{1/2} \phi , \qquad
m^2_0 = Z_{m^2} m^2 ,\qquad
\lambda_0 = Z_{\lambda} \lambda ,
\end{equation}
where the renormalization constants can be conveniently computed in the unbroken phase of the standard model.
Their expressions at one loop  in the $\overline{\rm{MS}}$ scheme read (see e.g.~\cite{Chetyrkin:2012rz,Mihaila:2012pz}):
\begin{align}
\label{ZphiFermi}
&Z_{\phi}^{1/2}= 1 +
\frac{\Delta_\epsilon}{(4\pi)^2}
\left(
\frac{3}{8} g'^2 + \frac{9}{8} g^2 -\frac{3}{2} y_t^2 -\frac{1}{8} \xi _B g'^2-\frac{3}{8} \xi _W g^2
\right) \\
\label{Zm2Fermi}
&Z_{m^2} = 1+ \frac{\Delta_\epsilon}{(4\pi)^2}
\left(
-\frac{3}{4} g'^2 -\frac{9}{4} g^2 +3 y_t^2 +6 \lambda
\right) \\
\label{ZlamFermi}
&Z_{\lambda} = 1+ \frac{\Delta_\epsilon}{(4\pi)^2}
\left(
 -\frac{3}{2} g'^2 -\frac{9}{2} g^2 +6 y_t^2 +12 \lambda
+\frac{3 }{16} \frac{g'^4}{\lambda }
+\frac{3}{8} \frac{g'^2 g^2}{\lambda }
+\frac{9}{16}\frac{g^4}{\lambda }
-3 \frac{y_t^4}{\lambda }
\right).
\end{align}
It is a simple exercise to check that the renormalization of the tree-level potential,
via the renormalization constants in \eqs{ZphiFermi}{ZlamFermi}, cancels the
$\phi$-dependent poles in \eq{1loopdivFermi}. We point out that in the
Fermi gauge the field $\phi$
gets only multiplicatively renormalized by the wavefunction
of the Higgs field, but this feature does not necessarily hold in other gauges.

After renormalization,
the one-loop contribution to the effective potential
in the $\overline{\rm{MS}}$ scheme reads
\begin{align}
\label{1loopEPFermi}
V_1(\phi)  &= \frac{1}{4 (4 \pi)^2} \left[
-12 \bar{m}_t^4 \left( \log\frac{\bar{m}_t^2}{\mu^2} - \frac{3}{2} \right)
+6 \bar{m}_W^4 \left( \log\frac{\bar{m}_W^2}{\mu^2} - \frac{5}{6}  \right) \right.  \\
&\left.+3 \bar{m}_Z^4 \left( \log\frac{\bar{m}_Z^2}{\mu^2} - \frac{5}{6}  \right)
+\bar{m}_h^4 \left( \log\frac{\bar{m}_h^2}{\mu^2} - \frac{3}{2}  \right)
+2 \bar{m}_{A^+}^4 \left( \log\frac{\bar{m}_{A^+}^2}{\mu^2} - \frac{3}{2} \right)  \right. \nonumber \\
&\left. +2 \bar{m}_{A^-}^4 \left( \log\frac{\bar{m}_{A^-}^2}{\mu^2} - \frac{3}{2}  \right)
+ \bar{m}_{B^+}^4 \left( \log\frac{\bar{m}_{B^+}^2}{\mu^2} - \frac{3}{2}  \right)
+ \bar{m}_{B^-}^4 \left( \log\frac{\bar{m}_{B^-}^2}{\mu^2} - \frac{3}{2}  \right)
\right], \nonumber
\end{align}
where the $\phi$-dependent mass terms are defined in
\eqs{mhphiFermi}{mtphiFermi} and \eqs{defmassZFermi}{defmassBpm}.
Note that at the tree-level minimum,
$m^2 = -\lambda \phi^2$, we have
$\bar{m}_{\chi} = 0$ and $\bar{m}_{A^\pm} = \bar{m}_{B^\pm} = 0$,
and the gauge dependence drops from $V_1(\phi)$.
The meaning of the gauge dependence of the standard model effective potential
will be discussed in \sect{gaugedep}.
In particular, for $\xi_W = \xi_B = 0$ one has $\bar{m}_{A^+} = \bar{m}_{B^+}
= \bar{m}_{\chi}$ and $\bar{m}_{A^-} = \bar{m}_{B^-} = 0$,
so that \eq{1loopEPFermi} reproduces the well-known one-loop result in the Landau gauge \cite{Coleman:1973jx}.

\subsection{Large-field behaviour of the standard model effective potential}

We shall see that the analysis of the stability of the electroweak ground state requires
the knowledge of the effective potential $V_{\rm eff}(\phi)$ at large values of the classical field
$\phi$. In this limit, the terms proportional to $\log \frac{\phi}{\mu}$ appearing in the one-loop correction to the effective potential grow large, and eventually spoil the reliability of perturbation theory. Such large logarithms need resummation in this case.
The standard technique to resum such logarithms is the renormalization group, which we review here.

For fixed values of the bare parameters, the effective
potential $V_{\rm eff}(\phi)$, does not depend on the renormalization scale $\mu$
(see~\eq{defep}).
This is expressed by the renormalization group equation
(RGE)
\begin{equation}
\label{RGE}
\left( \mu \frac{\partial}{\partial \mu} + \beta_i \frac{\partial}{\partial \lambda_i}
- \gamma \phi \frac{\partial}{\partial \phi} \right) V_{\rm{eff}}(\mu,\phi) = 0,
\end{equation}
where we have made explicit the dependence of $V_{\rm eff}$ on the scale $\mu$
and on the parameters of the theory (couplings and masses), collectively denoted by
$\lambda_i$.

The formal solution of the RGE in \eq{RGE} can be obtained by the method of the characteristics~\cite{Ford:1992mv}: the effective potential, as a function of $\mu$,
is constant along the characteristic curves $\lambda_i=\lambda_i(\mu),
\phi=\phi_i(\mu)$, solutions of the differential equations
\begin{align}
\label{defbf}
&\mu \frac{d \lambda_i(\mu)}{d \mu}=\beta_i;\qquad \lambda_i(\mu_0)=\lambda_i
\\
\label{defanomdim}
&\mu\frac{d \phi(\mu)}{d \mu}=-\gamma\phi(\mu);\qquad \phi(\mu_0)=\phi.
\end{align}

The usefulness of the renormalization group is that $\mu$ can be chosen in such a way that
the convergence of perturbation theory is improved.
For instance, a standard choice in vacuum stability analyses is $\mu= \phi$.
Without sticking, for the time being, to any specific choice of scale,
the RG improved effective potential can be formally written as
\begin{equation}
\label{1loopEPimproved}
V_{\rm{eff}} (\mu,\phi) =
\Omega_{\rm{eff}}(\mu,\phi)
+\frac{m_{\rm{eff}}^2(\mu,\phi)}{2} \phi^2
+ \frac{\lambda_{\rm{eff}}(\mu,\phi)}{4} \phi^4 .
\end{equation}
In the case of the standard model the full expressions of
$\Omega_{\rm{eff}}$, $m_{\rm{eff}}^2$ and $\lambda_{\rm{eff}}$ in the Fermi gauge
can be read off \eq{1loopEPFermi}.
We do not report those expressions explicitly here;
we just observe that $\Omega_{\rm{eff}}$ and $m_{\rm{eff}}^2$ are proportional to
$m^4(\mu)$ and $m^2(\mu)$, respectively,
where $m(\mu)$ is the $\overline{\rm MS}$ renormalized parameter of the Higgs potential,
essentially an electroweak mass scale up to logarithmic running.
Hence, in the $\phi \gg m$ limit \eq{1loopEPimproved} takes the simplified form
\begin{equation}
\label{1loopEPimprovedapprox}
V_{\rm{eff}} (\phi) \simeq \frac{1}{4} \lambda_{\rm{eff}}(\mu,\phi)\phi^4,
\end{equation}
where
\begin{equation}
\label{lameffapprox}
\lambda_{\rm{eff}}(\mu,\phi) = e^{4 \Gamma(\mu)} \left[ \lambda(\mu)
+ \frac{1}{(4\pi)^2} \sum_p N_p \kappa_p^2 (\mu)
\left( \log \frac{\kappa_p(\mu) e^{2 \Gamma(\mu)} \phi^2}{\mu^2} - C_p \right) \right]
\end{equation}
and
\beq
\Gamma(\mu) = - \int_{\mu_0}^\mu \gamma \frac{d\mu'}{\mu'}.
\eeq
The coefficients $N_p$, $C_p$ and $\kappa_p(\mu)$ appearing in \eq{lameffapprox}
are listed in \Table{tab:pvaluesFermi} in the Fermi gauge (the corresponding
Landau-gauge values are obtained by taking $\xi_W = \xi_B = 0$).
\begin{table*}[h]
  \begin{center}
  \begin{footnotesize}
      \begin{tabular}{|c|cccccc|}
      \hline
        $p$ & $t$ & $W$ & $Z$ & $h$ & $A^{\pm}$ & $B^{\pm}$ \\
        \hline
        $N_p$ & $-12$ & $6$ & $3$ & $1$ & $2$ & $1$ \\
        $C_p$ & $3/2$ & $5/6$ & $5/6$ & $3/2$ & $3/2$ & $3/2$ \\
        $\kappa_p$ & $\frac{y_t^2}{2}$ & $\frac{g^2}{4}$ & $\frac{g^2+g'^2}{4}$ & $3 \lambda$
        & $\frac{1}{2} \left( \lambda \pm \sqrt{\lambda^2 - \lambda \xi_W g^2} \right)$
        & $\frac{1}{2} \left( \lambda \pm \sqrt{\lambda^2 - \lambda (\xi_W g^2 + \xi_B g'^2)} \right)$  \\
        \hline
        \end{tabular}
    \caption{\label{tab:pvaluesFermi} The $p$-coefficients entering the expression of $\lambda_{\rm{eff}}$
    in \eq{lameffapprox} for the Fermi gauge.
      }
      \end{footnotesize}
  \end{center}
\end{table*}
Note that the gauge dependence of the RG improved effective potential is twofold:
the gauge fixing parameters appear both in the couplings $\kappa_p$ (cf.~\Table{tab:pvaluesFermi}),
and in the anomalous dimension of $\phi$ (cf.~\eq{ZphiFermi}).

Eq.~(\ref{1loopEPimprovedapprox}) can be further simplified by choosing
$\mu\sim\phi$, so that the logarithms in the last term of
\eq{lameffapprox} are small, and neglecting the RGE evolution of the field $\phi$,
which amounts to taking $\Gamma(\mu)\sim 0$ (this is formally correct only at the LO). We hence obtain
\beq
V_{\rm eff}(\phi)\simeq \frac{1}{4}\lambda(\phi) \phi^4.
\label{Veffapp}
\eeq
The RG evolution of $\lambda$ in the $\overline{\rm MS}$ scheme is displayed in the left panel of \fig{Runlam}
for central values of the standard model parameters,
employing one-, two- and three-loop beta functions and two-loop matching at the $M_t$
scale \cite{Buttazzo:2013uya}. Similarly, in the right panel of \fig{Runlam} we show the extrapolation of the
SM effective potential at high energies in the approximation of \eq{Veffapp}.

Note that within such an approximation the gauge dependence drops out,
even though it is not perturbatively consistent to neglect the anomalous dimension and the
fixed-order contribution of the effective potential in \eq{lameffapprox} when the running of $\lambda$ is considered beyond one loop.
This notwithstanding, already at the one-loop level the above approximation captures the most important physical feature:
around $\mu = 10^{8}$ GeV the coupling $\lambda$ becomes negative, thus triggering the instability of the Higgs potential.
The electroweak minimum (whose fine structure is not visible in \fig{Runlam}) is not anymore the absolute
minimum of the Higgs potential after considering radiative corrections. Since $\lambda$ keeps running
negative at large field values, the effective potential must be ultraviolet completed in such a way that it is bounded from below and
a second, deeper minimum arises.
The latter might be due to Planck-scale physics or simply to the radiative self-completion of the SM,
given the fact that $\lambda$ crosses again the zero at trans-planckian energies and grows positive
due to the Landau pole of the hypercharge.
\begin{figure}[t]
\centering
\includegraphics[width=0.45\textwidth]{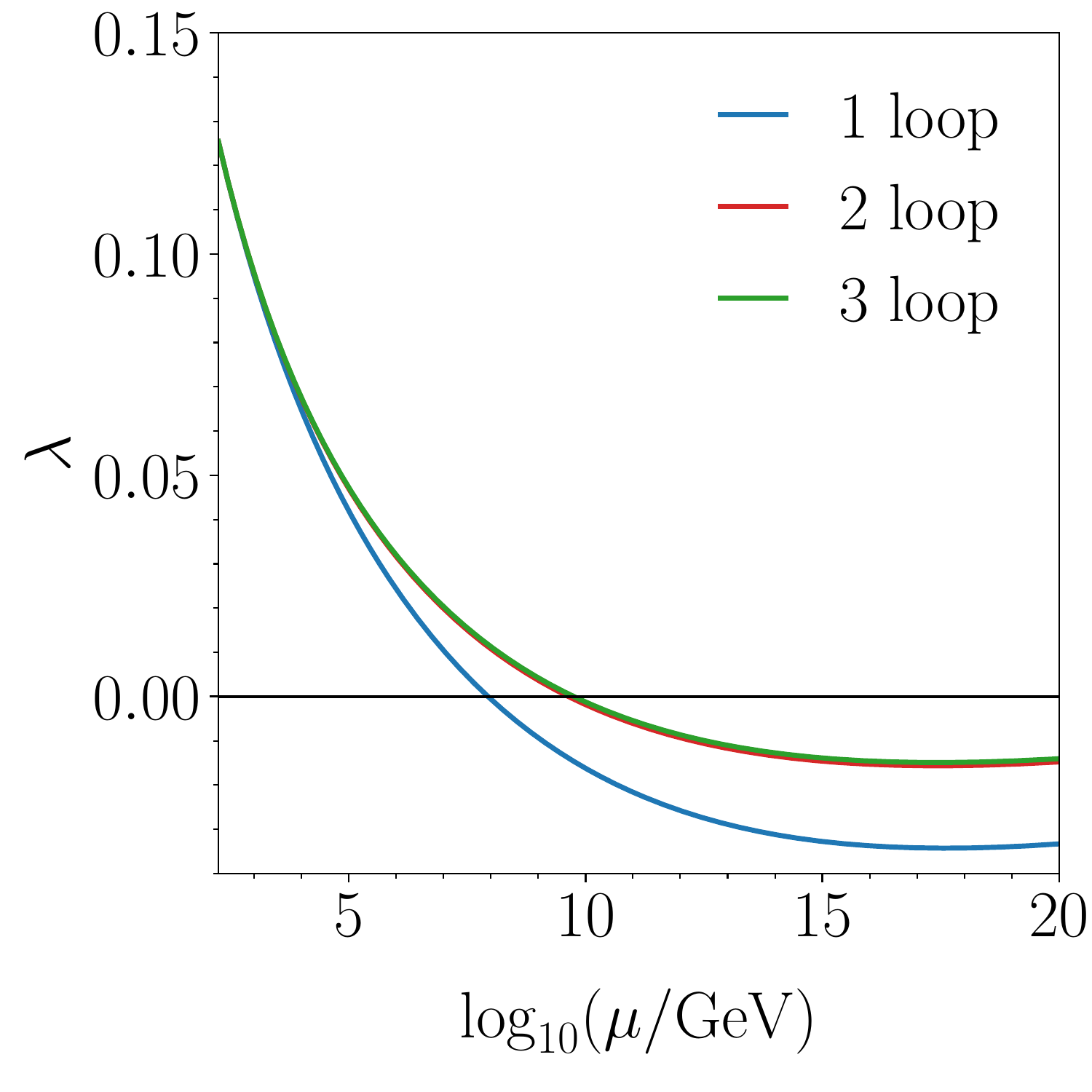} \qquad
\includegraphics[width=0.45\textwidth]{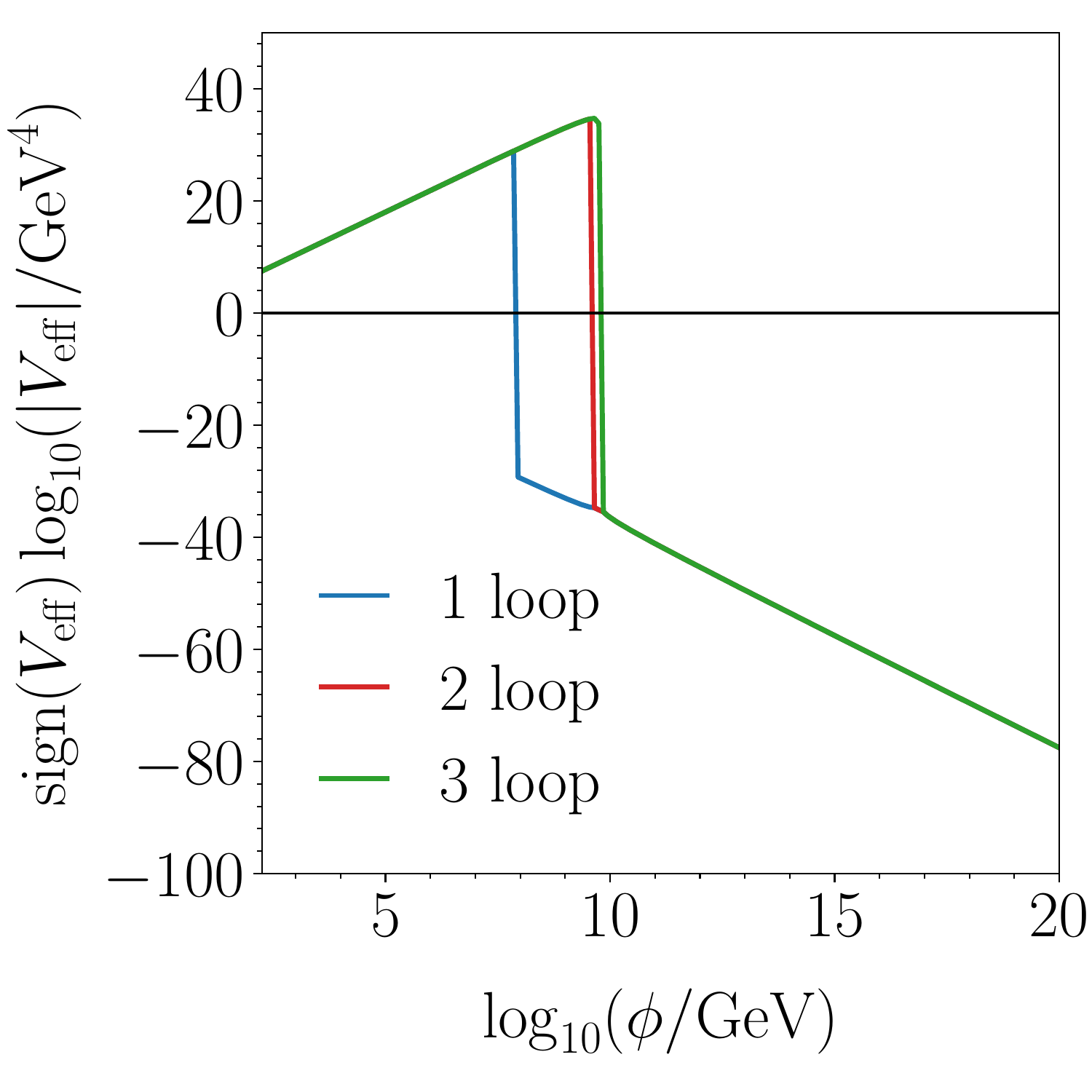}
\caption{\label{Runlam}
Left panel: RG evolution of $\lambda$ in the standard model at one, two and three loops (respectively in blue, red and green).
Right panel: standard model effective potential in the approximation $V_{\rm eff} \simeq \lambda(\phi) \phi^4 / 4$.
}
\end{figure}

\subsection{Gauge dependence}
\label{gaugedep}

Although the effective potential turns out to be a very useful tool for the study
of spontaneous symmetry breaking, there are some subtleties related to its
physical interpretation which are worth to be discussed.
In particular, we are going to review here the role of the gauge dependence
of the effective potential in the vacuum stability analysis
and, in the next section, the so-called convexity and reality issues.

The gauge dependence of the effective potential raises the question
of which are the physical observables entering the vacuum stability analysis.
To fix the ideas, let us assume that all the parameters of the standard model are exactly determined,
but the Higgs boson mass.
After choosing the renormalization scale $\mu$, the RG improved effective potential,
$V_{\rm{eff}} (\phi, M_h; \xi)$, is a function of $\phi$, the Higgs pole mass
$M_h$,
and the gauge fixing parameters, collectively denoted by $\xi$.
One can think of $M_h$ as an order parameter, whose variation modifies the shape of the effective potential, as sketched in \fig{Mchplot}.
\begin{figure}[t]
\centering
\hspace{-1cm}\includegraphics[width=0.75\textwidth]{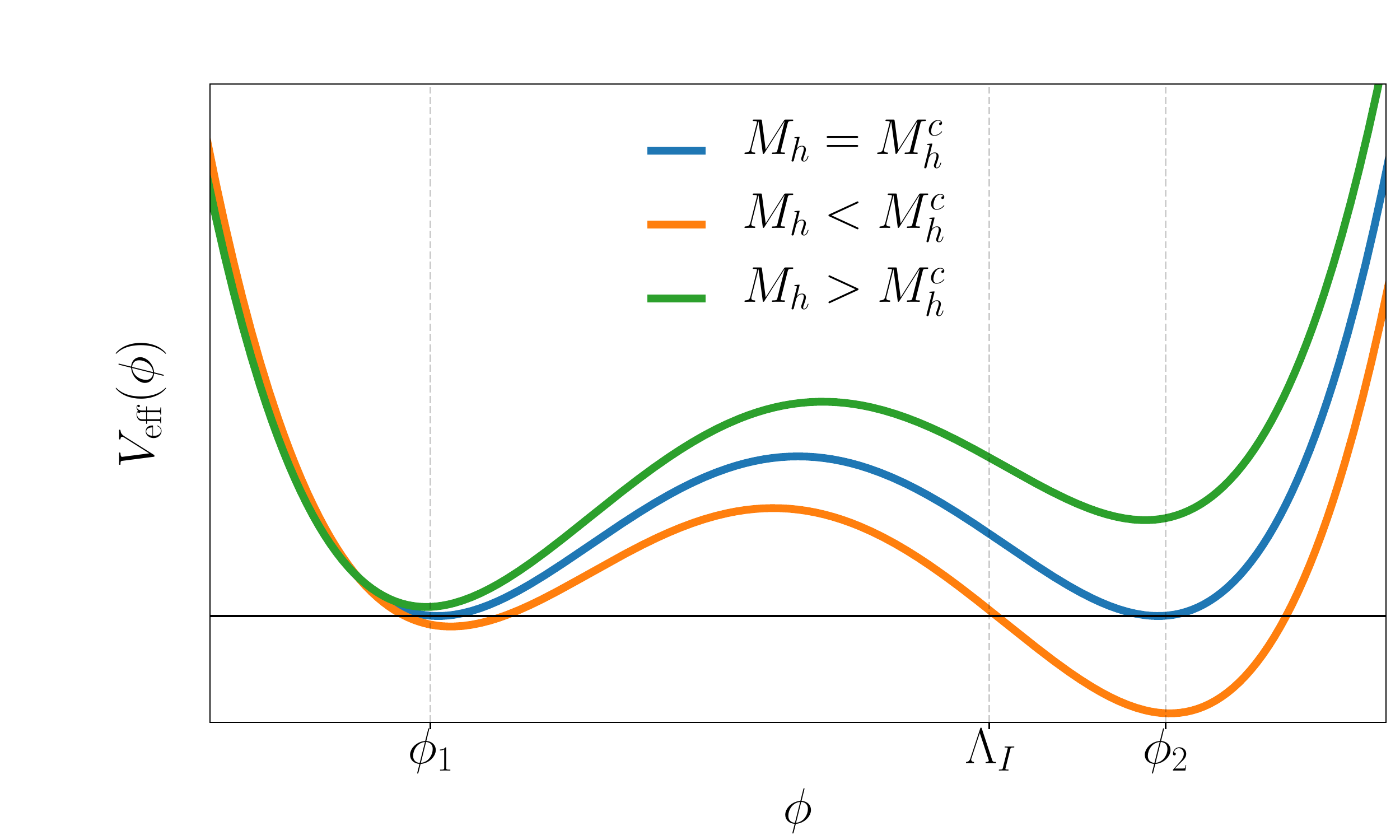}
\caption{\label{Mchplot}
Schematic representation of the standard model effective potential for different values of the Higgs boson mass.
For $M_h < M_h^c$, the electroweak vacuum is unstable.}
\end{figure}

The absolute stability bound on the Higgs boson mass is obtained by defining a ``critical'' mass, $M_h^c$,
for which the value of the effective potential at the electroweak minimum, $\phi_{1}$,
and at a second minimum, $\phi_2 > \phi_{1}$, are the same. Analytically,
this translates into the three conditions
\begin{align}
\label{absvacstab1}
& V_{\rm{eff}} (\phi_{1}, M_h^c; \xi) - V_{\rm{eff}} (\phi_2, M_h^c; \xi) = 0 ,\\
\label{absvacstab2}
& \left. \frac{\partial V_{\rm{eff}}}{\partial \phi} \right|_{\phi_{1}, M_h^c} = \left. \frac{\partial V_{\rm{eff}}}{\partial \phi} \right|_{\phi_2, M_h^c} = 0 .
\end{align}
We have seen in the previous section that, in the $\phi \gg \phi_{1}$ limit, the standard model effective potential is well approximated by \eq{Veffapp},
\begin{equation}
\label{Veffapprox}
V_{\rm{eff}} (\phi)
\simeq \frac{1}{4} \lambda_{\rm{eff}}(\phi) \phi^4 .
\end{equation}
Hence, the absolute stability conditions in \eqs{absvacstab1}{absvacstab2}
can be equivalently rewritten in the following way:
\begin{align}
\label{absvacstab1mod}
\lambda_{\rm{eff}} (\phi_2, M_h^c; \xi) = 0 ,\\
\label{absvacstab2mod}
\left. \frac{\partial \lambda_{\rm{eff}}}{\partial \phi} \right|_{\phi_2, M_h^c} = 0 ,
\end{align}
up to $(\phi_{1} / \phi_2)^2 \ll 1$ corrections.

On the other hand, due to the explicit presence of $\xi$ in the vacuum
stability condition, it is not obvious a priori which are
the physical (gauge-independent) observables entering the vacuum stability analysis.
The basic tool, in order to capture the gauge-invariant content of the effective potential is given by the
Nielsen identity \cite{Nielsen:1975fs}
\begin{equation}
\label{NielsenIdbis}
\left( \frac{\partial}{\partial \xi} + C(\phi, \xi)
\frac{\partial}{\partial \phi} \right) V_{\rm eff}(\phi,\xi) = 0 ,
\end{equation}
where $C(\phi, \xi)$ is a correlator involving the ghost fields and the
gauge-fixing functional, whose explicit expression
will not be needed for our argument. \eq{NielsenIdbis} is valid for the class
of linear gauges and can be derived
from the BRST non-invariance of a composite operator involving the ghost field and the gauge fixing functional
(see e.g.~\cite{Metaxas:1995ab}).

The identity in \eq{NielsenIdbis} has the following physical interpretation: the
effective potential is gauge independent where it is
stationary and hence spontaneous symmetry breaking is a gauge-invariant statement.
We can actually use the Nielsen identity, in combination
with the vacuum stability condition in \eqs{absvacstab1}{absvacstab2},
to formally prove that the critical Higgs boson mass, $M_h^c$, is a gauge-independent
quantity \cite{DiLuzio:2014bua}.
To this purpose, let us assume that simultaneously inverting \eqs{absvacstab1}{absvacstab2}
would yield gauge dependent field values and critical Higgs boson mass:
$ \phi_1 = \phi_1 (\xi)$, $\phi_2 = \phi_2 (\xi)$ and $M_h^c = M_h^c (\xi)$.
The total differential of \eq{absvacstab1} with respect to $\xi$ then reads
\begin{multline}
\label{totaldeiffxiA}
\left. \frac{\partial V_{\rm{eff}}}{\partial \phi} \right|_{\phi_1, M_h^c} \frac{\partial \phi_1}{\partial \xi}
+ \left. \frac{\partial V_{\rm{eff}}}{\partial M_h} \right|_{\phi_1, M_h^c} \frac{\partial M_h^c}{\partial \xi}
+ \left. \frac{\partial V_{\rm{eff}}}{\partial \xi} \right|_{\phi_1, M_h^c}  \\
=\left. \frac{\partial V_{\rm{eff}}}{\partial \phi} \right|_{\phi_2, M_h^c} \frac{\partial \phi_2}{\partial \xi}
+ \left. \frac{\partial V_{\rm{eff}}}{\partial M_h} \right|_{\phi_2, M_h^c} \frac{\partial M_h^c}{\partial \xi}
+ \left. \frac{\partial V_{\rm{eff}}}{\partial \xi} \right|_{\phi_2, M_h^c} .
\end{multline}
The first terms in both the l.h.s.~and the r.h.s.~of \eq{totaldeiffxiA} vanish because of the stationary conditions in \eq{absvacstab2}.
The third terms in both the l.h.s.~and the r.h.s.~of \eq{totaldeiffxiA} vanish for the same reason, after using the Nielsen identity.
Hence, we are left with
\begin{equation}
\label{leftwith1}
\left( \left. \frac{\partial V_{\rm{eff}}}{\partial M_h} \right|_{\phi_1, M_h^c}
- \left. \frac{\partial V_{\rm{eff}}}{\partial M_h} \right|_{\phi_2, M_h^c} \right)  \frac{\partial M_h^c}{\partial \xi} = 0 .
\end{equation}
Since the expression in the bracket of \eq{leftwith1} is in general different from zero, one concludes that
\begin{equation}
\label{leftwith}
\frac{\partial M_h^c}{\partial \xi} = 0 ,
\end{equation}
namely, the critical Higgs boson mass does not depend on the gauge-fixing parameters,
and hence telling whether the electroweak minimum is absolutely stable or not
is a physical statement.
Note, however, that a non-consistent use of perturbation
theory might still be responsible for a residual (spurious)
gauge dependence in the determination of $M_h^c$ \cite{Andreassen:2014eha,Andreassen:2014gha}.

On the other hand, field values (as for instance
the instability scale) are essentially gauge-dependent quantities.
The standard model vacuum instability scale is operatively defined as the field value $\phi = \Lambda_I$,
for which the effective potential has the same depth as the electroweak minimum (see \fig{Mchplot}).
This is analytically expressed by
\begin{equation}
\label{definstscale2}
V_{\rm eff} (\Lambda_I; \xi) = V_{\rm eff} (\phi_1; \xi) .
\end{equation}
The r.h.s.~of \eq{definstscale2} is a gauge-independent quantity, since $\phi_1$ is by definition a minimum
and we can apply the Nielsen identity.
Hence, by solving \eq{definstscale2}, one has in general
$\Lambda_I = \Lambda_I (\xi)$.
In particular, by taking the total differential of \eq{definstscale2}
with respect to $\xi$, we get
\begin{equation}
\label{totdiffdefinstscale2}
\left. \frac{\partial V_{\rm eff}}{\partial \phi} \right|_{\Lambda_I} \frac{\partial \Lambda_I}{\partial \xi}
+ \left. \frac{\partial V_{\rm eff}}{\partial \xi} \right|_{\Lambda_I} = 0.
\end{equation}
By using the Nielsen identity, we can substitute back the second term in \eq{totdiffdefinstscale2}, thus obtaining
\begin{equation}
\label{totdiffdefinstscale2bis}
\left( \frac{\partial \Lambda_I}{\partial \xi} - C(\Lambda_I, \xi) \right)  \left. \frac{\partial V_{\rm eff}}{\partial \phi} \right|_{\Lambda_I} = 0.
\end{equation}
Since, in general, $\Lambda_I$ is not an extremum of the effective potential, \eq{totdiffdefinstscale2bis} yields
\begin{equation}
\label{totdiffdefinstscale2bisyields}
\frac{\partial \Lambda_I}{\partial \xi} = C(\Lambda_I, \xi).
\end{equation}
It turns out that by varying $\xi$ within its perturbative domain,
the scale $\Lambda_I$ in the standard model suffers from a gauge-fixing uncertainty of up to one order of magnitude
in the Fermi gauge \cite{DiLuzio:2014bua}.
Although $\Lambda_I$ is clearly unphysical, one can still
identify some gauge-invariant scales related with the Higgs potential instability
(see e.g.~Ref.~\cite{Espinosa:2016nld}).

\subsection{Convexity and reality}
\label{convex}

The effective potential $V_{\rm eff}(\phi_c)$, being defined as the Legendre transform of
the generating functional of the connected Green's functions,
must be a convex function of its argument.\footnote{Some mathematical properties
of Legendre transforms are reviewed in \app{app:LegTra}.}
Moreover, as we are going to show,
it carries the following physical interpretation:
it is the expectation value of the energy density
$\langle \Psi | \mathcal{H} | \Psi \rangle$,
for the lowest-energy state $|\Psi\rangle$, subject to the constraints $\langle \Psi | \Psi \rangle = 1$ and
$\langle \Psi | \phi(x) | \Psi \rangle = \phi_c$. This also implies that $V_{\rm eff}$ must be real.

These two properties of the effective potential are in apparent contrast with what done so far,
since any scalar potential relevant for spontaneous symmetry breaking has
non-convex regions. Moreover, the convexity and reality problems are
closely related. In fact, the one-loop correction to $V_{\rm eff}$ is proportional to
$[V''(\phi_c)]^2 \log V''(\phi_c)$,
so that non-convex regions ($V''(\phi_c)<0$) lead to an imaginary part.

It is commonly understood that this paradox has nothing to do
with the breakdown of perturbation theory, but it is rather an issue of terminology
(see e.g.~\cite{Fujimoto:1982tc,Dannenberg:1987fw,Sher:1988mj}).
In the derivation of $V_{\rm eff}$ in \sect{sec:effpot} we have implicitly assumed that
$\phi_c[J]$ (defined via the relation $\phi_c = \delta W / \delta J$) is a
single-valued function. It can be shown that whenever the tree-level potential has non-convex regions,
the relation between $\phi_c$ and $J$ becomes multi-valued and the standard one-loop
derivation of what we improperly called the effective potential gets modified \cite{Fujimoto:1982tc}.
Specifically, the saddle point approximation employed in the background field method
of \sect{sec:bkgd} features multiple stationary points with respect to which the effective action should be expanded,
and the true effective potential is obtained by summing over all of them.

It is therefore more appropriate to distinguish between two objects:
$V_{\rm eff}$, defined in terms of the Legendre transform of the generating functional of connected Green's functions,
and $V_{\rm 1PI}$,
defined as the negative of the spatially independent part of the generating functional of the 1PI Green's functions
(cf.~\eq{defep}).
The two generating functionals are related by (see e.g.~\cite{Brandenberger:1984cz})
\beq
\exp{\left[\frac{i}{\hbar}\left(W[J]+\mathcal{O}(\hbar)\right)\right]} = \mathcal{N} \int \mathcal{D}\phi_c
\exp{\left[\frac{i}{\hbar}\left(\Gamma_{\rm 1PI}[\phi_c]+\int d^4x \, J(x) \phi_c(x) \right)\right]} ,
\eeq
for some irrelevant normalization constant $\mathcal{N}$.
In the semi-classical limit, $\hbar \to 0$, we can use the saddle point approximation to obtain
\beq
W[J] = \left. \left[ \Gamma_{\rm 1PI}[\phi_c] + \int d^4x \, J(x) \phi_c(x) \right] \right|_{\frac{\delta \Gamma_{\rm 1PI}}{\delta \phi_c}=-J} .
\eeq
Hence, $W[J]$ is the Legendre transform of $\Gamma_{\rm 1PI}[\phi_c]$.
On the other hand, the effective action was defined as the Legendre transform of $W[J]$
\beq
\label{effact2}
\Gamma_{\rm eff} [\phi_c] = \left. \left[ W[J] - \int d^4x \, J(x) \phi_c(x) \right] \right|_{\frac{\delta W}{\delta J}=\phi_c} .
\eeq
Therefore $\Gamma_{\rm eff} [\phi_c]$ is the double Legendre transform of $\Gamma_{\rm 1PI}[\phi_c]$.
From the properties of Legendre transforms it follows that $\Gamma_{\rm eff} [\phi_c]$ is the
convex envelope of $\Gamma_{\rm 1PI}[\phi_c]$. In turn, this implies that $V_{\rm eff}$ is
the convex envelope of $V_{\rm 1PI}$.

Note that the formalism behind $V_{\rm eff}$
is defined in the thermodynamical limit, where the system has relaxed on the ground state
after an infinite amount of time.
On the other hand, in most applications, we are rather interested
in metastable regimes described by field configurations localized
around a local minimum. The goal of this section is to show that
$V_{\rm 1PI}$ is actually the appropriate object in order to describe this physical situation.

Following Coleman \cite{Coleman:1985rnk}, we start by considering the quantum mechanical problem of constructing a
state $|a\rangle$ that is a stationary state of the quadratic form $\langle a|H|a \rangle$, subject to the constraints
$\langle a | a \rangle = 1$ and $\langle a | A | a \rangle = A_c$, for some Hermitian operator $A$.
This can be cast in the variational problem
$\delta \langle a | H - E - J A | a \rangle = 0$,
where $E$ and $J$ are Lagrangian multipliers.
The variational equation leads to
\beq
\label{vareqpert}
[H - E - J A] | a \rangle = 0 ,
\eeq
from which we see that $| a \rangle$ is an eigenstate of the perturbed Hamiltonian $H-JA$,
with energy $E$. \eq{vareqpert} allows us to express $| a \rangle = | a (E,J) \rangle$ and,
further using the normalization condition, we can express $E$ in terms of $J$.
Since we are eventually interested in the dependence on $A_c$,
the relation between $J$ and $A_c$ can be obtained via
standard perturbation theory\footnote{This relation holds true in general (see e.g.~\cite{Sher:1988mj}).
However, it is sufficient for the sake of the argument to treat $J$ as a perturbation.}
\beq
A_c (J)= \langle a (J) | A | a (J) \rangle = - \frac{dE}{dJ} .
\eeq
Hence, we conclude that $|a \rangle$ is the stationary state of the quantity
\beq
\langle a (J)|H|a (J) \rangle = E (J)+ J A_c = E (J) - J \frac{dE}{dJ} .
\eeq
The generalization to quantum field theory is obtained by replacing:
$|a \rangle \to |\Psi \rangle$, $H\to \mathcal{H}$, $A \to \phi$, $A_c \to \phi_c$, $E \to W$,
which corresponds to the definition of the effective action in \eq{effact2}.
Hence, we conclude that
\beq
V_{\rm eff} (\phi_c) = \langle \Psi | \mathcal{H} | \Psi \rangle ,
\eeq
for a state $| \Psi \rangle$ such that $\delta \langle \Psi | \mathcal{H} | \Psi \rangle = 0$,
under the constraints $\langle \Psi | \Psi \rangle = 1$ and
$\langle \Psi | \phi(x) | \Psi \rangle = \phi_c$.

To investigate this further, let us consider for definiteness the case of the double-well potential
$V(\phi) = -\frac{1}{2}\mu^2\phi^2 + \frac{1}{4} \lambda \phi^4$ with minima in
$\phi = \pm \sigma \equiv \sqrt{\mu^2/\lambda}$ (cf.~\fig{fig:WeinbergWu}).
\begin{figure}[t]
\centering
\includegraphics[width=0.5\textwidth]{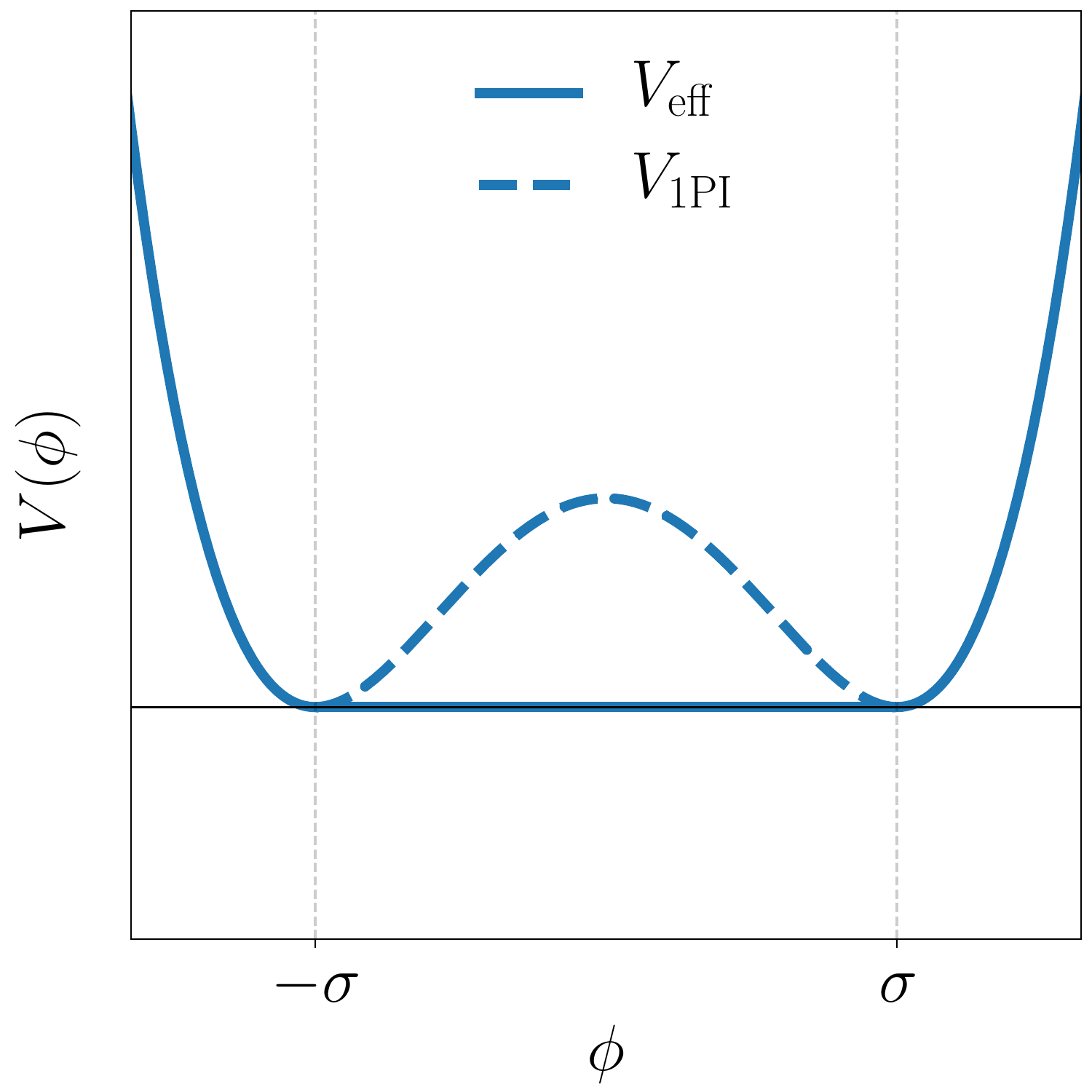}
\caption{\label{fig:WeinbergWu}
Example of a non-convex tree-level potential.
The true effective potential ($V_{\rm eff}$) is the convex envelope of $V_{\rm 1PI}$.
}
\end{figure}
The classical potential is clearly non-convex in the region $\phi < \abs{\sigma}$.
Following Weinberg and Wu \cite{Weinberg:1987vp}, the main features of the effective potential
can be understood via a classical argument. Let us consider
the classical analog of $\langle \Psi | \phi(x) | \Psi \rangle$
to be the spatial average of $\phi$, denoted by $\bar \phi$.
Using the physical interpretation of the effective potential,
the classical analog of the effective potential, $V_c$,
is the minimum value of the energy density
among all the states with a given value of $\bar \phi$.
For $\phi > |\sigma|$ the energy density
\beq
\mathcal{H} = \frac{1}{2} \dot{\phi}^2 + \frac{1}{2} (\bm\nabla \phi)^2 + V(\phi) ,
\eeq
is mimimized by a static homogeneous
configuration with $\phi(x) = \bar \phi$ everywhere, and $V_c$ is equal to $V(\phi)$.
If $\phi < |\sigma|$, there is a non-trivial way in which the energy density can be minimized,
namely via an \emph{inhomogeneous} state
\beq
\phi(x) = \left\{
\begin{array}{l}
+\sigma \quad \text{for a fraction $f$ of space} \\
-\sigma \quad \text{otherwise}
\end{array}
\right.
\eeq
where $0 \leq f \leq 1$ is fixed by requiring that the spatial average of $\phi(x)$,
$\sigma f - \sigma (1-f)$, is equal to $\bar \phi$, which yields
$f = (\bar\phi +\sigma) / (2\sigma)$.
Hence, in the $\phi < |\sigma|$ region the energy density is $V(\pm \sigma) < V(\phi)$.
Note that the gradient term relevant at the boundaries of the two domains can be neglected
in the infinite volume limit.
$V_c$ is thus flat in the region between the two minima,
and corresponds to the convex envelop of the $V(\phi)$.
In the quantum theory,
the main difference is that the state which minimizes the energy density is not a spatially
inhomogeneous mixture of two phases but rather a quantum superposition of two vacuum states.

If one is interested only in homogeneous states,
i.e.~states in which the field is concentrated about a specific
value,\footnote{This is clearly the case in the physical problem considered in the present review:
we always start from a homogeneous field configuration localized around the electroweak minimum
$\phi(x) \approx v$ and ask about its stability.}
then the effective potential is not the appropriate quantity to describe the region $\phi < |\sigma|$.
In that case one would need a modified effective potential which might be defined
as the expectation value of the energy density in a state $|\Psi\rangle$ which minimizes
$\langle \Psi | \mathcal{H} | \Psi \rangle$ subject to the condition $\langle \Psi | \phi(x) | \Psi \rangle = \phi_c$
and to the further restriction that the wave functional for $| \Psi \rangle$ be concentrated
on configurations with $\phi(x) \approx \phi_c$. Hence, to leading order the minimum energy density is $V(\phi_c)$
which does not need to be convex. Zero point fluctuations are further taken into account by
$V_{\rm 1PI}(\phi_c)$, whose real part provides the relevant potential to address localized field
configurations prior to a phase transition.
Note that the restriction to localized configurations
does not commute with the Hamiltonian
(e.g.~in the case if the double-well potential the energy eigenstates are
even/odd linear combinations of the states localized in $\pm \sigma$),
thus these configurations are unstable and can be interpreted as imaginary contributions to the energy \cite{Weinberg:1987vp}.
This instability, related to the imaginary part of the perturbatively computed $V_{\rm 1PI}$,
should not to be confused with the non-perturbative decay of the false vacuum itself (for instance,
one could consider a non-convex $V_{\rm 1PI}$ with field configurations localized around the true vacuum).

Following a standard terminology, we will keep to refer to $V_{\rm eff}$ as the
(localized) effective potential.

\section{Decay of the standard model vacuum}
\label{sec:SMstability}

The stability of the electroweak ground state is a time-honoured subject
\cite{Krive:1976sg,Krasnikov:1978pu,Maiani:1977cg,Politzer:1978ic,Hung:1979dn,Cabibbo:1979ay,Linde:1979ny,Lindner:1988ww,Sher:1988mj,Arnold:1989cb,Arnold:1991cv}.
Prior to the Higgs discovery in 2012 \cite{Aad:2012tfa,Chatrchyan:2012xdj}
it served as a theoretical tool to set a lower bound on the Higgs boson mass \cite{Sher:1993mf,Altarelli:1994rb,Casas:1994qy,Espinosa:1995se,Isidori:2001bm,Espinosa:2007qp,Ellis:2009tp}.
Recently, the analysis of the standard model vacuum stability has reached a higher level of complexity
(see e.g.~\cite{Bezrukov:2012sa,Degrassi:2012ry,Masina:2012tz,Buttazzo:2013uya,Bednyakov:2015sca,Iacobellis:2016eof,Andreassen:2017rzq,Chigusa:2017dux}).
In this section we describe the basic aspects of the calculation of the standard model vacuum decay rate
and we introduce the so-called standard model phase diagram, namely the parameter space region in the standard model for which the electroweak vacuum is absolutely stable, metastable (i.e.~with a sufficiently 
long lifetime compared to the age of the Universe) or worryingly unstable.

\subsection{The origin of the standard model instability}

The case of the standard model of electroweak interactions is different from the simple examples described in \sect{tunnellingQFT}. Indeed, the tree-level scalar potential of the standard model
\beq
V_0(\phi)=\frac{1}{2}m^2\phi^2+\frac{1}{4}\lambda\phi^4
\eeq
with $m^2<0$ and $\lambda>0$ has a local maximum at $\phi=0$, an absolute minimum at
$\phi=\sqrt{-\frac{m^2}{\lambda}}=v$ (the electroweak minimum), with $v\sim 246$~GeV, and no other stationary points. Hence, at the classical level, there is no instability of the ground state.

The situation changes when loop corrections are included. 
As we have seen in the previous
section, the effective potential behaves as $\lambda(\phi)\phi^4$ for large values
of the classical field $\phi$, and the running coupling $\lambda(\mu)$ becomes negative
at $\mu=\Lambda_I\sim 10^{10}$~GeV if the top quark is sufficiently heavy. If this is the case, the electroweak
minimum is not the absolute minimum of the effective potential: a second, deeper
minimum, located
somewhere between $10^{10}$~GeV and the position of the Landau singularity,
must therefore exist, and quantum fluctuations may induce tunnelling from the
electroweak minimum to the absolute minimum.

The study of the instability of the electroweak ground state must therefore be carried on consistently
at the level of loop corrections.

Since such large energy scales and field values are involved, the calculation of the tunnelling
probability is in general influenced by the ultraviolet completion of the standard model (if any), and
by gravitational effects; it is however reasonable to take a conservative point of view,
and perform a calculation of the false vacuum lifetime under the assumption that the standard model is a correct description of our Universe up to scales of the order of the Planck mass.

\subsection{The standard model bounce}

We have seen in \sect{tunnellingQFT} that the calculation of the decay rate of an unstable ground state in quantum field theory is based on an $\O(4)$-invariant bounce, $\phi(x)=\phi_b(r)$, solution of
the field equation
\beq
\phi''_b(r)+\frac{3}{r}\phi_b'(r)=V'(\phi_b)
\label{bounce}
\eeq
where $V(\phi)$ is the scalar potential expressed in terms of renormalized parameters,
with boundary conditions
\begin{align}
\label{boundary1}
\lim_{r\to\infty}\phi_b(r)&=v, \\
\label{boundary2}
\phi_b'(0)&=0.
\end{align}
In the case of the standard model the value of $\phi$
(which denotes the Higgs background field, see \eqs{eq:Higgspot}{Hshift})
at the true minimum of the scalar potential is typically many orders of magnitudes larger that the
electroweak order parameter, $v\sim 246$~GeV. It seems therefore appropriate
to simply neglect the finite value of $v$ with respect to the typical energy scales
involved in the problem of the quantum stability of the electroweak ground state.
We will therefore adopt the boundary condition
\beq
\lim_{r\to\infty}\phi_b(r)=0,
\label{boundary1sm}
\eeq
and we will approximate the scalar potential in the unstable region by
\beq
V(\phi)=\frac{1}{4}\lambda(\phi) \phi^4,
\label{SMpot}
\eeq
with the renormalized coupling $\lambda(\mu)$ computed at a scale $\mu\sim\phi$ to reduce the impact of large logarithms.
Eq.~(\ref{bounce}) takes the form
\beq
\phi''_b+\frac{3}{r}\phi'_b=\phi_b^3\left[\lambda(\phi_b)+\frac{1}{4}\beta(\lambda(\phi_b))\right],
\label{bounce2}
\eeq
where
\beq
\beta(\lambda(\mu))=\mu\frac{d\lambda(\mu)}{d\mu}.
\eeq
(Although not explicitly indicated, the function $\beta(\lambda)$ actually depends on all
coupling constants in the theory, and in particular on the top Yukawa coupling, which drives
$\lambda$ to negative values, as we have seen in \sect{effatceffpot}.)
A bounce solution of \eq{bounce2}, that is, a solution with $\phi'_b(0)=\phi_b(\infty)=0$,
certainly exists, because the scalar potential \eq{SMpot} has a local minimum in $\phi=0$
and an absolute minimum at some very large value of $\phi$, and the undershoot-overshoot
argument of \sect{overunder} can be applied. We recall that, by the same argument,
the value of the bounce at $r=0$ is larger than the position of the first zero of
$\lambda(\mu)$, $\mu\sim\Lambda_I=10^{10}$~GeV. At such large values of the classical field, the use of the approximation
(\ref{SMpot}) is largely justified; the mass term would induce a correction 
\beq
\frac{\Delta\phi(0)}{\phi(0)}<\frac{v}{\Lambda_I}\sim 10^{-8}.
\eeq
An analytic solution of \eq{bounce2} cannot be found. However, it can be checked that the dependence of 
the combination
\beq
\lambda(\phi)+\frac{1}{4}\beta(\lambda(\phi))
\eeq
on $\phi$ in the region where
$\lambda(\phi)<0$ is very weak: it is only logarithmic, and it is suppressed by 
powers of the coupling constants. An approximate solution
can therefore be found by replacing the r.h.s.~of \eq{bounce2} by a negative constant $\lambda$. 
As we are going to show below, a good approximation is obtained by fixing 
$\lambda=\lambda(\bar\mu)$ 
where $\bar\mu$ is such that $\beta(\lambda(\bar\mu))=0$.
Eq.~(\ref{bounce2}) then becomes
\beq
\phi''_b(r)+\frac{3}{r}\phi'_b(r)=\lambda\phi_b^3(r).
\label{bouncexSM}
\eeq
(Incidentally, we note that we are now precisely in the situation described in \sect{barrierQFT}, namely a scalar potential with no barrier.) As anticipated in \sect{sec:zeromodes} and shown explicitly in \app{app:FLinstanton},
Eq.~(\ref{bouncexSM}) with  the
boundary conditions (\ref{boundary2}, \ref{boundary1sm}) can be solved analytically.
There is in fact an infinity of solutions:
\beq
\phi_b(r)=\frac{\phi_b(0)}{1+\frac{|\lambda|}{8}\phi_b^2(0) r^2},
\label{SMh}
\eeq
parametrized by $\phi_b(0)$, which can take any positive value. Eq.~(\ref{SMh}) is often 
rewritten in terms of a length scale $R$, rather than by its value at $r=0$:
\beq
\phi_b(r)=\sqrt{\frac{8}{|\lambda|}}\frac{R}{r^2+R^2}.
\label{SMhR}
\eeq
The parameter $R$ can be physically interpreted as the approximate radius of the bubble, in three dimensional space,
where the system has jumped from the false-vacuum configuration $\phi_b=0$
to a non-zero value beyond the barrier. Indeed, $\phi_b(r)$ has its
maximum in $r=0$:
\beq
\phi_b(0)=\sqrt{\frac{8}{|\lambda|}}\frac{1}{R},
\label{phi-0}
\eeq
and drops
to zero for $r\gg R$. Note that $\phi_b(R)=\frac{\phi_b(0)}{2}$, which justifies our definition of $R$ as the radius of the bubble.

As already noted in \sect{sec:zeromodes}, this degeneracy is a consequence of scale invariance. Indeed, the underlying action functional
\beq
S[\phi]=\int d^4x\,\left[\frac{1}{2}(\partial_\mu\phi)^2+\frac{1}{4}\lambda\phi^4\right]
\eeq
is invariant upon scale transformations:
\beq
\phi(r)\to a\phi(ar),
\label{scale}
\eeq
since it does not contain any dimensionful parameter.
The proof is straightforward: 
\beq
S[a\phi(ar)]
=\int d^4x\,\left[\frac{a^2}{2}\left(\frac{\partial\phi(ar)}{\partial x_\mu}\right)^2+\frac{a^4}{4}\lambda\phi^4(ar)\right]
=S[\phi]
\eeq
after the integration variable change $y=ax$. As a consequence,
if $\phi(r)$ is a solution of
\eq{bouncexSM}, than $a\phi(ar)$ is also a solution, for any choice of the scale factor $a$:
\begin{equation}
\left(\frac{d^2}{dr^2}+\frac{3}{r}\frac{d}{dr}\right)[a\phi(ar)]
=a^3\left[\phi''(ar)+\frac{3}{ar}\phi'(ar)\right]
=\lambda [a\phi(ar)]^3.
\end{equation}
The transformation $\phi(r)\to a\phi(ar)$ on \eq{SMh}
amounts to rescaling 
\beq
\phi_b(0)\to a\phi_b(0).
\eeq
However, the scale invariance of \eq{bouncexSM} is not a feature of our original problem,
but rather a consequence of our approximations. The original field
equation, \eq{bounce2}, is {\it not} scale-invariant, even when mass terms are neglected,
because of renormalization; therefore, its bounce solution has a definite value at $r=0$.
Unfortunately, the correct value of $\phi_b(0)$ cannot be determined analytically. A numerical study of the bounce solution to \eq{bounce2} was 
presented for instance in Ref.~\cite{DiLuzio:2015iua} based on the shooting method reviewed in \app{app:nummethods}. The results show that the commonly adopted approximation in \eq{bouncexSM} is rather accurate. However, its application in the calculation of the first quantum corrections requires some care, as discussed below in \sect{sec:qc}.

We conclude this section by computing the action of the Fubini-Lipatov bounces \eq{SMh}, which enters
the calculation of the decay rate of the metastable ground state.
The action is easily computed using \eq{eq:flatvirial}:
\begin{equation}
\label{eq:virial}
S[\phi_b]=-\int d^4x\,V(\phi_b(x)) = - 2 \pi^2 \int_0^{+\infty} dr\, r^3 V(\phi_b(r)),
\end{equation}
where in the last step we have used the $\O(4)$ invariance of the bounce solution.
We find
\beq
\label{BactionSM}
S[\phi_b]=\frac{8\pi^2}{3|\lambda|},
\eeq
independently of the value of $\phi_b(0)$, as expected.

\subsection{A digression on scale transformations}
\label{sec:scaletr}

In this section we collect some considerations about the role of scale invariance (or non-invariance) in the search for bounce solutions of the field equations.
Let us consider a generic scalar field theory, with scalar potential $V(\phi)$,
and let us assume that
the action $S[\phi]$ is stationary around $\phi=\phi_b$. Then, in particular, it will be stationary
upon infinitesimal scale transformations, that is we expect 
\beq
\Delta B =\Delta \left(S[\phi_b]-S[\phi_{\rm FV}]\right)=0
\eeq
upon a variation
\beq
\Delta\phi=(1+x_\rho\partial_\rho )\phi_b
\eeq 
around a solution of the field equation. A straightforward calculation gives
\beq
\Delta B=\int d^4x\,\left[\phi V'(\phi_b)-4V(\phi_b)+4V(\phi_{\rm FV})\right].
\label{scaleS}
\eeq
The r.h.s.~of \eq{scaleS} is obviously zero for 
\beq
V(\phi)=\frac{1}{4}\lambda\phi^4+V(\phi_{\rm FV}),
\label{Vscaleinv}
\eeq
because the integrand vanishes identically in this case. We already know that in this case bounce solutions of the field equation exist
if $\lambda<0$.

It is easy to find examples of scalar potentials that do not admit any bounce solution of the field equation. Consider for instance
\beq
V(\phi)=\frac{1}{4}\lambda\phi^4+\frac{1}{2}m^2\phi^2
\label{Vscalenoninv}
\eeq
with $\lambda<0$ and constant $m^2$, either positive or negative. In this case, 
\beq
\Delta B=\Delta S[\phi]=-m^2\int d^4x\,\phi^2(x),
\label{nonscaleS}
\eeq
which vanishes only for $\phi(x)=0$.

There are however intermediate cases, that is, non-scale-invariant scalar potentials that admit a bounce solution of the field equation.
For example
\beq
V(\phi)=g\left[
\frac{1}{4}\phi^4
-\frac{1}{3}\phi^3(M+\phi_{\rm TV})
+\frac{1}{2}\phi^2M\phi_{\rm TV}
\right]
\label{V432}
\eeq
with $g>0$ and
\beq
0<M<\frac{\phi_{\rm TV}}{2}
\eeq
has a local minimum for $\phi=0$ and an absolute minimum for $\phi=\phi_{\rm TV}$, separated by a barrier with its maximum for  $\phi=M$. 
A plot of $\frac{V(\phi)}{g\phi_{\rm TV}^4}$ as a function  of $h=\frac{\phi}{\phi_{\rm TV}}$ for different values of
$m=\frac{M}{\phi_{\rm TV}}$ is shown
in Fig.~\ref{fig:FDpot}.
\begin{figure}[h]
\centering
\includegraphics[scale=0.5]{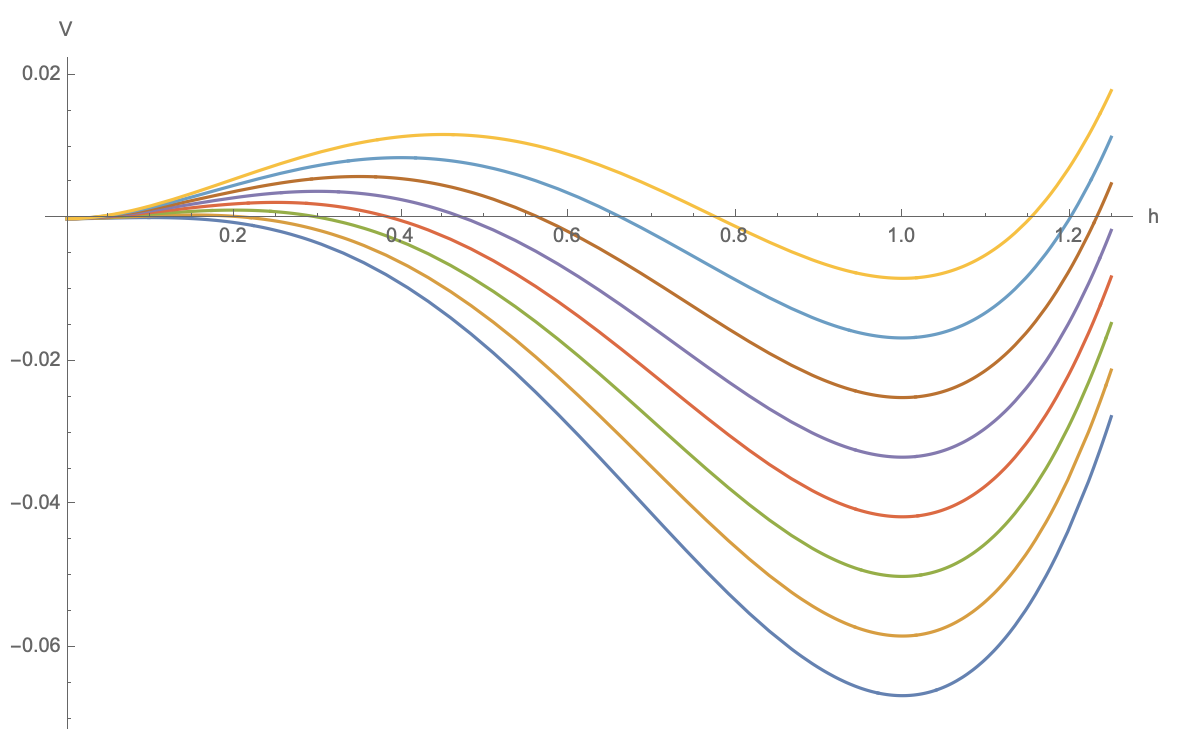}
\caption{The scalar potential \eq{V432}, normalized to $g\phi_{\rm TV}^4$, for different values of $m$. From top to bottom, $m=0.45,0.4,0.35,0.3,0.25,0.2,0.15,0.1$.}
\label{fig:FDpot}
\end{figure}
In this case the field equation
\beq
\phi''+\frac{3}{r}\phi'=g\phi(\phi-M)(\phi-\phi_{\rm TV})
\eeq
cannot be solved analytically. We know, however, that it must have a bounce solution, as a consequence of the overshoot-undershoot
argument illustrated in \sect{overunder}. It is possible to solve the field equation numerically; the bounce $h(x)$ as a function of $x=\sqrt{g}\phi_{\rm TV} r$ is plotted in Fig.~\ref{FDbounces} for different values of $m$.
\begin{figure}[h]
\centering
\includegraphics[scale=0.35]{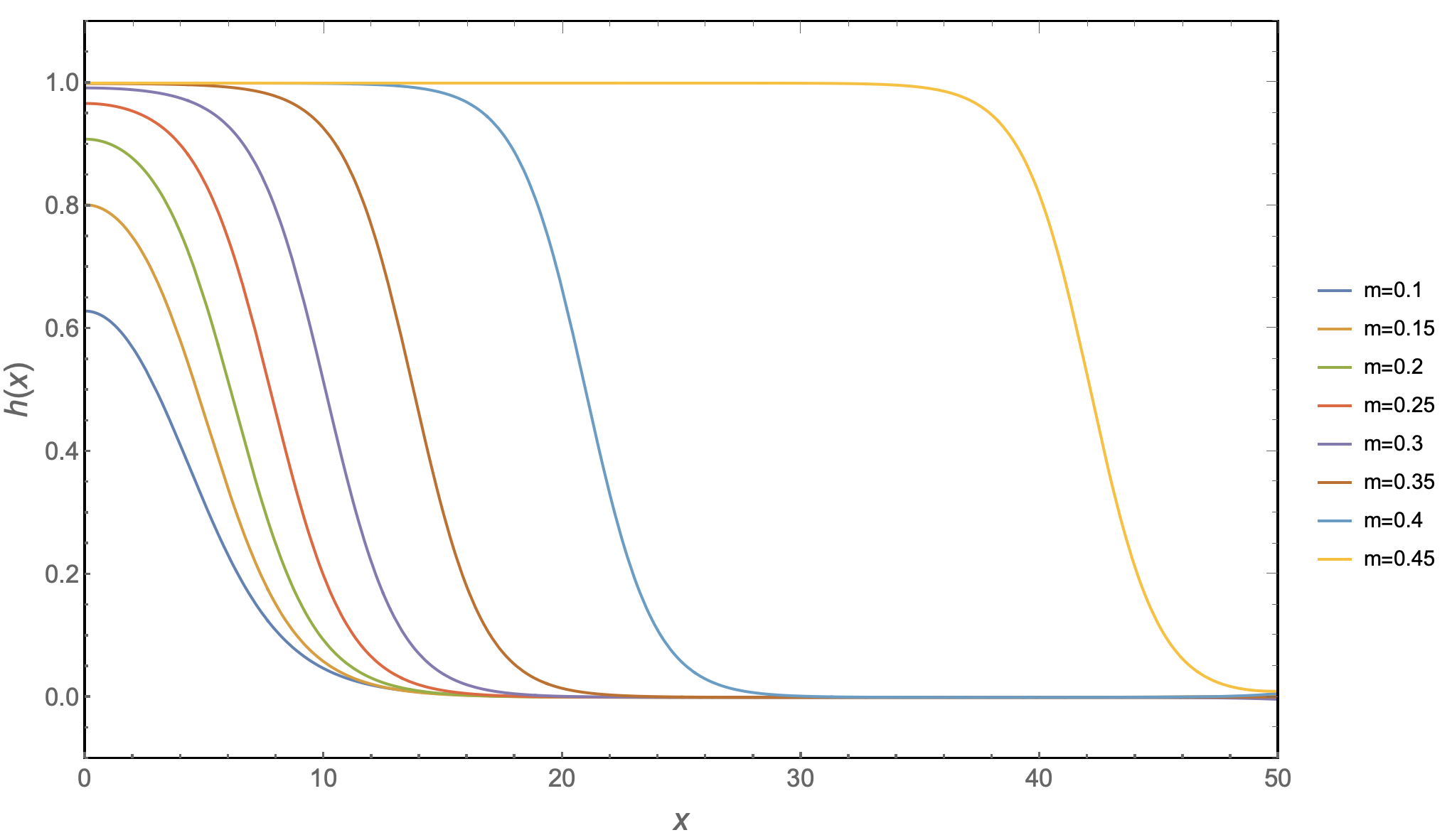}
\caption{Bounce solutions relative to the scalar potential in \eq{V432} for different values of $m$.}
\label{FDbounces}
\end{figure}
Note that $h(0)$ gets close to 1, i.e.\ $\phi(0)$ close to $\phi_{\rm TV}$, as $m\to\frac{1}{2}$, where the two minima are degenerate. Also note that the solution is unique, as a consequence of the explicit breaking of scale invariance, contrary to what happens for the scale-invariant potential \eq{Vscaleinv}: the condition
\beq
\Delta B=g\int d^4x\,\left[\frac{1}{3}\phi_b^3(M+\phi_{\rm TV})-\phi_b^2M\phi_{\rm TV}\right]=0
\label{DeltaB}
\eeq
fixes the value of the bounce at $r=0$. It can be shown that in this case the bounce is well approximated by a gaussian
\beq
\phi_b(r)=\phi_0\e^{-\frac{r^2}{R^2}},
\eeq
in the sense that values of the parameters $\phi_0$ and $R$ can be found, such that the true bounce action is accurately reproduced.

The standard model scalar potential with the mass term neglected and the running coupling computed at $\mu\sim\phi$,
\beq
V(\phi)=\frac{1}{4}\lambda(\phi)\phi^4;\qquad \mu\frac{d\lambda}{d\mu}=\beta(\lambda)
\eeq
also belongs to the same class: it has a local minimum at $\phi=0$, increases with $\phi$, then starts decreasing when $\lambda(\phi)$ becomes negative. We have argued in the previous section that in this case the bounce is well approximated by the Fubini-Lipatov bounce; the value of $\phi(0)$ is determined by the condition of stationarity of the action upon scale transformations,
\beq
\Delta B=\frac{1}{4}\int d^4x\,\beta(\lambda(\phi_b))\phi_b^4=0.
\label{statB}
\eeq
An analytical estimate of $\phi_b(0)$ in this case is presented in Ref.~\cite{Espinosa:2020qtq}: for values of $\phi$ not too far from the value
of the scale $\bar\mu$ where $\lambda(\bar\mu)$ has a minimum we have
\beq
\lambda(\phi)=\lambda(\bar\mu)
+\frac{1}{2}\left.\frac{d\beta(\lambda(\phi))}{d\log\phi}\right|_{\phi=\bar\mu}\log^2\frac{\phi}{\bar\mu}+O\left(\log^3\frac{\phi}{\bar\mu}\right)
\eeq
and therefore
\beq
\beta(\lambda(\phi))=
\left.\frac{d\beta(\lambda(\phi))}{d\log\phi}\right|_{\phi=\bar\mu}\log\frac{\phi}{\bar\mu}+O\left(\log^2\frac{\phi}{\bar\mu}\right).
\eeq
Neglecting terms of order $\log^2\frac{\phi}{\bar\mu}$, the stationarity condition in \eq{statB} yields
\beq
\int d^4x\,\phi_b^4(r)\log\phi_b(r)=\log\bar\mu\int d^4x\,\phi_b^4(r).
\eeq
The two integrals are easily computed with $\phi(r)$ replaced by
 the Fubini-Lipatov bounce \eq{SMh}. We get
\beq
\log\frac{\phi_b(0)}{\bar\mu}\int\frac{r^3dr}{\left(1+\frac{|\lambda|}{8}\phi^2_b(0) r^2\right)^4}
=\int\frac{r^3dr}{\left(1+\frac{|\lambda|}{8}\phi_b^2(0) r^2\right)^4}
\log\left(1+\frac{|\lambda|}{8}\phi_b^2(0) r^2\right)
\eeq
or
\beq
\phi_b(0)=\bar\mu\e^{\frac{5}{6}}.
\label{esp}
\eeq
The result in \eq{esp} can be turned into an estimate of the bounce size $R$, defined by \eq{phi-0}. We obtain
\beq
R=\sqrt{\frac{8}{|\lambda(\bar\mu)|}}\frac{1}{\phi_b(0)}\sim\frac{10}{\bar\mu}
\eeq
for $|\lambda(\bar\mu)|\sim 0.013$.
Because $\bar\mu\sim10^{17}$ GeV, \eq{esp} is an a-posteriori confirmation that the mass term in the potential can be safely neglected.

For earlier work discussing the breaking of scale invariance due to the running of $\lambda$ in the context of the standard model vacuum
tunnelling calculation see also Refs.~\cite{Arnold:1989cb,Arnold:1991cv}.

\subsection{First quantum corrections}
\label{sec:qc}

The calculation of the first quantum corrections to the decay rate of the standard model vacuum is presented in Ref.~\cite{Isidori:2001bm}.
The calculation is performed by the procedure illustrated in \sect{quantumcorr}, which produces the result
\begin{equation}
\frac{\gamma}{{\mathcal V}_3}=
 \frac{\left(S[\phi_b]-S[\phi_{\rm FV}]\right)^2}{(2\pi)^2}
\sqrt{
    \frac{  \Det S''[\phi_{\rm FV}]   }
           { |\Det'S''[\phi_b] |}
           }\e^{-\left(S[\phi_b]-S[\phi_{\rm FV}]\right)}
           \e^{-\left(S_{\rm ct}[\phi_b]-S_{\rm ct}[\phi_{\rm FV}]\right)},
\label{QFTdecayratebis}
\end{equation}
for the decay rate per unit volume of the metastable ground state.

In the case of the standard model, the calculation of the pre-exponential factor is complicated by the presence of fermion and vector degrees of freedom in the theory, which contribute fluctuations around the bounce configuration. We will not reproduce here the full calculation; we limit ourselves to pointing out a few important features.

The first point is the choice of a suitable basis for the space of eigenfunctions of $S''[\phi]$ for
 the evaluation of
ratios of functional determinants
\beq
\label{eq:detratio}
\frac{\Det S''[\phi_b]}{\Det S''[\phi_{\rm FV}]}. 
\eeq
This requires solving eigenvalue equations of the type 
\beq
\label{eq:eigeq}
S''[\phi]\,u=[-\partial^2 + W(\phi)]u=\lambda u,
\eeq
where $u$ generically denotes scalar, fermion or gauge fields, and the function $W(\phi)$ takes different forms for each contribution. 

Because $W(\phi)$ in~(\ref{eq:eigeq}) depends only on the radial
coordinate $r$, it is convenient to decompose the various fields in eigenstates of the
four-dimensional angular momentum operator $L^2=L_{\mu\nu}L_{\mu\nu}$, where
\beq
\label{eq:Lmunu}
L_{\mu\nu}=\frac{i}{\sqrt{2}}\left(x_\mu\partial_\nu-x_\nu\partial_\mu\right).
\eeq
The Laplace operator takes the form
\beq
\label{eq:laplace}
\partial^2=\partial_\mu \partial_\mu=
\frac{d^2}{dr^2}+\frac{3}{r}\frac{d}{dr}-\frac{L^2}{r^2}.
\eeq
In the case of scalar fields, the eigenfunctions of $L^2$ are the
four-dimensional spherical harmonics $Y_{j}(\theta)$ 
(where $\theta$ collectively denotes the 3 polar angles)
with eigenvalues
$4j(j+1)$, and degeneracy $(2j+1)^2$, where $j$ takes integer and
semi-integer values. In this basis, $S''[\phi_b]$ is block-diagonal, with blocks $\left(S''[\phi_b]\right)_j$ labelled by $j$; therefore
\beq
\label{eq:jsum}
\log\Det S''[\phi_b]=\sum_j \log\Det(S''[\phi_b])_j.
\eeq
The situation is slightly more complicated for fermion and vector
fields; details can be found in Ref.~\cite{Isidori:2001bm}.

As illustrated in \sect{sec:funcdet}, in order to compute the ratio in \eq{eq:detratio}
it is only necessary to solve \eq{eq:eigeq} for $\lambda=0$:
\beq
\label{eq:thdet}
\rho_j \equiv 
\frac{\Det\left[-\partial^2 + W(\phi_b) \right]}{\Det\left[-\partial^2+ W(\phi_{\rm FV}) \right]} 
=\lim_{r\to\infty} \frac{\det u^j_{\phi_b}(r)}{\det u^j_{\phi_{\rm FV}}(r)},
\eeq
where $u^j_\phi(r)$ are solutions of
\beq
[-\partial^2 + W(\phi)]\, u^j_\phi(r)=0.
\eeq
The symbol det in \eq{eq:thdet} stands for the ordinary
determinant over residual (spinorial, gauge group, etc.) indices of
these solutions.

The second important point is the subtraction of ultraviolet divergences.
The ultraviolet
behaviour of the functional determinant is encoded in the behaviour of the determinants for $j\to
\infty$;
as a consequence, the sum
over $j$ in Eq.~(\ref{eq:jsum}) is not convergent, and the usual renormalization procedure is needed. 
The counterterm action $S_{\rm ct}$ removes
the ultraviolet divergences that arise in the calculation of the functional determinants.
Indeed,
\beq
\sqrt{ \Det S''[\phi]}\e^{S[\phi]+S_{\rm ct}[\phi]}
=\e^{S[\phi]+S_{\rm ct}[\phi]+\frac{1}{2}\log{\rm Det}\, S''[\phi]}
=e^{\Gamma[\phi]},
\eeq
where $\Gamma[\phi]$ is the one-loop effective action, defined in \sect{effatceffpot}.  
For example, in 
the $\overline{\rm MS}$ scheme 
(dimensional regularization
with $\overline{\rm MS}$ subtraction) the renormalized one-loop action
can be written as
\beq
\label{subtr}
\Gamma[\phi]=S^{\overline{\rm MS}}[\phi]
+ \left[\Delta S[\phi]-\left(\Delta S[\phi]\right)_{\rm pole} \right],
\eeq
where $S^{\overline{\rm MS}}[\phi]$ is the lowest-order action 
expressed in terms of the renormalized $\overline{\rm MS}$ couplings,
and $\left(\Delta S\right)_{\rm pole}$ is the divergent 
part of $\Delta S$ defined according to the $\overline{\rm MS}$
renormalization prescription. The calculation of the full determinant in $d=4-2\epsilon$ dimensions 
can be avoided by observing that the ultraviolet divergences are all contained in 
$\Delta S^{[2]}$, defined as 
the expansion of $\Delta S$ up to second order in  $W$:
\beq
\Delta S^{[2]}[\phi] =
\frac{1}{2}\left[\log \SDet \left( -\partial^2 + W(\phi)  \right)  \right] _{O(W^2)}, 
\label{eq:defS2}
\eeq
where ${\rm SDet}\,O={\rm Det}\,O$ or ${\rm SDet}\,O=({\rm Det}\,O)^{-2}$ depending on 
whether the operator $O$ acts on boson or fermion fields. This is easily seen in the case of scalar degrees of freedom, when
\beq
W_{\rm scalar}(\phi)=V''(\phi).
\eeq
In renormalizable theories, the second derivative of the scalar potential is a polynomial of degree 2 in the field, while counterterms are a polynomial of degree 4. Hence the difference
$\Delta S - \Delta S^{[2]}$ is ultraviolet-finite.
Eq.~(\ref{subtr}) can be rewritten as 
\begin{equation}\label{eq:+-+}
\Gamma[\phi]= S^{\overline{\rm MS}} [\phi]
+\bigg[\Delta S[\phi]-\Delta S^{[2]}[\phi]\bigg]
+\bigg[\Delta S^{[2]}[\phi] -\left(\Delta S[\phi]\right)_{\rm pole}\bigg],
\end{equation}
where the two terms in square brackets are separately finite.
The advantage of this last expression is that 
$\Delta S^{[2]}[\phi]$ can be computed either as a (divergent) sum of terms
corresponding to different values of the angular momentum, which gives
a finite result when subtracted from $\Delta S[\phi]$, 
or by standard diagrammatic techniques in $4-2\epsilon$ dimensions.

The details of the calculation of one-loop contributions to $\Gamma[\phi_b]$ from field fluctuations in the different sectors of the theory described in Ref.~\cite{Isidori:2001bm} is performed in terms of the Fubini-Lipatov bounce, \eq{SMh} or \eq{SMhR}, with 
$\phi_b(0)$, or equivalently $R$, fixed as described at the end of \sect{sec:scaletr}. In this approximation, the zero mode of scale invariance
in the scalar sector is not normalizable, as shown \sect{sec:zeromodes}. This is however an artifact of our approximations: the bounce is a solution of \eq{bouncexSM}, rather than \eq{bounce2}. This difficulty is overcome by removing the zero eigenvalue of dilatations  from the functional determinant, and replacing it by a slightly negative eigenvalue, induced by radiative corrections.

Additional zero modes of $S''[\phi_b]$ arise because of the gauge symmetry; the corresponding contributions to the path integral must be removed and replaced by an integration over suitable collective coordinates. This procedure is described in Ref.~\cite{Kusenko:1996bv}, and involves a jacobian factor proportional to
\beq
\int d^4x\,\phi_b^2(r)
\label{bouncenorm}
\eeq
(see \eq{zmint}),
which is infrared divergent in the case of the Fubini-Lipatov bounce. Again, this is a consequence of our approximation of neglecting mass terms, and hence the nonvanishing vacuum expectation value of the Higgs field, in the scalar potential; in the real world, the Higgs mass acts as a regulator of the infrared divergence of \eq{bouncenorm}.
To see how this works we consider the field equation
\beq
\phi''_b(r)+\frac{3}{r}\phi'_b(r)=V'(\phi_b)
\label{larger}
\eeq
in the large-$r$ limit. In this region, $\phi_b(r)\to\phi_{\rm FV}$, which is different from zero if the mass term is not neglected. In this limit, the field equation 
can be linearized by expanding the scalar potential in powers of $\phi_b-\phi_{\rm FV}$:
\beq
V(\phi_b)=V(\phi_{\rm FV})+\frac{1}{2}V''(\phi_{\rm FV})(\phi_b-\phi_{\rm FV})^2+\O\left((\phi_b-\phi_{\rm FV})^3\right),
\eeq
where $V''(\phi_{\rm FV})=m^2$ is a positive constant, related to the physical Higgs squared mass. Neglecting terms of order $(\phi_b-\phi_{\rm FV})^3$, \eq{larger} becomes
\beq
(\phi_b(r)-\phi_{\rm FV})''+\frac{3}{r}(\phi_b(r)-\phi_{\rm FV})'=m^2(\phi_b(r)-\phi_{\rm FV}),
\label{larger2}
\eeq
and can be solved analytically. In terms of the dimensionless quantities
\beq
x=rm;\qquad f(x)=r(\phi_b(r)-\phi_{\rm FV})
\eeq
\eq{larger2} reads
\beq
x^2f''(x)+xf'(x)-(1+x^2)f(x)=0,
\eeq
which is solved by a linear combination of modified Bessel functions
\beq
f(x)=\alpha K_1(x)+\beta I_1(x),
\eeq
where
\begin{align}
I_1(x)&=\frac{x}{2}\sum_{n=0}^\infty\frac{1}{n!(n+1)!}\left(\frac{x^2}{4}\right)^n
\\
K_1(x)&=\left(\log\frac{x}{2}+\gamma\right)I_1(x)+\frac{1}{x}
+\frac{x}{4}\sum_{n=0}^\infty\frac{1}{n!(n+1)!}\left(\sum_{k=1}^{n+1}\frac{1}{k}+\sum_{k=1}^n\frac{1}{k}\right)\left(\frac{x^2}{4}\right)^n.
\end{align}
The large-$x$ behaviors of $I_1(x),K_1(x)$ are as follows:
\begin{align}
I_1(x)&\sim\frac{1}{\sqrt{2\pi x}}e^x
\\
K_1(x)&\sim\sqrt{\frac{\pi}{x}}e^{-x}.
\end{align}
The condition $\phi_b(r)-\phi_{\rm FV}\to 0$ for $r\to\infty$ gives $\beta=0$, and therefore in this limit
\beq
\phi_b(r)-\phi_{\rm FV}\simeq \alpha m\frac{K_1(rm)}{rm}\sim\frac{1}{(rm)^{3/2}}\e^{-rm}.
\eeq
This shows that $\phi^2_b(r)$ is integrable, if the finite value of the Higgs mass is taken into account.

The final result for the tunnelling rate of the metastable ground state can now be written as
\begin{equation}
\frac{\gamma}{{\mathcal V}_3}=\frac{1}{{\mathcal V}_3T}\frac{\e^{-\Gamma[\phi_b]} }{\e^{-\Gamma[\phi_{\rm FV}]}};
\end{equation}
the space-time volume factor ${\mathcal V}_3T$ is cancelled by the exclusion of zero eigenvalues, as we have seen.

The final result is expressed in terms of the renormalized parameters
$\lambda,~y_t$ and $g_i$, whose definitions depend on the
renormalization scheme. Of course, the scheme dependence disappears
once these couplings are re-expressed in terms of physical
observables, such as Higgs and top pole masses. In practice, however, in
the case of the gauge couplings it turns out to be more convenient to
directly use the $\overline{\rm MS}$ definitions, since these
parameters are accurately determined by fitting multiple observables.

An updated next-to-leading order analysis of the tunnelling rate of the standard model vacuum
yields \cite{Andreassen:2017rzq}
\beq
\tau_{\rm SM} = \(\frac{\gamma_{\rm SM}}{\mathcal{V}_3}\)^{-1/4} = 10^{161^{+160}_{-59}} \ \text{yr}
\eeq
at $68\%$ confidence level, where the uncertainty is roughly due in equal parts
to the uncertainty on the value of the top quark mass and the strong-coupling constant
and the theory uncertainty from threshold corrections at the top mass scale.
Even taking into account these uncertainties, the lifetime of the standard model vacuum easily complies
with the age of the Universe $\tau_U \approx 13.8 \times 10^9$ yr.

\subsection{The standard model phase diagram}
\label{sec:SMphasedia}

We now come to a discussion of the stability of the standard model ground state in the light of our present knowledge of the relevant parameters. The most relevant parameters in the calculation of the decay rate of the standard model vacuum are the value of 
the top quark mass $M_t$, which is responsible for the negative value of the beta function for $\lambda(\mu)$, the Higgs boson mass $M_h$,
which determines the value of $\lambda(\mu)$ at the electroweak scale,  and the strong coupling $\alpha_s (M_Z)$, which affects indirectly the calculation of the beta function. Traditionally, the result of the calculation if the vacuum decay rate are displayed
 in the $(M_h, M_t)$ plane (cf.~\fig{fig:SMphasedia}), often referred to as the standard model phase diagram.\footnote{Similar analyses have been performed also in
Refs.~\cite{Bezrukov:2012sa,Degrassi:2012ry,Masina:2012tz,Andreassen:2014gha,Bednyakov:2015sca,Iacobellis:2016eof,Andreassen:2017rzq,Chigusa:2017dux}.}
\begin{figure}[ht]
\centering
\includegraphics[width=0.45\textwidth]{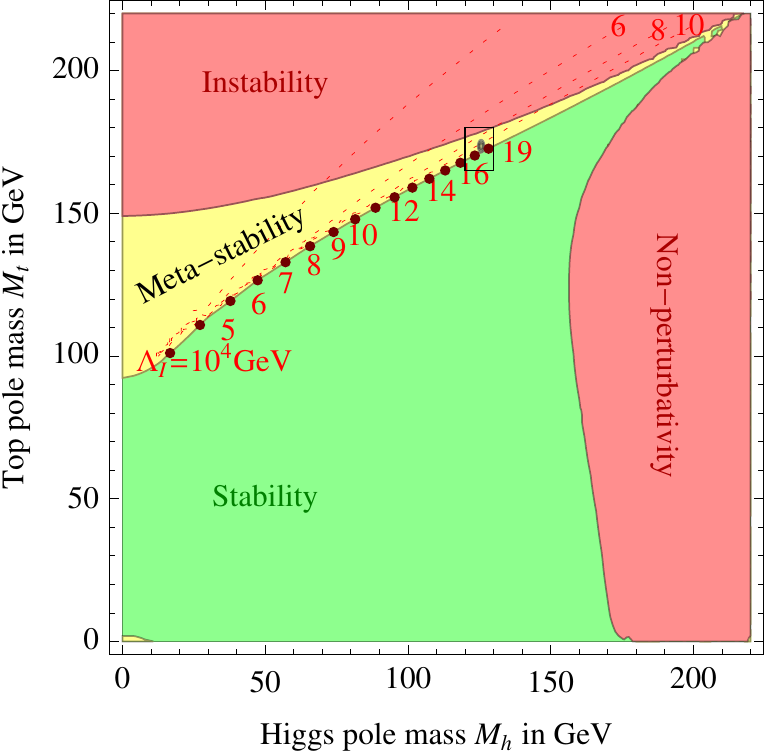} \qquad
\includegraphics[width=0.45\textwidth]{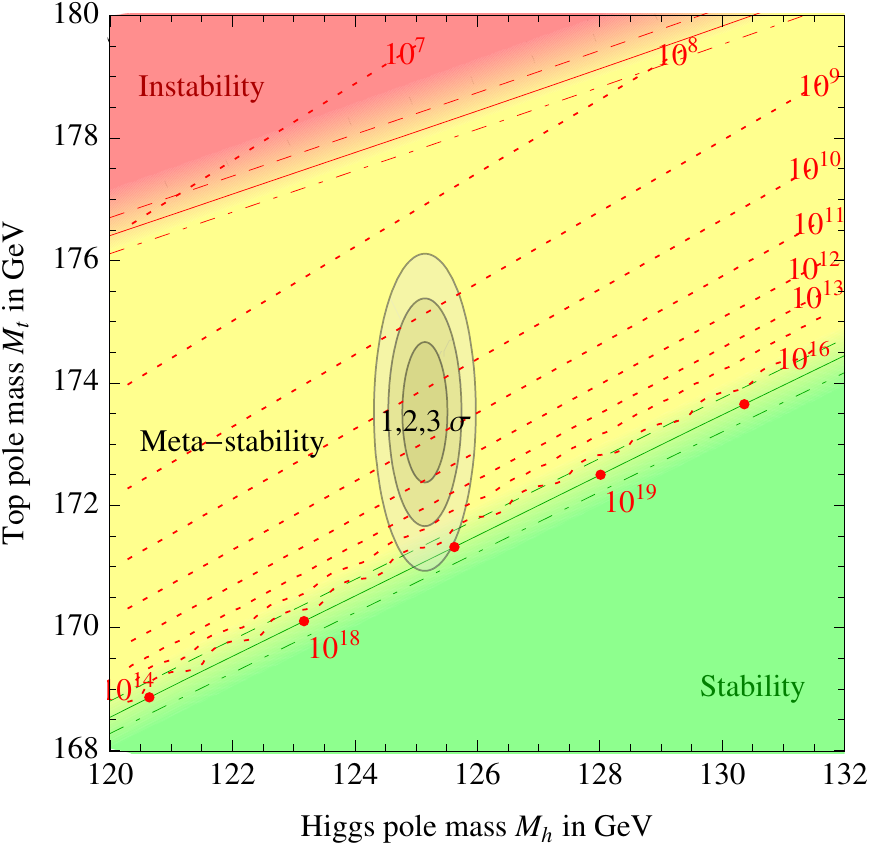}
\caption{\label{fig:SMphasedia}
The standard model phase diagram in the $(M_h,M_t)$ plane.
Images taken from Ref.~\cite{Buttazzo:2013uya}.
}
\end{figure}

Let us describe how the standard model phase diagram
(which by definition assumes the standard model in isolation)
is obtained and comment
on possible interpretations of the fact that the experimental values of the
standard model parameters seem to lie at the boundary between the
stability and metastability region.

\subsubsection*{Instability/Metastability boundary}

The probability of nucleating a bubble of true vacuum
during the history of the Universe can be estimated as
\beq
p = \frac{\gamma}{{\mathcal V}_3} \({\mathcal V_3}T \)_{\rm light-cone},
\eeq
with the space-time volume of our past light-cone given by \cite{Buttazzo:2013uya}
\beq
\( {\mathcal V_3}T \)_{\rm light-cone} \approx \frac{0.15}{H_0^4},
\eeq
and $H_0 = 1.44 \times 10^{-42}$ GeV being the present Hubble constant.
The boundary region between instability (red) and metastability (yellow) in
\fig{fig:SMphasedia} is obtained by setting $p=1$ as a function of the top and Higgs pole masses.

\subsubsection*{Metastability/Stability boundary}

The absolute stability condition was already discussed in \sect{gaugedep} (cf.~\eqs{absvacstab1}{absvacstab2}
and \eqs{absvacstab1mod}{absvacstab1mod}) and reads
\beq
\label{eq:absstabcond}
\lambda_{\rm eff} (\mu_c) =  \beta_{\lambda_{\rm eff}} (\mu_c) = 0,
\eeq
where $\beta_{\lambda_{\rm eff}} = \mu \frac{d \lambda_{\rm eff}}{d\mu}$ and
we have neglected small corrections of order $(v / \mu_c)^2 \approx 10^{-30}$ which corresponds
to approximating $V_{\rm eff} (\phi) = \frac{1}{4} \lambda_{\rm eff} (\phi) \phi^4$.
A more accurate stability criterium, which takes into account the consistent use of the
$\hbar$ expansion of the standard model effective potential, was formulated in Ref.~\cite{Andreassen:2014gha}.

The critical line between metastability (yellow) and stability (green) in \fig{fig:SMphasedia}
is obtained by selecting the region in the $(M_h, M_t)$ plane
where \eq{eq:absstabcond} holds.
The technical tools entering the high-energy extrapolation of the standard model effective potential
have been developed dramatically after the Higgs discovery and involve the calculation
of beta functions \cite{Mihaila:2012fm,Mihaila:2012pz,Chetyrkin:2012rz,Chetyrkin:2013wya}, threshold corrections at the $M_t$ mass scale \cite{Bezrukov:2012sa,Degrassi:2012ry,Buttazzo:2013uya,Kniehl:2015nwa}
and the fixed-order standard model effective potential \cite{Martin:2013gka,Martin:2015eia,Martin:2018emo}.
With these ingredients at hand,
one arrives to the remarkable conclusion that the electroweak vacuum is near-critical.
While the central values of the standard model parameters point to the metastability region,
with a lifetime way larger than the age of the Universe,
absolute stability
is excluded only between 2 and 3$\sigma$ confidence level (cf.~right panel in \fig{fig:SMphasedia}).
At the moment the leading uncertainty is of experimental nature and is due to the uncertainty
on $M_t$ and $\alpha_s (M_Z)$.

The main message of the standard model phase diagram is that the standard model can be consistently extrapolated
up to the Planck scale.
Pushing its interpretation even further,
the apparent near-criticality of the standard model vacuum has triggered many speculations
about the possible dynamics
behind that (see e.g.~\cite{Buttazzo:2013uya,Cline:2018ebc}).

\subsection{Ultraviolet sensitivity and new physics}
\label{sec:UVandNP}

Physical degrees of freedom beyond those present in the standard model are expected, both on the basis  of some
experimental observations (massive neutrinos, dark matter, etc.) and of theoretical considerations (e.g.~the naturalness problem).
The presence of extra degrees of freedom can in general affect the standard model effective potential,
possibly modifying, or even invalidating, the conclusions of the previous sections about the stability of the 
electroweak ground state.
In such cases, the study of the electroweak vacuum lifetime is still a useful tool
in order to constrain extensions of the standard model: non-standard physics should not make the electroweak
vacuum too unstable.

The lifetime of the electroweak vacuum is obviously sensitive to new physics
if its mass scale is $M<\bar\mu\approx 10^{17}$ GeV,
since the standard model bounce gets directly affected.
On the other hand, also new physics with $M\gg \bar\mu$ can enhance the tunnelling rate, 
e.g.~by opening new decay channels for the electroweak vacuum \cite{Branchina:2013jra,Branchina:2014usa,Branchina:2014rva,Branchina:2015nda,DiLuzio:2015iua,Andreassen:2016cvx}.\footnote{For the moment, we are neglecting
Planck-scale physics in order to highlight some general features of tunnelling in flat space-time.
The inclusion of gravitational effects will be discussed in \sect{sec:gravity}.}

Such non-decoupling behaviour of ultraviolet physics might appear counter-intuitive.
In quantum mechanics the shape of the potential beyond the barrier does not affect the tunnelling rate through the barrier itself.
The case of quantum field theory is different. As shown in \sect{barrierQFT},
the generalization of the potential barrier in quantum field theory involves
a space-time integral of the scalar potential evaluated at the bounce configuration $\phi_b(r)$; as a consequence, all energy scales between 0 and $\phi_b(0)$ are relevant.

The impact of non-standard physics on the ground state lifetime can be assessed parametrizing new physics effects on the scalar potential by non-renormalizable operators. For example, in Ref.~\cite{Branchina:2013jra} it was proposed to modify the standard model scalar potential
as
\begin{equation}
V(\phi)=\frac 14 \lambda \phi^4 + \frac {\lambda_6}{6M^2}\phi^6+ \frac {\lambda_8}{8M^4}\phi^8,
  \label{eq:branchina}
\end{equation}
where $\lambda$ is the standard Higgs self-coupling, $\lambda_6,\lambda_8$ are assumed to be of order one, and $M$ is the energy scale at which new physics effects become relevant.
The potential in \eq{eq:branchina} is not meant to be a realistic ultraviolet completion of the standard model.
From an effective field theory perspective, for instance, a more realistic picture would be obtained
by considering a full tower of higher dimensional operators, including derivative operators as well.
In the present discussion, we do not address the question of how specific models can affect the standard model vacuum lifetime,
but just use the potential in \eq{eq:branchina} as a toy model in order to parametrize new physics effects.

The inclusion of two non-renormalizable terms in \eq{eq:branchina} does not allow us anymore to find an analytical
solution of the bounce equation and one has to proceed
numerically.\footnote{Numerical techniques based on the shooting algorithm
for the determination of the bounce are reviewed in \app{app:nummethods}.}
The sensitivity of the tunnelling rate to new physics at very large energy scales can be assessed by
comparing the potential energy barrier obtained with the potential in \eq{eq:branchina}
with the one in the pure standard model.
Following the definitions in \eqs{eq:Uphi}{eq:Vphi},
for the standard model case with $\lambda<0$ one finds:
\beq
U_\nabla[\phi_b]
=\phantom{-}\frac{2\pi^2}{\left|\lambda\right|R}\left(\frac 1{1+\frac{\tau^2}{R^2}}\right)^\frac 32, \qquad
U_V[\phi_b]
=-\frac{2\pi^2}{\left|\lambda\right|R}\left(\frac 1{1+\frac{\tau^2}{R^2}}\right)^\frac 52,
\eeq
which have opposite sign, while their sum
\beq
U[\phi_b] = U_\nabla[\phi_b] + U_V[\phi_b]
=\frac{2\pi^2}{\left|\lambda\right|R}\left(\frac 1{1+\frac{\tau^2}{R^2}}\right)^\frac 32\frac{\frac {\tau^2}{R^2}}{1+\frac {\tau^2}{R^2}},
\eeq
is positive, thus providing a true potential barrier.

The potential barrier in the presence of non-renormalizable terms in the scalar potential as in \eq{eq:branchina}
can be computed by solving numerically the bounce equation. In \fig{fig:barriernp} we compare
the barrier $U[\phi_b]$ in the presence of new physics (orange) and in the pure standard model (blue).
We have chosen $\lambda=\lambda(\bar\mu)\simeq -0.01345$ (corresponding to the minimum of $\lambda(\mu)$ in the standard model),
$\lambda_6=-2$ and $\lambda_8=2.1$, and we have set $M = M_{\rm Pl}$ for the scale of new physics.
While the two barriers have comparable heights,
the area under the curve which corresponds to the inclusion of new physics in this case is considerably smaller, which results in a much shorter lifetime of the metastable ground state for this particular choice of the parameters.
\begin{figure}[ht]
\centering
\includegraphics[width=0.75\textwidth]{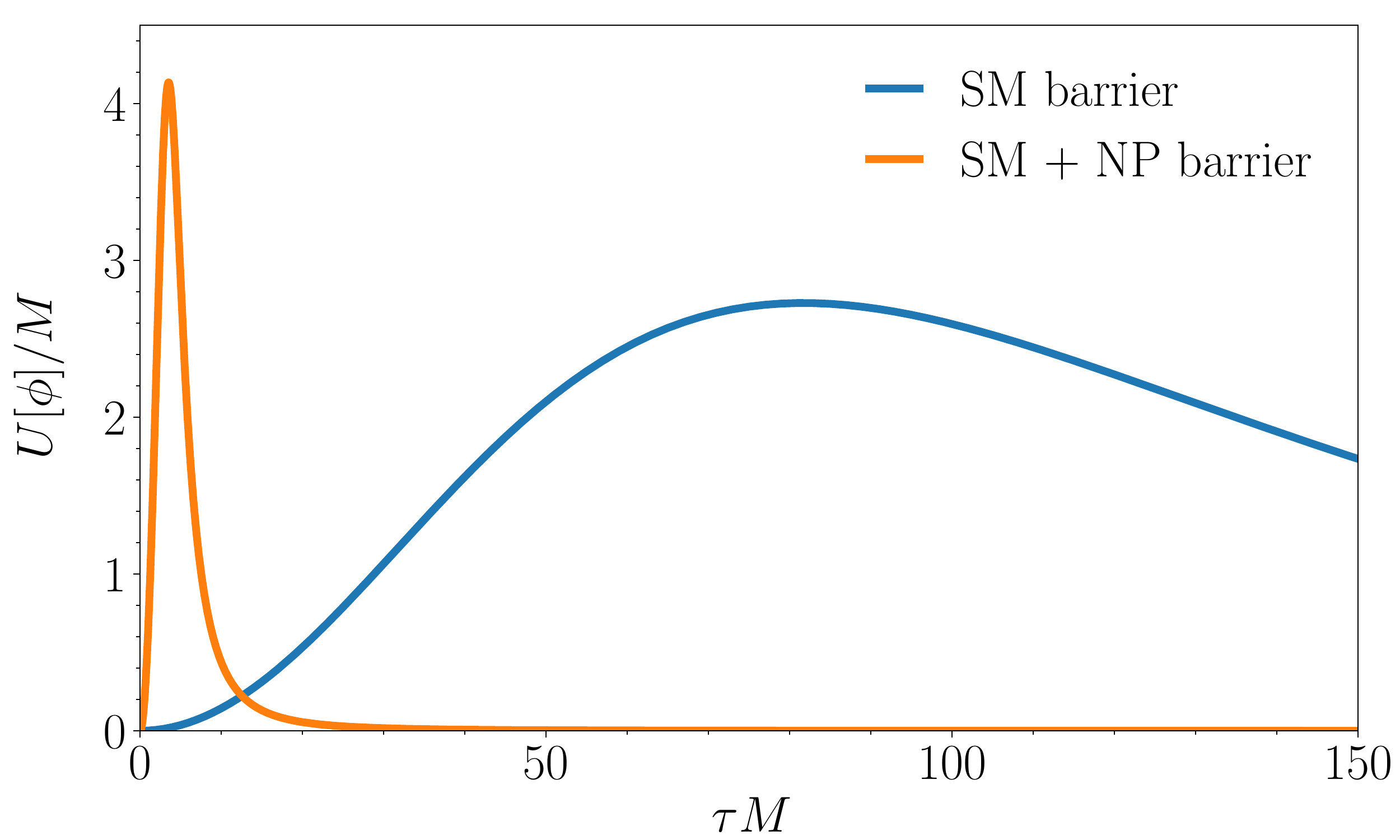}
\caption{Potential barrier with and without the contribution of new physics.
The values chosen for the new physics parameters are $\lambda_6=-2$, $\lambda_8=2.1$.}
\label{fig:barriernp}
\end{figure}

The above discussion can be generalized to the case of more non-renormalizable terms,
including derivative operators, or even renormalizable ultraviolet completions of the standard model
(see also \cite{Branchina:2015nda,Patel:2017aig}). 
On the other hand, specific embeddings of the standard model into ultraviolet theories 
at the Planck scale can help in taming new instabilities of the Higgs potential \cite{Branchina:2018xdh}. 

Summarizing, while new physics at scales $M < \bar\mu \approx 10^{17}$ GeV can affect
the decay rate of the standard model vacuum in any direction (possibly stabilizing the electroweak vacuum),
modifications of the standard model potential at scales $M \gg \bar\mu \approx 10^{17}$
can only shorten the lifetime of the electroweak vacuum,
whenever they  lead to a new bounce solution which is related to
a new, deeper minimum around the scale $M$.

\section{Gravitational effects on vacuum decay}
\label{sec:gravity}

We have seen that very high energy scales are involved in the study of the metastability
of the standard model ground state: the running coupling constant of the Higgs boson,
$\lambda(\mu)$ becomes negative at $\mu\sim 10^{10}$~GeV,
and reaches its maximum negative value (which determines the ground state lifetime)
at $\mu=\bar\mu\approx 10^{17}$~GeV. It is therefore natural to ask whether
gravitational effects become relevant in this regime. Indeed, the typical scale of gravitational interactions, the Planck mass $M_{\rm Pl}=1.22\times 10^{19}$~GeV, is only
a couple of orders of magnitude larger than $\bar\mu$. 

This issue was first addressed  by Coleman and De Luccia in
Ref.~\cite{Coleman:1980aw}, where they considered the effects of gravity on the decay
of a metastable vacuum in quantum field theory.
In this section we will review their work, which allows us to introduce the relevant formalism.
Next, following Ref.~\cite{Isidori:2007vm}, we will apply the
formalism of Coleman and De Luccia to the standard model.
More recent developments on vacuum decay in the presence of gravity can be 
found e.g.~in Refs.~\cite{Ghosh:2021lua,Espinosa:2021tgx}.

\subsection{The tunnelling rate in the presence of gravity}

The inclusion of gravity in the calculation of the decay rate of the standard model vacuum 
is in principle straightforward: one must find
a configuration $\phi_b(x)$ of the Higgs field which is a stationary point
of the Euclidean Einstein-Hilbert action
\beq
S_g=\int d^4x\,\sqrt{g}\left[\frac{1}{2}\partial_\mu \phi_b \, \partial_\nu \phi_b\,g^{\mu\nu}+V(\phi_b)-\frac{\mathcal R}{2\kappa}
\right],
\label{Seucl}
\eeq
where $\mathcal R$ is the Riemann curvature, and
\beq
\kappa=8\pi G_N=\frac{8\pi}{M_{\rm Pl}^2},
\eeq
where $G_N$ is the Newton constant and $M_{\rm Pl} = 1.22 \times 10^{19}$ GeV 
is the Planck mass. We will limit ourselves to the case of minimal coupling to gravity, which formally corresponds to
 the leading order in the $M_{\rm Pl}$ expansion. The impact on the tunnelling rate of unknown non-minimal coupling 
terms in the Einstein-Hilbert-Higgs action was  addressed e.g.~in Refs.~\cite{Rajantie:2016hkj,Salvio:2016mvj}. 

A difficulty immediately arises: contrary to the case of a flat spacetime, there is no guarantee that the bounce solution is $\O(4)$ invariant.
We will make this assumption, which simplifies considerably the problem,
but it should be kept in mind that a bounce solution with a smaller value of
the action and no $\O(4)$ symmetry might in principle exist, although
no example has been exhibited so far. We will therefore look for an $\O(4)$
invariant metric
\beq
ds^2=dr^2+\rho^2(r)d\Omega^2_3,
\eeq
where $d\Omega_3^2$ is the element of distance on a unit three-sphere,
\begin{equation}
d\Omega^2_3 = d\chi^2 + \sin^2 \chi  d\theta^2 + \sin^2 \theta d\phi^2,
\end{equation}
and an $\O(4)$ invariant bounce $\phi_b(r)$, which obeys
the field equations\footnote{Some details on the derivation of the field equations in the 
presence of gravity are provided in \app{section:graveom}.}
\begin{align}
&\phi_b''+\frac{3\rho'}{\rho}\phi_b'=V'(\phi_b),
\label{heq}
\\
&{\rho'}^2=1+\frac{1}{3}\kappa\rho^2\left(\frac{1}{2}{\phi_b'}^2-V(\phi_b)\right). 
\label{rhoeq}
\end{align}
The boundary conditions for the field equations depend on the topology of the solution, which cannot be 
specified in advance. Since we are mainly interested in the problem of the stability of the standard model vacuum, we 
can neglect the small positive cosmological constant and assume a Minkowski-like 
background metric in the false vacuum. 
In such a case the bounce solution is non-compact, and $r \to \infty$. 
The boundary conditions leading to a finite bounce action are then
\beq
\phi_b'(0) = 0,  \qquad 
\phi_b(\infty) = 0, \qquad 
\rho(0) = 0, \qquad 
\rho(\infty) = r,
\label{gravbound}
\eeq
where both the false vacuum and the value of the potential on the false vacuum are set to zero 
and the metric is asymptotically Minkowski-like. The condition $\rho (0) = 0$ is less obvious and comes from the 
fact that $\rho(r)$ must have at least one zero which can be set to $r=0$ by translational invariance \cite{Weinberg:2012pjx}. 
The Euclidean action \eq{Seucl} takes the form
\beq
S_g[\phi_b,g]=2\pi^2\int_0^\infty dr\,\left[\rho^3\left(\frac{{\phi_b'}^2}{2}+V(\phi_b)\right)
+\frac{3}{\kappa}\left(\rho^2\rho''+\rho{\rho'}^2-\rho\right)
\right].
\label{SeuclO4}
\eeq

The bounce action can be rewritten in a simpler form by a generalization of the
 virial theorem, proven in \sect{sec:QFT} in the flat-space case. Let us assume that the action is stationary upon variations of $\phi_b(x)$ and
$g^{\mu\nu}(x)$; then in particular it should be stationary
upon the transformations
\begin{align}
\label{transfh}
&\phi_b(x)\to \phi_b(bx),
\\
\label{transfg}
&g^{\mu\nu}(x)\to g^{\mu\nu}(bx),
\end{align}
which imply
\beq
{\mathcal R}(x)\to b^2 {\mathcal R}(bx).
\eeq
We find
\beq 
S_g\to S_g(b),
\eeq
where
\begin{align}
S_g(b)&=
\int d^4x\,\sqrt{g(bx)}\Bigg[\frac{1}{2}\frac{\partial \phi_b(bx)}{\partial x^\mu}
\frac{\partial \phi_b(bx)}{\partial x^\nu}g^{\mu\nu}(bx) 
-\frac{b^2{\mathcal R}(bx)}{2\kappa}
+V(\phi_b(bx))\Bigg]
\nonumber\\
&=
\int d^4x\,\sqrt{g(x)}\Bigg[\frac{1}{2b^2}\frac{\partial \phi_b(x)}{\partial x^\mu}
\frac{\partial \phi_b(x)}{\partial x^\nu}g^{\mu\nu}(x)  -\frac{{\mathcal R}(x)}{2b^2\kappa}
+\frac{1}{b^4}V(\phi_b(x))\Bigg].
\end{align}
The stationarity condition
\beq
\left.\frac{dS_g(b)}{db}\right|_{b=1}=0
\eeq
gives a relationship between the derivative terms and the potential term 
in the action:
\beq
\label{Svirial}
\int d^4x\,\sqrt{g(x)}\left[\frac{1}{2}\frac{\partial \phi_b(x)}{\partial x^\mu}
\frac{\partial \phi_b(x)}{\partial x^\nu}g^{\mu\nu}(x)
-\frac{{\mathcal R}(x)}{2\kappa}
\right]
=
-2\int d^4x\,\sqrt{g(x)}V(\phi_b(x)).
\eeq
Hence
\beq
\label{Svirialized}
S_g[\phi_b,g]=-\int d^4x\,\sqrt{g(x)}V(\phi_b(x)).
\eeq
In the flat-space limit this expression reproduces the standard virialized form of the action \cite{Callan:1977pt}. 
Assuming an $\O(4)$ invariant metric, \eq{Svirialized} becomes (see also~\cite{Salvio:2016mvj}) 
\beq
\label{SvirializedO4}
S_g[\phi_b,g]=-2 \pi^2\int_0^\infty dr \, \rho^3(r) V(\phi_b(r)).
\eeq
\eq{SvirializedO4} turns out to be much more useful than the equivalent form \eq{SeuclO4} in numerical estimates of the tunnelling rate,
because the last two terms in \eq{SeuclO4} are both divergent for $r \to \infty$, 
while their difference is finite. 

\subsection{Perturbative solution of the field equations}
The field equations (\ref{heq}, \ref{rhoeq})
cannot be solved analytically. Since the effects of
gravity are typically small, it is tempting to
look for a solution as an expansion in powers of $\kappa$,
as suggested in Ref.~\cite{Isidori:2007vm},
and to neglect systematically all terms of order $\kappa^2$ and higher.
To this purpose, one must expand $\phi_b$ to order $\kappa$ and $\rho$ to order
$\kappa^2$, because of the $\kappa^{-1}$ coefficient in the Einstein-Hilbert action:
\begin{align}
\label{expkh}
&\phi_b=\phi_{b0}+\kappa \phi_{b1}+O(\kappa^2),
\\
\label{expkrho}
&\rho=\rho_0+\kappa \rho_1+\kappa^2\rho_2+O(\kappa^3),
\end{align}
where $\phi_{b0}$ is the bounce in the absence of gravity, and
\beq
\rho_0(r)=r,
\eeq
as appropriate for a flat metric. The two equations for the
order-$\kappa$ corrections:
\begin{align}
&\phi_{b1}''+\frac{3}{r}\phi_{b1}'+\frac{3}{r}\left(\rho_1'-\frac{\rho_1}{r}\right)\phi_{b0}'
=V''(\phi_{b0})\phi_{b1}
\label{phib1}
\\
&\rho_1'=\frac{1}{12} r^2
\left({\phi_{b0}'}^2-2V(\phi_{b0})\right)
\label{rho1}
\end{align}
are no longer coupled to each other.

It is interesting to note that, to order $\kappa$, the correction
terms $\phi_{b1}$ and $\rho_2$ do not affect the Euclidean action. Indeed,
the first term in Eq.~(\ref{SeuclO4}) becomes
\begin{align}
&\rho^3\left(\frac{{\phi_b'}^2}{2}+V(\phi_b)\right)
=r^3\left(\frac{{\phi_{b0}'}^2}{2}+V(\phi_{b0})\right)
\nonumber\\
&\qquad+\kappa\left[3r^2\rho_1\left(\frac{{\phi_{b0}'}^2}{2}+V(\phi_{b0})\right)
+r^3\left(\phi_{b1}'\phi_{b0}'+\phi_{b1}V'(\phi_{b0})\right)
\right]+O(\kappa^2).
\label{S1}
\end{align}
We now observe that
\beq
r^3\left(\phi_{b1}'\phi_{b0}'+\phi_{b1}V'(\phi_{b0})\right)=
\frac{d}{dr}\left(r^3 \phi_{b1} \phi_{b0}'\right),
\eeq
where we have used the flat-space field equation for $\phi_{b0}$
\beq
\phi_{b0}''+\frac{3}{r}\phi_{b0}'=V'(\phi_{b0})
\eeq
to eliminate $V'(\phi_{b0})$. Hence, the last term in Eq.~(\ref{S1})
gives no contribution to the action, provided
\beq
\lim_{r\to 0}\left(r^3 \phi_{b1} \phi_{b0}'\right)
=\lim_{r\to +\infty}\left(r^3 \phi_{b1} \phi_{b0}'\right)=0.
\eeq
Furthermore, expanding $\rho$ to order $\kappa^2$ we find
\beq
\rho^2\rho''+\rho{\rho'}^2-\rho
=\kappa\frac{d}{dr}\left(r^2\rho_1'\right)
+\kappa^2\frac{d}{dr}\left(r^2\rho_2'\right)
+\kappa^2\left(2\rho_1\rho_1'+2r\rho_1\rho_1''+r{\rho_1'}^2\right).
\eeq
The only $\rho_2$-dependent  term vanishes upon integration.
Therefore, the bounce action
depends only on $\phi_{b0},\rho_0,\rho_1$, up to corrections of order $\kappa^2$:
\begin{align}
S_g[\phi_b,g]&=2\pi^2\int_0^\infty dr\,\Bigg[
r^3\left(\frac{{\phi_{b0}'}^2}{2}+V(\phi_{b0})\right)
+3\kappa r^2\rho_1\left(\frac{{\phi_{b0}'}^2}{2}+V(\phi_{b0})\right)
\nonumber\\
&+3\kappa\left(2\rho_1\rho_1'+2r\rho_1\rho_1''+r{\rho_1'}^2\right)
\Bigg]+O(\kappa^2).
\label{SeuclO4k2}
\end{align}

\subsection{Decay of the standard model vacuum in the presence of gravity}
The formalism outlined in the previous sections can be applied to the standard model~\cite{Isidori:2007vm}.
As in the case of a flat space-time, we approximate the standard model scalar potential by
\beq
V(\phi)=\frac{1}{4}\lambda\phi^4,
\label{phi4}
\eeq
with $\lambda=\lambda(\bar\mu)<0$.
The linearized field equations, \eqs{phib1}{rho1}, take the form
\begin{align}
&\phi_{b1}''+\frac{3}{r}\phi_{b1}'+\frac{3}{r}\left(\rho_1'-\frac{\rho_1}{r}\right)\phi_{b0}'
=3\lambda \phi_{b0}^2\phi_{b1}
\label{phib1sm}
\\
&\rho_1'=\frac{1}{12} r^2
\left({\phi_{b0}'}^2-\frac{\lambda}{2}\phi_{b0}^4\right),
\label{rho1sm}
\end{align}
where $\phi_{b0}$ is the bounce in the absence of gravity, the Fubini-Lipatov bounce \eq{SMh} or (\ref{SMhR}).
We have shown in the previous section that the first order correction to the bounce, $\phi_{b1}(r)$, does not affect the value of the bounce action.
On the other hand, \eq{rho1} is immediately integrated, to give
\beq
\rho_1(r)=\frac{1}{3|\lambda|R}\frac{(1+x^2)^2\arctan x+x(x^2-1)}{(1+x^2)^2};\qquad x=\frac{r}{R}.
\eeq
This is sufficient to make an estimate of the impact of gravitational interactions on the decay rate of the standard model vacuum, to first order in $\kappa$.
One finds that the value of the effective bounce action is shifted by
\beq
\Delta\Gamma[\phi_b]=\frac{256\pi^3}{45(RM_{\rm Pl}\lambda)^2}.
\eeq

There is however a difficulty that arises in the computation of higher-order corrections, as pointed out in Ref.~\cite{Branchina:2016bws}, which is again related to the transformation properties of the action upon scale transformations.
Adopting the approximate form of the scalar potential \eq{phi4} a scale transformation
\begin{align}
\label{scaletransfh}
&\phi_b(x)\to a\phi_b(ax),
\\
\label{scaletransfg}
&g^{\mu\nu}(x)\to g^{\mu\nu}(ax),
\\
\label{scaletransfR}
&{\mathcal R}(x)\to a^2 {\mathcal R}(ax)
\end{align}
transforms the action as
\beq 
S_g\to S_g(a),
\eeq
where
\begin{align}
S_g(a)&=
\int d^4x\,\sqrt{g(ax)}\left[\frac{a^2}{2}\frac{\partial \phi_b(ax)}{\partial x^\mu}
\frac{\partial \phi_b(ax)}{\partial x^\nu}g^{\mu\nu}(ax)+\frac{a^4\lambda}{4}\phi_b^4(ax)
-\frac{a^2{\mathcal R}(ax)}{2\kappa}\right]
\nonumber\\
&=\int d^4x\,\sqrt{g(x)}\left[\frac{1}{2}\frac{\partial \phi_b(x)}{\partial x^\mu}
\frac{\partial \phi_b(x)}{\partial x^\nu}g^{\mu\nu}(x)
+\frac{\lambda}{4}\phi_b^4(x)
-\frac{{\mathcal R}(x)}{2a^2\kappa}
\right]. 
\end{align}
The stationarity condition
\beq
a\left.\frac{dS_g(a)}{da}\right|_{a=1}=\int d^4x\,\sqrt{g(x)}
\frac{{\mathcal R}(x)}{\kappa}=0
\eeq
can only be satisfied in the flat limit $\mathcal{R}(y)=0$, or equivalently
$\rho(r)=r$. But in this case \eq{rhoeq} gives
\beq
\frac{1}{2}{\phi_b'}^2-V(\phi_b)=\frac{1}{2}{\phi_b'}^2+\frac{|\lambda|}{4}\phi_b^4=0,
\eeq
which is only possible if $\phi_b(r)=0$, similarly to what happens in the flat-space-time case with the introduction of a mass term in the scalar potential \eq{phi4}: the breaking of scale invariance leads to the disappearance of the bounce.
This is confirmed by an explicit calculation of the order-$\kappa$ correction to the bounce. \eq{phib1sm} can be solved analytically~\cite{Rajantie:2016hkj}, with the following result:
\begin{align}
\phi_{b1}(r)&=\frac{1}{3R^3|\lambda|^{3/2}}\frac{4\sqrt{2}}{(1+x^2)^2}\Bigg[
c_1(x^2-1)
+c_2\frac{1-17x^2-x^2(1-x^2)(x^2+12\log x)}{2x^2}
\nonumber\\
&+\frac{1}{15x^2}
\frac{1+x^2(7-9x^2)}{1+x^2}-x\arctan x
+\frac{2}{5}
(1-x^2)\log\frac{1+x^2}{x^2}\Bigg],
\end{align}
where $c_1$ and $c_2$ are arbitrary constants.
Since $\phi_{b0}(r)$ obeys the required asymptotic 
conditions for $\phi_b(r)$, namely $\phi_b'(0)=\phi_b(\infty)=0$, the same
conditions must also apply to $\phi_{b1}(r)$. We find
\begin{equation}
\lim_{r\to\infty}\phi_{b1}(r)=
\frac{2\sqrt{2}}{3R^3|\lambda|^{3/2}}c_2
\end{equation}
and
\begin{equation}
\phi_{b1}(r)\sim \frac{1}{R^3|\lambda|^{3/2}}\frac{4\sqrt{2}}{3}
\left(\frac{c_2}{2}+\frac{1}{15}\right)\left(\frac{1}{x^2}-12\log x\right)
+\text{regular terms}
\end{equation}
for $r$ close to zero. Hence, the two boundary conditions cannot be satisfied simultaneously.

\begin{figure}[t!]
 \centering
  {\includegraphics[width=0.45\textwidth]{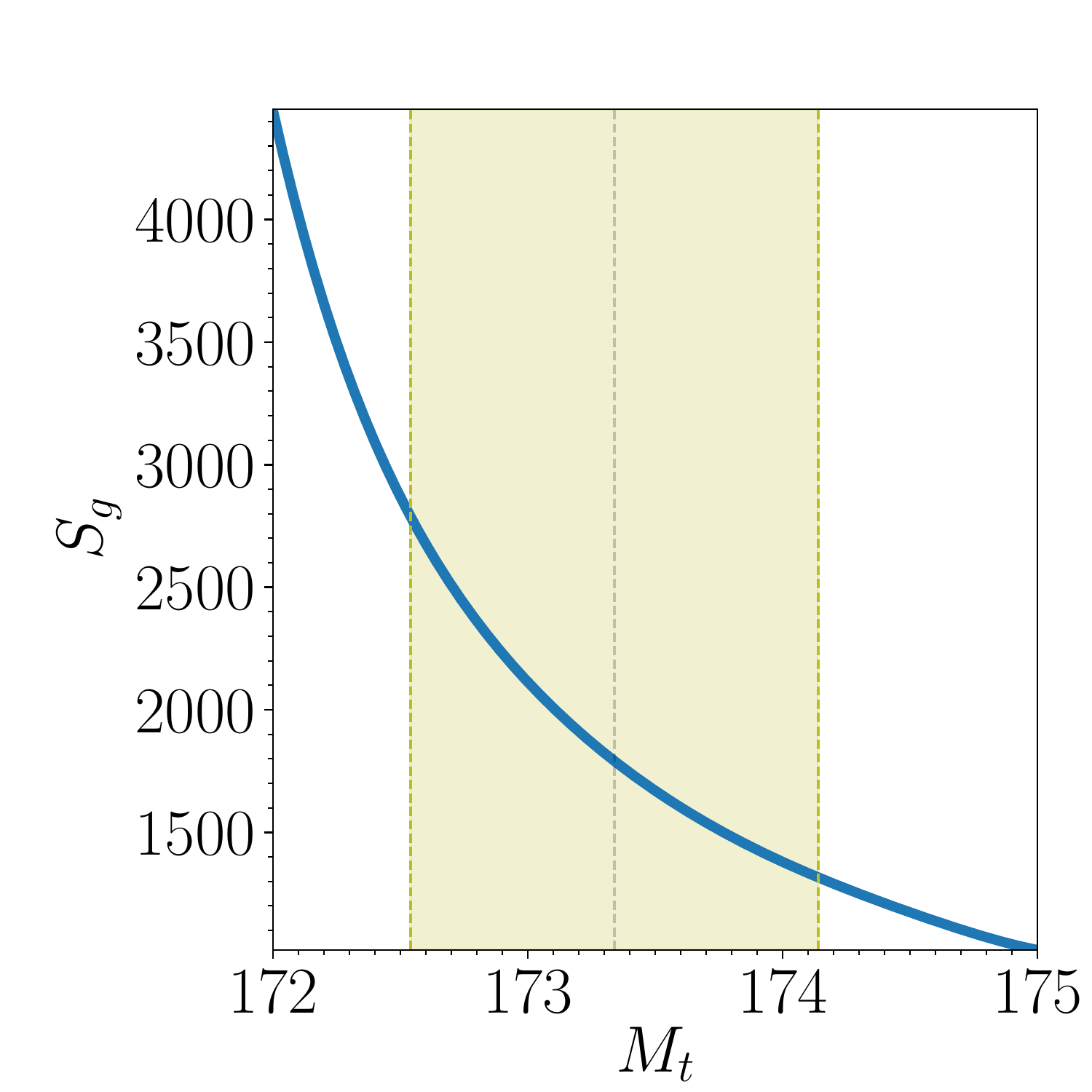}} \qquad
  {\includegraphics[width=0.45\textwidth]{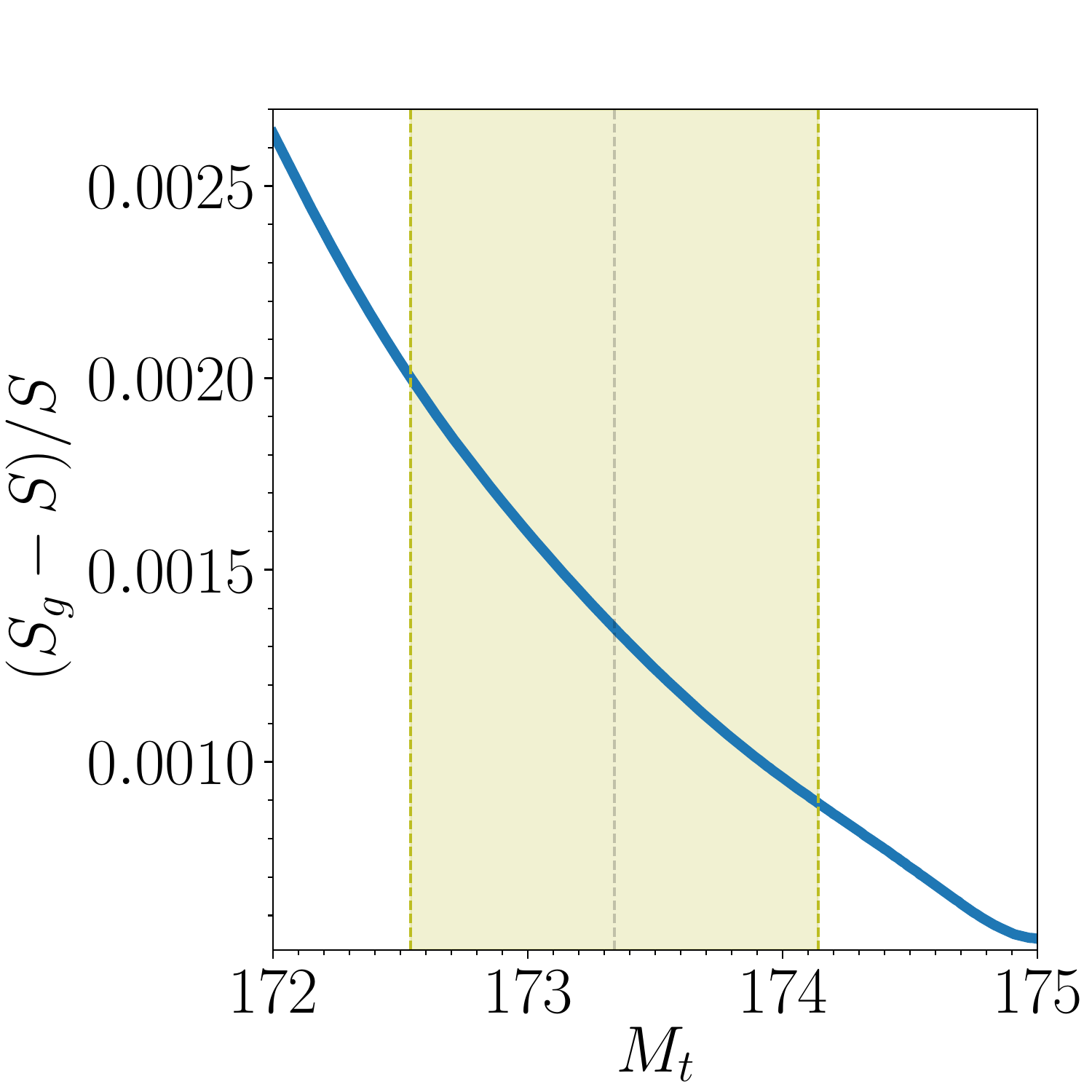}}
    \caption{Left panel: value of the action as a function of the top quark mass, including gravitational effects. Right panel: fractional difference between the action including or neglecting gravity. Displayed are also the central value of the top mass and its $1$-$\sigma$ uncertainty, $M_t =  173.34 \pm  0.8$.}\label{fig:Stop1}
\end{figure}

On the other hand, 
when the scale dependence of the renormalized coupling is taken into account, the stationarity condition upon scale transformations becomes
\beq
a\left.\frac{dS_g(a)}{da}\right|_{a=1}=\int d^4x\,\sqrt{g(x)}
\left[\frac{{\mathcal R}(x)}{\kappa}+\frac{1}{4}\beta(\lambda(\phi_b))\phi_b^4\right]=0,
\eeq
and a non-vanishing bounce solution of the field equation can be found.
We have checked that this is indeed the case by a numerical calculation 
(numerical methods to find the bounce in the presence of gravity are also reviewed in \app{app:nummethods}). Comparing the action computed taking gravity into account, $S_g$, with the action computed in a flat space-time, $S$, we notice that gravity stabilizes the ground state, thus making its lifetime slightly longer.

Since one of the most important parameters describing the running of the coupling constant $\lambda$ is the mass of the top quark $M_t$, it is interesting to study how the lifetime of the ground state is modified by varying $M_t$ within its experimental uncertainty. In the left panel of \fig{fig:Stop1} we represent the value of the action 
in the presence of gravity 
as a function of the top quark mass. As in the flat space-time case, a lower value for the top quark stabilizes the potential, making the ground state lifetime longer.

Furthermore, from the fractional difference between the action including or neglecting gravity 
in the right panel of \fig{fig:Stop1} one can see that  
(especially for lower values of the top quark mass)  
the bounce action becomes larger in the presence of gravity and hence the vacuum lifetime is longer. 
The reason of this phenomenon can be better understood by thinking at the running of $\lambda$. 
A lower mass for the top quark means a larger energy scale involved in the tunnelling process since the zero of $\beta_\lambda$ is shifted 
at higher values, and hence gravitational effects are more important.

Finally, we remark that in the presence of new Planck-scale suppressed effective operators 
(cf.~\sect{sec:UVandNP})
the tunnelling rate might still be strongly enhanced, even after taking into account 
gravitational effects \cite{Bentivegna:2017qry}.

\section{Conclusions}

In this work, we have reviewed the phenomenon of quantum tunnelling 
both in ordinary quantum mechanics and quantum field theory. 
In particular, we have applied the formalism to the relevant case of the 
stability of the electroweak ground state in the standard model, 
which presents several subtle aspects related to the radiative instability of the Higgs potential 
and the approximate scale invariance of the electroweak theory at high-energy scales.  

The present situation regarding the phase diagram of the standard model vacuum is 
represented in \fig{fig:SMphasedia} (from Ref.~\cite{Buttazzo:2013uya}). 
Given the present value of the top quark mass, the standard model vacuum is metastable 
(with a lifetime much larger than the age of the Universe), although remarkably 
close to the absolute stability 
boundary which can be reached at the 2-3$\sigma$ level, depending on the 
uncertainty on the top quark mass. 
Future improvements on the measurement of the top quark mass, 
which may be decisively achieved with a dedicated top-threshold lepton collider 
\cite{Franceschini:2022veh}, 
will be crucial in order to establish 
whether the electroweak ground state is metastable or 
absolutely stable. 

The fate of the standard model vacuum is also deeply intertwined with the 
cosmological evolution of 
the Universe, a fascinating subject that we have 
completely omitted in this introductory review. 
For instance, the electroweak vacuum instability might have played an important 
role for the 
evolution of the 
Higgs field during inflation (see e.g.~\cite{Espinosa:2015qea}). 
Moreover, finite temperature effects
might enhance the decay rate due to thermal fluctuations which should be taken 
into account for the vacuum stability analysis in the early Universe (see e.g.~\cite{DelleRose:2015bpo}).

Although the vacuum decay rate in quantum field theory 
is inherently ultraviolet sensitive (new physics in the deep ultraviolet can always 
enhance the decay rate), the apparent near-criticality 
of the standard model vacuum 
(under the conservative hypothesis of no new physics between the electroweak and 
the Planck scale)
hints to a deeper structure, that is possibly related to the 
hidden meaning of the standard model itself.

\section*{Acknowledgments}

We thank Gino Isidori, Alessandro Strumia and Lorenzo Ubaldi for discussions. The work of FD is partially supported by ERC Starting Grant 804394 hipQCD. 
The work of SD is supported by the Italian Ministry of Research (MIUR) under grant PRIN 201719AVICI-01.
The work of GR is partially supported by the Italian Ministry of Research (MIUR) under grant PRIN 20172LNEEZ.

\appendix

\section{Appendices}
\label{app:Appendices}

\subsection{The semi-classical approximation in quantum mechanics}
\label{app:WKB}

We consider a particle of mass $m$ moving in one space dimension: $x$
is the space coordinate, $p$ its conjugate momentum, and $U(x)$ the
potential energy. The Hamiltonian is
\begin{equation}
H(p,x)=\frac{p^2}{2m}+U(x).
\end{equation}
In classical physics, the particle's motion is only allowed in the region
defined by $E\geq U(x)$, where $E$ is the total particle energy.
The values of $x$ for which $U(x)=E$ are usually called
turning points, for obvious reasons.

In ordinary quantum mechanics one looks for stationary states $\psi(x)$,
solutions
of the time-independent Schr\"odinger equation
\begin{equation}
-\frac{\hbar^2}{2m}\psi''(x)+U(x)\psi(x)=E\psi(x)
\label{sse}
\end{equation}
(primes denote differentiation with respect to the argument.)
Apart from a few simple cases, the stationary state wave functions can be only be determined
in the context of some approximation framework. In the case of barrier penetration,
the so-called semi-classical approximation proves especially useful \cite{landau2013quantum,Konishi:2009qva,griffiths2018introduction}.

The classical limit of quantum mechanics, $\hbar\to 0$,
is a singular limit of \eq{sse}: for $\hbar=0$ the derivative term
disappears, and the equation has a different space of solutions.
The correct way to take the classical limit is the following: the wave function
is written
\begin{equation}
\psi(x)=\exp\left[i\frac{\sigma(x)}{\hbar}\right],
\end{equation}
where $\sigma(x)$ is a complex function.
In this way, all terms in \eq{sse} are of the same
order as $\hbar\to 0$, because
\begin{equation}
\psi''(x)=
-\frac{1}{\hbar^2}\left[{\sigma'}^2(x)-i\hbar\sigma''(x)\right]\psi(x).
\end{equation}
\eq{sse} takes the form
\begin{equation}
{\sigma'}^2(x)-i\hbar\sigma''(x)=2m\left[E-U(x)\right].
\label{ssesigma}
\end{equation}
The complex function $\sigma(x)$ is assumed to have a power expansion
in $\hbar$:
\begin{equation}
\sigma(x)=\sigma_0(x)+\frac{\hbar}{i}\sigma_1(x)+\left(\frac{\hbar}{i}\right)^2
\sigma_2(x)+O(\hbar^3).
\label{sigmaexp}
\end{equation}

Replacing \eq{sigmaexp} in \eq{ssesigma} we obtain a tower of differential equations
for the coefficients $\sigma_n(x)$. To lowest order we get
\begin{equation}
{\sigma_0'}^2(x)=2m\left[E-U(x)\right],
\end{equation}
which is immediately integrated:
\begin{equation}
\sigma_0(x)=\pm\int dx\,p(x),
\end{equation}
where
\begin{equation}
p(x)=+\sqrt{2m\left[E-U(x)\right]}
\end{equation}
is the classical particle momentum. Note that
the term proportional to $\hbar$ in the l.h.s.~of \eq{ssesigma}
can be neglected provided
\begin{equation}
|{\sigma'}^2(x)|\gg \hbar|\sigma''(x)|,
\end{equation}
or
\begin{equation}
\left|\hbar\frac{d}{dx}\frac{1}{\sigma'(x)}\right|
\simeq\left|\frac{d}{dx}\frac{\hbar}{p(x)}\right|\ll 1,
\label{sc}
\end{equation}
which means that the approximation is only valid where the particle
wavelength $\lambda=\frac{h}{p}$ does not vary too much over distances of
its same order.
This is clearly not the case in the vicinity of turning points, where $p(x)=0$
and $\lambda\to\infty$.

Including one more term in the $\hbar$ expansion we get
\begin{equation}
2\sigma_1'(x)\sigma_0'(x)+\sigma_0''(x)=0,
\end{equation}
which gives
\begin{equation}
\sigma_1'(x)=-\frac{1}{2}\frac{\sigma_0''(x)}{\sigma_0'(x)}=-\frac{p'(x)}{2p(x)};
\qquad \sigma_1(x)=\log\frac{1}{\sqrt{|p(x)|}}
\end{equation}
up to an arbitrary integration constant. Hence, the wave function
in the semi-classical approximation, including
corrections up to order $\hbar$ in $\sigma(x)$, is given by
\begin{equation}
\psi(x)=\frac{C_1}{\sqrt{p(x)}}\e^{\frac{i}{\hbar}\int_a^x p(x')dx'}
+\frac{C_2}{\sqrt{p(x)}}\e^{-\frac{i}{\hbar}\int_a^x p(x')dx'}
\label{psiallowed}
\end{equation}
in the classically allowed regions, where $p(x)$ is real and positive, and
\begin{equation}
\psi(x)=\frac{D_1}{\sqrt{|p(x)|}}\e^{-\frac{1}{\hbar}\int_a^x |p(x')|dx'}
+\frac{D_2}{\sqrt{|p(x)|}}\e^{\frac{1}{\hbar}\int_a^x |p(x')|dx'}
\label{psiforbidden}
\end{equation}
in the classically forbidden ones, where $p(x)$ is purely
imaginary.
In the latter case, when $x\to\pm\infty$ only one of the two terms
must be retained, to ensure normalizability of the wave function.
Otherwise, one of the two terms is exponentially suppressed
with respect to the other, and must be consistently neglected.
In Eqs.~(\ref{psiallowed}, \ref{psiforbidden}),
$a$ is an arbitrary reference point on the $x$ axis.
A different choice of $a$ can be absorbed in a redefinition of the constants
$C_i,D_i$. Higher-order terms in the expansion of $\sigma(x)$ in powers of $\hbar$
are very rarely needed.

The constants $C_1,C_2,D_1,D_2$ must be fixed on the basis of normalization
and continuity of the wave function and its first
derivative at turning points.
This is not a trivial task, because the semi-classical
approximation does not hold in the vicinity of
a turning point. A way to circumvent this difficulty
is the following. To be definite, let us assume that $x=a$
is a turning point, with $E<U(x)$ for $x>a$.
Then
\begin{equation}
\psi(x)=\left\{
\begin{array}{ll}
\frac{C_1}{\sqrt{p(x)}}\e^{\frac{i}{\hbar}\int_a^x p(x')dx'}
+\frac{C_2}{\sqrt{p(x)}}\e^{-\frac{i}{\hbar}\int_a^x p(x')dx'}
&
x\ll a
\\
\frac{D}{\sqrt{|p(x)|}}\e^{-\frac{1}{\hbar}\int_a^x |p(x')|dx'}
& x\gg a
\end{array}
\right..
\label{psi1}
\end{equation}
Our task is relating $D$ to $C_1$ and $C_2$ by requiring the continuity
of the wave function and its first derivative at $x=a$. These
conditions cannot be imposed directly on $\psi(x)$ as given in
\eq{psi1}, because the semi-classical
approximation fails in the vicinity of a turning point.
We will instead continue analytically $\psi(x)$ in the complex plane
$x$ from the real axis to the right of the turning point,
${\rm Re\,}x>a$, to the real axis to the left ${\rm Re\,} x<a$, following a path which is never
too close to the turning point.
For example, we may choose a half-circle with center $a$
and radius $r$, in the upper or in the lower complex half-plane,
provided $r$ is large enough for the semi-classical approximation
to hold. We further assume that we can choose $r$ so that the potential
energy $U(x)$ can be reliably expanded in Taylor series to first
order in $x-a$ for all complex $x$ with $|x-a|=r$:
\begin{equation}
U(x)=U(a)+U'(a)(x-a)+O(r^2)=E-F(x-a)+O(r^2),
\end{equation}
with $F=-U'(a)<0$ by assumption.
Neglecting terms of order $r^2$ and higher in $U(x)$ the exponent of
$\psi(x)$ for $x>a$ reads
\begin{equation}
\frac{i}{\hbar}\sigma_0(x)=-\frac{1}{\hbar}
\int_a^x |p(x')|dx'=-\frac{2}{3\hbar}\sqrt{2m|F|}(x-a)^{\frac{3}{2}}.
\end{equation}
We now set
\begin{equation}
x=a+r\e^{i\theta}
\end{equation}
and continue analytically the wave function
following the half-circle in the upper complex plane. We get
\begin{equation}
\frac{i}{\hbar}\sigma_0(x)=-\frac{2}{3\hbar}\sqrt{2m|F|}r^{\frac{3}{2}}
\e^{\frac{3}{2}i\theta};\qquad 0\leq\theta\leq \pi.
\end{equation}
For $\theta=\pi$
\begin{equation}
\frac{i}{\hbar}\sigma_0(x)=\frac{2i}{3\hbar}\sqrt{2m|F|}r^{\frac{3}{2}}
=-\frac{i}{\hbar}\int_a^x p(x')dx',
\end{equation}
which is the exponent of the term proportional to $C_2$ in \eq{psi1}.
On the other hand,
\begin{equation}
\frac{1}{\sqrt{|p(x)|}}=
\frac{1}{(2m|F|)^{\frac{1}{4}}}(x-a)^{-\frac{1}{4}}
\end{equation}
which, by the same argument, is continued to
\begin{equation}
\frac{1}{(2m|F|)^{\frac{1}{4}}}r^{-\frac{1}{4}}\e^{-\frac{1}{4}i\pi}
=\frac{1}{\sqrt{p(x)}}\e^{-\frac{1}{4}i\pi}.
\end{equation}
We therefore conclude that
\begin{equation}
C_2=D\e^{-\frac{1}{4}i\pi}.
\end{equation}
Following the half-circle in the lower half-plane we obtain,
by an analogous argument,
\begin{equation}
C_1=D\e^{\frac{1}{4}i\pi}.
\end{equation}
Finally,
\begin{equation}
\psi(x)=\left\{
\begin{array}{ll}
\frac{2D}{\sqrt{p(x)}}
\cos\left[\frac{1}{\hbar}\int_a^x p(x')dx'+\frac{\pi}{4}\right]
&
x\ll a
\\
\frac{D}{\sqrt{|p(x)|}}\e^{-\frac{1}{\hbar}\int_a^x |p(x')|dx'}
& x\gg a
\end{array}
\right..
\label{psi2}
\end{equation}
This formula was obtained in the case when the region $x>a$
is classically forbidden; it is easy to show that in the general case
\begin{equation}
\psi(x)=\left\{
\begin{array}{ll}
\frac{2D}{\sqrt{p(x)}}
\cos\left[\frac{1}{\hbar}\left|\int_a^x p(x')dx'\right|-\frac{\pi}{4}\right]
&
U(x)<E
\\
\frac{D}{\sqrt{|p(x)|}}\e^{-\frac{1}{\hbar}\left|\int_a^x p(x')dx'\right|}
& U(x)>E
\end{array}
\right..
\label{psi3}
\end{equation}

\subsection{Double-well potential in the semi-classical approximation}
\label{app:doublewell}

In \sect{sec:doublewell} we have discussed how the double-well potential can be studied in the path integral formalism. In this Appendix, we review the standard formulation of the same problem in the semi-classical approximation. We will follow
the discussion presented in Ref.~\cite{landau2013quantum}.

We consider again the hamiltonian of a particle moving in one dimension
under the effect of a potential with two degenerate minima:
\begin{equation}
H=\frac{1}{2}m\left(\frac{dx}{dt}\right)^2+U(x),
\end{equation}
where $U(x)$ is an even function, with two minima at $x=\pm a$ and a local
maximum at $x=0$.
We further assume $U(\pm a)=0$,  and define $U''(\pm a)=m\omega^2$. An example
of a potential energy with these features is
\begin{equation}
\label{example}
U(x)=\frac{m\omega^2}{8a^2}(x^2-a^2)^2,
\end{equation}
but the discussion can be kept at a general level, and reference to a specific form of
the potential is not needed.

If the potential energy barrier separating the two minima
were infinitely high and broad, there would be
two identical towers of energy eigenvalues and eigenstates, corresponding
to the two potential wells at $x=\pm a$.
The wave functions $\psi(\pm x)$ of the fundamental states, with energy
$E_0$,  are strongly
peaked around $x=\pm a$, and decrease exponentially away from the potential
minima. The function $\psi(x)$ is a solution of
\begin{equation}
\label{notunnel}
-\frac{\hbar^2}{2m}\psi''(x)+U(x)\psi(x)=E_0\psi(x).
\end{equation}

We now take quantum tunnelling into account, under the assumption
that the semi-classical approximation is
reliable. Because the potential is parity-even, the wave functions
of the two lowest-energy states are
\begin{align}
&\psi_1(x)=\frac{1}{\sqrt{2}}\left[\psi(x)+\psi(-x)\right]
\label{dw1}
\\
&\psi_2(x)=\frac{1}{\sqrt{2}}\left[\psi(x)-\psi(-x)\right]
\label{dw2}
\end{align}
with energies $E_1$ and $E_2$ respectively:
\begin{align}
&-\frac{\hbar^2}{2m}\psi_1''(x)+U(x)\psi_1(x)=E_1\psi_1(x)
\label{tunnel1}
\\
&-\frac{\hbar^2}{2m}\psi_2''(x)+U(x)\psi_2(x)=E_2\psi_2(x),
\label{tunnel2}
\end{align}
and $E_1<E_2$ because
$\psi_2(x)$ has a node in $x=0$. Furthermore, the two wave functions
are approximately normalized, because $\psi(x)$ has almost no overlap
with $\psi(-x)$.
Eqs.~(\ref{notunnel}, \ref{tunnel1}, \ref{tunnel2}) can be combined and
integrated, to obtain
\begin{align}
-\frac{\hbar^2}{2m}\int_0^{+\infty}dx\,
\left[\psi_1(x)\psi''(x)-\psi_1''(x)\psi(x)\right]
&=(E_0-E_1)\int_0^{+\infty}dx\,\psi(x)\psi_1(x)
\\
-\frac{\hbar^2}{2m}\int_0^{+\infty}dx\,
\left[\psi_2(x)\psi''(x)-\psi_2''(x)\psi(x)\right]
&=(E_0-E_2)\int_0^{+\infty}dx\,\psi(x)\psi_2(x).
\end{align}
The integrals are readily performed by means of
Eqs.~(\ref{dw1}, \ref{dw2}), which give
$\psi'_1(0)=\psi_2(0)=0$. We obtain
\begin{align}
E_1&=E_0-\frac{\hbar^2}{m}\psi(0)\psi'(0)
\\
E_2&=E_0+\frac{\hbar^2}{m}\psi'(0)\psi(0)
\end{align}
and therefore
\begin{equation}
E_2-E_1=\frac{2\hbar^2}{m}\psi(0)\psi'(0).
\end{equation}
It is at this point that the semi-classical approximation
comes into play.
The value of the wave function at the origin can be computed
using the approximate expression \eq{psi3}
for the wave function; we find
\begin{equation}
\psi(0)=\frac{D}{\sqrt{|p(0)|}}\e^{-\frac{1}{\hbar}\int_0^{x_-}dx\,|p(x)|}
,\qquad \psi'(0)=
\frac{D\sqrt{|p(0)|}}{\hbar}
\e^{-\frac{1}{\hbar}\int_0^{x_-}dx\,|p(x)|}
\end{equation}
where $x_\pm$ are the classical turning points corresponding
to the energy $E_0$ for the potential well at $x=a$:
\begin{equation}
U(x_\pm)=E_0;\qquad x_\pm>0.
\end{equation}
Therefore
\begin{equation}
E_2-E_1=\frac{2\hbar}{m}D^2\e^{-\frac{1}{\hbar}\int_{-x_-}^{x_-}dx\,|p(x)|}
\end{equation}
where we have used $U(x)=U(-x)$.

The constant $D$ can be obtained from the normalization condition of
the wave function. Neglecting the exponentially suppressed
contributions from the classically forbidden regions, we get
\begin{equation}
\int_{-\infty}^{+\infty}|\psi(x)|^2dx
=4D^2\int_{x_-}^{x_+}dx\,\frac{1}{p(x)}
\cos^2
\left[\frac{1}{\hbar}\left|\int_{x_-}^x p(x')dx'\right|-\frac{\pi}{4}\right]=1.
\end{equation}
We may replace the cosine squared in the integrand by its average
over a period of the classical motion, $1/2$, to obtain
\begin{equation}
2D^2\int_{x_-}^{x_+}\frac{dx}{p(x)}=1.
\end{equation}
Furthermore,
\begin{equation}
\int_{x_-}^{x_+}\frac{dx}{p(x)}=\int_{t_1}^{t_2}\frac{dx}{dt}\frac{dt}{p(x)}=
\frac{1}{m}(t_2-t_1)=\frac{\pi}{m\omega},
\end{equation}
which gives
\begin{equation}
D^2=\frac{m\omega}{2\pi}
\end{equation}
and
\begin{equation}
E_2-E_1=\frac{\hbar\omega}{\pi}
\e^{-\frac{1}{\hbar}\int_{-x_-}^{x_-}dx\,|p(x)|};\qquad |p(x)|=\sqrt{2m(U(x)-E_0)};\qquad U(x_-)=E_0.
\label{splitWKB}
\end{equation}
This is how the result is usually presented in textbooks (see e.g.~\cite{landau2013quantum}).

Since
\begin{equation}
U(x)=\frac{1}{2}m\omega^2(x-a)^2+O((x-a)^3)
\label{Uapp}
\end{equation}
in the vicinity of $x=a$, we have
\begin{equation}
E_0\simeq\frac{\hbar\omega}{2};\qquad
x_\pm\simeq a\pm\sqrt{\frac{\hbar}{m\omega}},
\end{equation}
and the exponent of \eq{splitWKB} reads
\begin{equation}
F(\hbar)=-\frac{2}{\hbar}
\int_0^{a-\sqrt{\frac{\hbar}{m\omega}}}dx\,\sqrt{2mU(x)-\hbar m\omega}
\label{expWKB}
\end{equation}
The exponent $F(\hbar)$
is singular as $\log\hbar$ as $\hbar\to 0$, as can be seen
by observing that, for small $\hbar$, the integration variable $x$ is close to
$a$ at the upper integration bound, and $U(x)\sim \frac{1}{2}m\omega^2(a-x)^2$
in this region.
In order take this singularity into account, we split the integration
range in \eq{expWKB}:
\begin{equation}
F(\hbar)=-\frac{2}{\hbar}
\int_0^{a-\epsilon}dx\,\sqrt{2mU(x)-\hbar m\omega}
-\frac{2}{\hbar}\int_{a-\epsilon}^{a-\sqrt{\frac{\hbar}{m\omega}}}dx\,
\sqrt{2mU(x)-\hbar m\omega},
\label{expWKB2}
\end{equation}
where $\epsilon$ is an arbitrary constant
in the range $\sqrt{\frac{\hbar}{m\omega}}<\epsilon<a$. The integration
range in the first integral now does not contain the singular point,
and we may expand the integrand in powers of $\hbar$
to first order, while in the second integral we may approximate $U(x)$ as in \eq{Uapp} if $\epsilon$ is not too large.
We get
\begin{align}
F(\hbar)
=&-\frac{2}{\hbar}\int_0^{a-\epsilon}dx\,\sqrt{2mU(x)}
+m\omega\int_0^{a-\epsilon}\frac{dx}{\sqrt{2mU(x)}}
\nonumber\\
&-\frac{2m\omega}{\hbar}\int_{a-\epsilon}^{a-\sqrt{\frac{\hbar}{m\omega}}}dx\,
\sqrt{(a-x)^2-\frac{\hbar}{m\omega}}+O(\hbar).
\end{align}
The last integral can be computed analytically. Neglecting $O(\hbar)$ corrections, and using \eq{Uapp} again whenever allowed, we get
\begin{align}
F(\hbar)=&
-\frac{2}{\hbar}\int_0^a dx\,\sqrt{2mU(x)}
-\frac{2m\omega}{\hbar}\int_a^{a-\epsilon}dx\,(a-x)
\nonumber\\
&+\int_0^{a-\epsilon}dx\,\left[\frac{m\omega}{\sqrt{2mU(x)}}-\frac{1}{a-x}\right]
+\int_0^{a-\epsilon}dx\,\frac{1}{a-x}
\nonumber\\
&-\log\left[\sqrt{\frac{m\omega}{\hbar}}
\left(\epsilon-\sqrt{\epsilon^2-\frac{\hbar}{m\omega}}\right)\right]
-\frac{m\omega}{\hbar}\epsilon\sqrt{\epsilon^2-\frac{\hbar}{m\omega}}
\nonumber\\
=&
-\frac{2}{\hbar}\int_0^a dx\,\sqrt{2mU(x)}
+\int_0^{a-\epsilon}dx\,\left[\frac{m\omega}{\sqrt{2mU(x)}}-\frac{1}{a-x}\right]
\nonumber\\
&-\log\left[\frac{\epsilon^2}{a}\sqrt{\frac{m\omega}{\hbar}}
\left(1-\sqrt{1-\frac{\hbar}{m\omega\epsilon^2}}\right)\right]
+\frac{m\omega\epsilon^2}{\hbar}\left(1-\sqrt{1-\frac{\hbar}{m\omega\epsilon^2}}\right).
\end{align}
In the semi-classical limit $\hbar\ll\epsilon^2m\omega$ we have
\begin{equation}
1-\sqrt{1-\frac{\hbar}{m\omega\epsilon^2}}
=\frac{\hbar}{2m\omega\epsilon^2}+O(\hbar^2),
\end{equation}
so that
\begin{equation}
F(\hbar)=
-\frac{2}{\hbar}\int_0^a dx\,\sqrt{2mU(x)}
+\int_0^{a-\epsilon}dx\,\left[\frac{m\omega}{\sqrt{2mU(x)}}-\frac{1}{a-x}\right]
+\log\left[2a\sqrt{\frac{m\omega}{\hbar}}\right]
+\frac{1}{2}.
\end{equation}
The only residual dependence on the arbitrary parameter $\epsilon$ is in the upper integration bound of the second integral,
where it has a negligible effect because the integrand is essentially zero in the range $a-\epsilon<x<a$.
We have therefore
\begin{equation}
E_2-E_1=\frac{\hbar\omega}{\pi}\exp F(\hbar)=\sqrt{\frac{e}{\pi}}\hbar\omega \left[2\hat a\sqrt{\frac{m\omega}{\pi\hbar}}\right]
\exp\left[-\frac{1}{\hbar}\int_{-a}^a dx\,\sqrt{2mU(x)}\right],
\label{splitWKBfinal}
\end{equation}
where 
\begin{equation}
\hat a=a\exp\int_0^a dx\,\left[\frac{m\omega}{\sqrt{2mU(x)}}-\frac{1}{a-x}\right].
\label{Adw0}
\end{equation}

The result Eqs.~(\ref{splitWKBfinal}, \ref{Adw0}) must be compared with the result Eqs.~(\ref{splitPI}, \ref{SPI}) with $U_0=0$, obtained with the path integral method. We see that the two expressions differ by a factor
\begin{equation}
\sqrt{\frac{\e}{\pi}}\sim0.93.
\end{equation}
This small discrepancy can be removed by a more accurate treatment of the matching
of the wave function at classical turning points in the semi-classical
approximation, see for example \cite{Konishi:2009qva}.

\subsection{The Maupertuis principle}
\label{app:maup}

The Maupertuis principle is a modified version of Hamilton's principle of classical mechanics,
which allows one to determine the trajectory of the actual motion (with no reference
to the time dependence of the coordinates), when energy is conserved.

We consider the action of a system with $N$ degrees of freedom
as a function of the generalized coordinates at the final time $t$, $\v x(t)$
and of $t$ itself:
\begin{equation}
S(\v x(t),t)=\int_{t_0}^t dt'\,L(\v x(t'),\dot{\v x}(t'))
=\int_{t_0}^t dt'\,\left[\frac{1}{2}\dot x_i\dot x_i-U(\v x)\right],
\end{equation}
where a sum over the repeated index $i$ is understood. Here $\v x(t')$ is a solution of the equation of motion, with $\v x(t_0)=\v a$. We find
\begin{align}
\frac{\partial}{\partial x_i(t)}S(\v x(t),t)&=
\int_{t_0}^t dt'\,\left[\frac{\partial L}{\partial x_j(t')}
\frac{\partial x_j(t')}{\partial x_i(t)}
+\frac{\partial L}{\partial \dot x_j(t')}
\frac{d}{dt'}\frac{\partial x_j(t')}{\partial x_i(t)}\right]
\\
&=
\int_{t_0}^t dt'\,\left[\frac{\partial L}{\partial x_j(t')}
-\frac{d}{dt'}\frac{\partial L}{\partial \dot x_j(t')}\right]
\frac{\partial x_j(t')}{\partial x_i(t)}
+
\left.\frac{\partial L}{\partial\dot x_j(t')}
\frac{\partial x_j(t')}{\partial x_i(t)}
\right|_{t_0}^t.
\nonumber
\end{align}
Since $\v x(t')$ is assumed to be a solution of the Euler-Lagrange equations, the
integrand in the first term vanishes identically, and we are left with
the result
\begin{equation}
\frac{\partial}{\partial x_i(t)}S(\v x(t),t)=p_i(t)
\end{equation}
where
\begin{equation}
p_i(t)=\frac{\partial L}{\partial\dot x_i(t)}=\dot x_i(t)
\end{equation}
is the conjugate momentum of $x_i$ at time $t$.
Also,
\begin{align}
\frac{d}{dt}S(\v x(t),t)
&=\frac{\partial}{\partial t}S(\v x(t),t)
+\frac{\partial S(\v x(t), t)}{\partial x_i(t)}\frac{dx_i(t)}{dt}
\nonumber\\
&=\frac{\partial}{\partial t}S(\v x(t),t)
+p_i\dot x_i(t).
\end{align}
On the other hand, by definition,
\begin{equation}
\frac{d}{dt}S(\v x(t),t)=L(\v x(t),\dot{\v x}(t)).
\end{equation}
By comparison,
\begin{equation}
\frac{\partial}{\partial t}S(\v x(t),t)
=L(\v x(t),\dot{\v x}(t))-p_i(t)\dot x_i(t)=-H(\v p(t),\v x(t)).
\end{equation}
The action can therefore be written
\begin{equation}
S(\v x(t),t)=S_0(\v x(t))-\int_{t_0}^tdt'\,H(\v p(t'),\v x(t')),
\label{newS}
\end{equation}
where
\begin{equation}
S_0(\v x(t))=\sum_i\int_{a_i}^{x_i(t)}p_idx_i
\end{equation}
is sometimes called the reduced action, and $\v a$ is a reference point in configuration space. When $H$ is constant in time,
\begin{equation}
H(p_i(t),x_i(t))=\frac{1}{2}\sum_i\frac{dx_i}{dt}\frac{dx_i}{dt}+U(\v x)=E,
\label{energycons}
\end{equation}
we get
\begin{equation}
S(\v x(t),t)=S_0(\v x(t))-E(t-t_0).
\label{newS2}
\end{equation}
We know that the action $S$ is stationary in correspondence
of the solutions of the equations of motion, provided the variation
is performed with $\v x(t_0)$, $\v x(t)$ and $t-t_0$ fixed. If we now allow
the time interval to vary as well, we obtain, from the above argument,
\begin{equation}
\delta S=-E\delta t.
\label{ds}
\end{equation}
On the other hand, from \eq{newS2} we get
\begin{equation}
\delta S=\delta S_0-E\delta t,
\label{ds0}
\end{equation}
and hence, comparing Eqs.~(\ref{ds}) and (\ref{ds0}),
\begin{equation}
\delta S_0=0.
\end{equation}
Thus we have shown that the reduced action is stationary in
correspondence of the solutions of the equations of motion
if energy is conserved.

We may now choose a parametrization of the trajectory, $\v x=\v x(s)$,
with $0\leq s\leq s_f$, $\v x(0)=\v x_0,\v x(s_f)=\v X$, such that
\begin{equation}
\left|\frac{d\v x(s)}{ds}\right|^2=
\sum_i\frac{dx_i(s)}{ds}\frac{dx_i(s)}{ds}=1
\label{param}
\end{equation}
or equivalently
\begin{equation}
\sum_i\frac{dx_i}{dt}\frac{dx_i}{dt}
=\sum_i\frac{dx_i}{ds}\frac{dx_i}{ds}\left(\frac{ds}{dt}\right)^2
=\left(\frac{ds}{dt}\right)^2.
\end{equation}
Using \eq{energycons},
\begin{equation}
\frac{ds}{dt}=\sqrt{2[E-U(\v x)]}.
\label{dsdt}
\end{equation}
Then
\begin{align}
S_0(\v x(t),t)&=\sum_i\int_{a_i}^{x_i(t)}p_idx_i
=\int_0^{s_f}\left(\sum_i\frac{dx_i}{ds}\frac{dx_i}{ds}\right)\frac{ds}{dt}ds
\nonumber\\
&=\int_0^{s_f}\sqrt{2[E-U(\v x(s))]}ds,
\label{S0maup}
\end{align}
where we have used \eq{dsdt} and \eq{param}.

\subsection{Legendre transforms}
\label{app:LegTra}

Given a function $f$ of a single variable $x$, its Legendre transform,
denoted by $g(p)$, is defined as
\beq
g (p) = \underset{x}{\text{max}} \[px - f(x)\].
\eeq
Graphically this corresponds to drawing a line through the origin with slope $p$
and taking a point $x = x_0(p)$ such that
the vertical distance between $f(x)$ and the line $px$ is maximized.
The maximization of the function $px - f(x)$ implies:
\beq
\label{1and2derf}
\left. \frac{df}{dx} \right|_{x=x_0} = p \qquad \text{and} \qquad
\left. \frac{d^2f}{dx^2} \right|_{x=x_0} \geq 0,
\eeq
so that we can equivalently rewrite
\beq
g (p) = x_0(p) p -  f(x_0(p)),
\eeq
whose form resembles that of the effective action, after
the proper generalization to a functional space (cf.~\eq{effectiveaction}).
The first equation in (\ref{1and2derf}) implies that
the line tangent to $f$ at the point $x_0$ must have slope $p$, i.e.~it must be
parallel to the line $px$. Note that if $f(x)$ is convex ($f''(x) \geq 0$ everywhere)
there will be only one value of $x_0$, otherwise there might be several
and one has to pick up that which maximizes $px - f(x)$.

The Legendre transform has the following properties (which we report without proof):
\begin{itemize}
\item $g$ is always a convex function (even if $f$ is non convex).
\item the double Legendre transform of a convex function $f$ is $f$ itself.
If $f$ is non convex, the double Legendre transform of $f$ is its convex envelope.
\end{itemize}

\subsection{The Fubini-Lipatov bounce}
\label{app:FLinstanton}

Solutions of the second-order differential equation
\beq
\phi''(r)+\frac{3}{r}\phi'(r)=\lambda\phi^3(r)
\label{bounceapp}
\eeq
with $\phi(0)>0,\phi'(0)=0,\phi(\infty)=0$ can be found 
in the form of a Taylor expansion around $r=0$:
\beq
\phi(r)=\sum_{k=0}^\infty A_k r^k
\label{bounceTaylor}
\eeq
with $A_0>0$. Eq.~(\ref{bounceapp}) takes the form of a recurrence relation for the coefficients $A_k$:
\beq
\frac{3A_1}{r}+
\sum_{k=0}^\infty (k+2)(k+4)A_{k+2}r^k
=\lambda\sum_{k=0}^\infty r^k\sum_{i=0}^k\sum_{j=0}^{k-i}A_i A_j A_{k-i-j}.
\eeq
It follows that
\beq
A_1=0.
\eeq
The first derivative of the solution at $r=0$ vanishes
as a consequence of the assumption Eq.~(\ref{bounceTaylor}).
The remaining coefficients are given by 
\beq
A_{k+2}=\frac{\lambda}{(k+2)(k+4)}\sum_{i=0}^k\sum_{j=0}^{k-i}A_i A_j A_{k-i-j}.
\eeq
The coefficients $A_k$ with $k$ odd are zero: indeed, when $k$ is odd,
at least one of the three summation indices $i,j,k-i-j$ is also odd.
Hence $A_1=0$ implies $A_3=0$, and so on.
Thus, we may rewrite Eq.~(\ref{bounceTaylor}) as
\beq
\phi(r)=\sum_{k=0}^\infty a_k r^{2k};\qquad a_k=A_{2k},
\label{bounceTayloreven}
\eeq
and the recurrence relation for the coefficients becomes
\beq
a_{k+1}=\frac{\lambda}{8}\frac{2}{(k+1)(k+2)}
\sum_{i=0}^k\sum_{j=0}^{k-i}a_i a_j a_{k-i-j}.
\label{recurrence}
\eeq
The coefficients $a_k$ are determined by the single number $a_0$, the
value of $\phi(r)$ at $r=0$. We now show that
\beq
a_j=\left(\frac{\lambda}{8}\right)^j a_0^{2j+1}.
\label{aj}
\eeq
The proof is by induction. For $k=0$ Eq.~(\ref{recurrence})
gives
\beq
a_1=\frac{\lambda}{8}a_0^3.
\eeq
We now assume that Eq.~(\ref{aj}) holds for $0\leq j\leq k$. Then
\begin{align}
a_{k+1}=&\frac{\lambda}{8}\frac{2}{(k+1)(k+2)}
\sum_{i=0}^k\sum_{j=0}^{k-i}a_i a_j a_{k-i-j}
\nonumber\\
=&\frac{\lambda}{8}\frac{2}{(k+1)(k+2)}
\sum_{i=0}^k\sum_{j=0}^{k-i}\left(\frac{\lambda}{8}\right)^k a_0^{2k+3}
=\left(\frac{\lambda}{8}\right)^{k+1} a_0^{2k+3}
\end{align}
which is what we set out to prove. The Taylor expansion in
Eq.~(\ref{bounceTayloreven}) can now be summed. We find
\beq
\phi(r)=a_0\sum_{k=0}^\infty \left(\frac{\lambda}{8}\right)^k a_0^{2k}r^{2k}
=\frac{a_0}{1-\frac{\lambda}{8}a_0^2r^2}.
\label{bounceSM2}
\eeq
The series has convergence radius
\beq
R=\sqrt{\frac{8}{|\lambda|}}\frac{1}{a_0}.
\label{Rdef}
\eeq
If $\lambda<0$ the sum can be analytically continued to
the whole positive real axis, and vanishes as $r\to\infty$. Finally, we note
that for $\lambda<0$ the bounce in Eq.~(\ref{bounceSM2}) coincides
with the solution given in Eq.~(\ref{SMhR}), with $R$ as in
Eq.~(\ref{Rdef}). The above construction shows that, given the value
of the bounce at the origin and the requirement of regularity in the
range $0<r<\infty$, the solution is unique.
An alternative derivation of the result  \eq{bounceSM2} is given in 
Ref.~\cite{Lee:1985uv}.

\subsection{Numerical methods for the bounce equation}
\label{app:nummethods}

Some of the results presented in this review have been obtained by numerical methods.
This Appendix contains an illustration of the techniques adopted for the numerical determination of the bounce solutions for a generic potential, both in the flat and curved space-time cases.

\subsubsection{Flat space}

\label{Appendix:Numerical Methods I: Flat Space}
It is convenient to introduce dimensionless variables:
\begin{equation}
  \label{eq:adim}
  x=Mr\,;\hspace{1.5cm}\tilde \phi(x)=\frac {\phi(r)}M\,;\hspace{1.5cm}\tilde V(\tilde \phi)=\frac{V(\phi)}{M^4},
\end{equation}
where $M$ is an arbitrary mass scale. It will prove convenient to choose $M$ of the same order of magnitude as the value of the bounce in its centre, $M\simeq \phi(0)$,
in order to keep the value of $\tilde\phi$ of order one.

In terms of dimensionless variables, the field equation reads
\begin{equation}
  \label{eq:eomadim}
  \tilde \phi''(x)+\frac 3x\tilde \phi'(x)=\tilde V'(\tilde \phi),
\end{equation}
while the boundary conditions become
\begin{align}
\label{eq:boundary1adim}
 &\lim_{x\to\infty}\tilde \phi(x)=\frac {\phi_{\rm FV}}M
 \\
 \label{eq:boundary2adim}
 &\left.\frac{d\tilde\phi(x)}{dx}\right|_{x=0}=0,
\end{align}
$\phi_{\rm FV}$ being the value of the field in the false vacuum configuration.

Finding a numerical solution of \eq{eq:eomadim} is not straightforward since the boundary conditions \eqs{eq:boundary1adim}{eq:boundary2adim} do not define a Cauchy problem, but rather a boundary value problem.
The solution can be found via a \emph{shooting} method. We define a Cauchy problem choosing as boundary conditions \eq{eq:boundary2adim} and
\begin{equation}
  \tilde \phi(0)=a.
\end{equation}
This problem has a unique solution, which can be found by means of available numerical routines, as a function of the parameter $a$.  Next, the
value of $a$ is adjusted so that the boundary condition \eq{eq:boundary1adim} is fulfilled.

Note that the origin $x=0$ is a singular point for \eq{eq:eomadim}: thus, the numerical solution must be looked for in a range $x_{\rm min} \le x\le x_{\rm max}$ with $x_{\rm min}>0$.
As a consequence, we need the initial conditions at $x=x_{\rm min}$ which correspond to those at $x=0$. To this purpose, we relate $\phi(x_{\rm min})$ to
$\phi(0)$ by a Taylor expansion:
\begin{equation}
  \label{eq:expansion1}
  \tilde \phi(x_{\rm min})=a+\tilde \phi'(0)x_{\rm min}+\frac 12 \tilde \phi''(0)x_{\rm min}^2+O(x_{\rm min}^3)
  =a+\frac 12 \tilde \phi''(0)x_{\rm min}^2+O(x_{\rm min}^3),
\end{equation}
where we have used \eq{eq:boundary2adim}.
The field equation \eq{eq:eomadim} gives
\begin{equation}
  \label{eq:h''(0)}
 \tilde\phi''(0)=\tilde V(\tilde \phi(0))-\left.\frac 3x \tilde \phi'(x)\right|_{x=0}
 =\tilde V(a)-3\tilde\phi''(0),
\end{equation}
which gives
\begin{equation}
\tilde\phi''(0)=\frac 14 \tilde V'(a).
\end{equation}
So finally
\begin{align}
  \label{eq:numericalboundary1}
  & \tilde \phi(x_{\rm min})=a+\frac 18 \tilde V'(a)x_{\rm min}^2+O(x_{\rm min}^3)
\\
  &\tilde \phi'(x_{\rm min})=\frac 14 \tilde V'(a) x_{\rm min}+O(x_{\rm min}^3).
\end{align}
The Cauchy problem which determines the bounce can now be solved numerically.

From the numerical point of view, the calculation of the Euclidean action is most conveniently performed by means of \eq{eq:flatvirial}. Exploiting the $\O(4)$ symmetry of the bounce, we find
\begin{equation}
  S_E=-2\pi^2\int_{x_{\rm min}}^{x_{\rm max}} dx\, x^3  \tilde V(\tilde \phi(x)).
\end{equation}
As a final step, one iterates the procedure decreasing $x_{\rm min}$ and increasing
$x_{\rm max}$ until the result is numerically stable.

\subsubsection{Curved space}

\label{Appendix:Numerical Methods II: Curved Space}

In the presence of gravitational effects, the algorithm closely follows the one described in the case of flat space-time.
We have now to solve the coupled \eqs{heq}{rhoeq}. A complication immediately arises,
due to the square root ambiguity in \eq{rhoeq}.
To get rid of it we differentiate \eq{rhoeq} with respect to $r$ \cite{Rajantie}. We get
\begin{equation}
  \rho''=-\frac \kappa 3 \rho(\phi_b'^2+V(\phi_b)).
\end{equation}
Following  the same strategy as in the flat case, we introduce the dimensionless variables
\begin{equation}
  \label{eq:adim2}
  x=Mr\,; \qquad \tilde \phi(x)=\frac {\phi_b(r)}M\,; \qquad
   \tilde V(\tilde \phi)=\frac{V(\phi_b)}{M^4}; \qquad \tilde \rho(x)=\rho(r),
\end{equation}
where $M$ is an arbitrary mass scale.
In terms of the new variables, the system of differential equations becomes
\begin{align}
\label{eq:g1}
&  \tilde \phi''=-3\frac {\tilde \rho'}{\tilde \rho}\tilde\phi'+\frac{d\tilde V}{d\tilde\phi},
\\
\label{eq:g2}
& \tilde \rho''=-\frac {8\pi}3\epsilon^2\tilde\rho\left(\tilde\phi'^2+\tilde V\right),
\end{align}
where $\kappa=8\pi/M_{\rm Pl}^2$, $\epsilon=M/M_{\rm Pl}$.

The boundary conditions for $\phi$ are the same as in flat  space. Thus it is not a Cauchy problem and we have to apply a shooting algorithm similarly to the case in absence of gravity.

On the other hand the  boundary condition for the function $\rho$ are
\begin{align}
& \rho'(0)=1,\\
& \rho(0)=0
\end{align}
from \eq{rhoeq} and \eq{gravbound}.

The boundary conditions in terms of dimensionless variables read
\begin{align}
  \label{eq:bg1}
   &\lim_{x\to\infty}\tilde \phi(x)=\frac {\phi_{\rm FV}}M,\\
   \label{eq:bg2}
  &\frac{d\tilde\phi(x)}{dx}\Bigg|_{x=0}=0,\\
  &\tilde \rho(0)=0,\\
  &\tilde \rho'(0)=1.
\end{align}
Proceeding as in the previous case, we solve numerically the two differential equations \eqs{eq:g1}{eq:g2} in the range
$(x_{\rm min},x_{\rm max})$ with initial conditions
\begin{align}
& \tilde\phi(x_{ \rm min}) = a + \frac{1}{8} \left. \frac{d \tilde V}{d \tilde\phi}\right|_{\tilde \phi = a} x_{\rm min}^2 + \mathcal{O} (x_{\rm min}^4),\\
& \tilde h' (x_{\rm min}) = \frac{1}{4} \left. \frac{d \tilde V}{d \tilde\phi}\right|_{\tilde h = a} x_{\rm min} + \mathcal{O} (x_{\rm min}^3),
\\
& \tilde \rho (x_{\rm min}) = x_{\rm min},\\
& \tilde \rho' (x_{\rm min}) = 1.
\end{align}
and we adjust the value of $a$ so that \eq{eq:bg1} is fulfilled.
The algorithm is iterated lowering $x_{\rm min}$ and increasing $x_{\rm max}$, until the desired level of accuracy is achieved.

\subsection{Gravitational field equations}
\label{section:graveom}

In this Appendix we provide some details about the field equations in the case of a scalar field minimally coupled to gravity.
Following Ref.~\cite{Coleman:1980aw}, we add the Euclidean version of the Einstein-Hilbert action to the action of the scalar field:\footnote{We neglect a non-minimal
coupling of gravity to the Higgs of the form $\xi \mathcal{R} |H|^2$. The effect of this operator on the tunnelling rate
of the electroweak vacuum has been considered e.g.~in \cite{Salvio:2016mvj,Branchina:2019tyy}.}
\beq
S[\phi,g]=S_g[g]+S_\phi[\phi,g],
\label{eq:einsteinhilbert}
\eeq
with
\begin{align}
S_g[g]&=\int d^4 x\, \sqrt{g} \left[ -\frac{1}{2\kappa} \mathcal{R} \right], \\
S_\phi[\phi,g]
&= \int d^4 x\, \sqrt{g} \left[\frac{1}{2} g^{\mu\nu} \partial_\mu \phi \partial_\nu \phi + V(\phi) \right],
\end{align}
where $\kappa = 8 \pi G_N = 8 \pi / M_{\rm Pl}^2$, $\mathcal{R}$ is the Riemann curvature,
and the metric tensor
$g_{\mu\nu}$ is understood to have a positive-definite Euclidean signature.
Note that, for a scalar field, there is no difference between ordinary and covariant derivative:
$\nabla_\mu \phi=\partial_\mu\phi$.

There is an important difference with respect to the flat space-time case:
without gravity, an additive constant to the scalar potential has no effect on the field equations. When gravity is taken into account, instead, a constant term in
the scalar potential contributes with a term proportional $\sqrt g$ to the
action integrand, which results in a contribution to the field equations via a
cosmological constant. Remarkably, vacuum decay affects gravitation itself: when tunnelling takes place,
the value of the potential at the minimum (and hence the cosmological constant)
gets modified.

We first derive the Euclidean field equations in the presence of gravity.
By taking the variation of $S[\phi,g]$ with respect to the metric we obtain the Einstein equations
\begin{equation}
\mathcal{G}_{\mu\nu}=-\kappa \mathcal{T}_{\mu\nu},
\label{eq:einst}
\end{equation}
where $\mathcal{G}_{\mu\nu}$ is the Einstein tensor
\begin{equation}
\mathcal{G}_{\mu\nu}=\mathcal{R}_{\mu\nu}-\frac 12 \mathcal{R}\, g_{\mu\nu},
\end{equation}
and $\mathcal{T}_{\mu\nu}$ the stress-energy tensor
\begin{equation}
\mathcal{T}_{\mu\nu} = - \frac{2}{\sqrt{g}} \frac{\delta S_\phi}{\delta g^{\mu\nu}}
= - \partial_\mu \phi \partial_\nu \phi + g_{\mu\nu} \left[ \frac{1}{2} g^{\rho\sigma} \partial_\rho \phi \partial_\sigma \phi + V(\phi) \right].
\end{equation}
The minus sign  in \eq{eq:einst} is due to the fact that the metric is Euclidean.

The field equation for the scalar field is obtained by taking the
variation with respect to $\phi$:
\begin{align}
\delta_\phi S[\phi,g] &= \int d^4 x\, \sqrt{g} \left[\frac{dV}{d\phi} \delta \phi
+\frac{1}{2} g^{\mu\nu} \frac{\partial (\partial_\mu\phi \partial_\nu \phi) }{\partial \partial_\rho\phi} \delta ( \partial_\rho\phi) \right]
\nonumber \\
&=\int d^4 x \,\sqrt{g} \left[ \frac{dV}{d\phi}
+ g^{\mu\nu} (\partial_\mu \phi) \partial_\nu  \right] \delta \phi
\nonumber \\
&=\int d^4 x\, \left[ \sqrt{g}  \frac{dV}{d\phi}
-\partial_\nu ( \sqrt{g} g^{\mu\nu} (\partial_\mu \phi))  \right] \delta \phi + \text{surface term} \nonumber \\
&= \int d^4 x \, \sqrt{g}  \left[ \frac{dV}{d\phi}
-\frac{1}{ \sqrt{g}} \partial_\mu ( \sqrt{g} (\partial^\mu\phi))  \right] \delta \phi + \text{surface term},
\end{align}
which yields
\begin{equation}
 \label{eq:gh}
 \frac{1}{ \sqrt{g}} \partial_\mu ( \sqrt{g} \partial^\mu\phi) = \frac{dV}{d\phi}.
\end{equation}
To find the bounce we thus have to solve the two coupled equations
\eq{eq:einst} and \eq{eq:gh} for the field $\phi$ and the metric $g$.
The solution, once replaced in \eq{eq:einsteinhilbert}, gives the value of the
bounce action, relevant for the calculation of the tunnelling rate.

\bibliographystyle{utphys}
\bibliography{bibliography}

\providecommand{\href}[2]{#2}\begingroup\raggedright\begin{thebibliography}{10}

\bibitem{Coleman:1977py}
S.~R. Coleman, ``{The Fate of the False Vacuum. 1. Semiclassical Theory},''
\href{http://dx.doi.org/10.1103/PhysRevD.15.2929,
  10.1103/PhysRevD.16.1248}{{\em Phys.Rev.} {\bfseries D15} (1977) 2929--2936}.

\bibitem{Callan:1977pt}
J.~Callan, Curtis~G. and S.~R. Coleman, ``{The Fate of the False Vacuum. 2.
  First Quantum Corrections},''
\href{http://dx.doi.org/10.1103/PhysRevD.16.1762}{{\em Phys.Rev.} {\bfseries
  D16} (1977) 1762--1768}.

\bibitem{Isidori:2001bm}
G.~Isidori, G.~Ridolfi, and A.~Strumia, ``{On the metastability of the standard
  model vacuum},'' \href{http://dx.doi.org/10.1016/S0550-3213(01)00302-9}{{\em
  Nucl.Phys.} {\bfseries B609} (2001) 387--409},
\href{http://arxiv.org/abs/hep-ph/0104016}{{\ttfamily arXiv:hep-ph/0104016
  [hep-ph]}}.

\bibitem{Branchina:2013jra}
V.~Branchina and E.~Messina, ``{Stability, Higgs Boson Mass and New Physics},''
  \href{http://dx.doi.org/10.1103/PhysRevLett.111.241801}{{\em Phys.Rev.Lett.}
  {\bfseries 111} (2013) 241801},
\href{http://arxiv.org/abs/1307.5193}{{\ttfamily arXiv:1307.5193 [hep-ph]}}.

\bibitem{Coleman:1980aw}
S.~R. Coleman and F.~De~Luccia, ``{Gravitational Effects on and of Vacuum
  Decay},''
\href{http://dx.doi.org/10.1103/PhysRevD.21.3305}{{\em Phys. Rev.} {\bfseries
  D21} (1980) 3305}.

\bibitem{Isidori:2007vm}
G.~Isidori, V.~S. Rychkov, A.~Strumia, and N.~Tetradis, ``{Gravitational
  corrections to standard model vacuum decay},''
  \href{http://dx.doi.org/10.1103/PhysRevD.77.025034}{{\em Phys. Rev.}
  {\bfseries D77} (2008) 025034},
\href{http://arxiv.org/abs/0712.0242}{{\ttfamily arXiv:0712.0242 [hep-ph]}}.

\bibitem{Weinberg:2012pjx}
E.~J. Weinberg, \href{http://dx.doi.org/10.1017/CBO9781139017787}{{\em
  {Classical solutions in quantum field theory}: {Solitons and Instantons in
  High Energy Physics}}}.
\newblock Cambridge Monographs on Mathematical Physics. Cambridge University
  Press, 9, 2012.

\bibitem{landau2013quantum}
L.~Landau and E.~Lifshitz, {\em Quantum Mechanics: Non-Relativistic Theory}.
\newblock Teoreticheskaia fizika. Elsevier Science, 2013.

\bibitem{Konishi:2009qva}
K.~Konishi and G.~Paffuti, {\em {Quantum mechanics}}.
\newblock Oxford Univ. Pr., Oxford, UK,
2009.
\newblock

\bibitem{Coleman:1985rnk}
S.~Coleman, \href{http://dx.doi.org/10.1017/CBO9780511565045}{{\em {Aspects of
  Symmetry}}}.
\newblock Cambridge University Press, Cambridge, U.K.,
1985.
\newblock

\bibitem{Weinberg:1996kr}
S.~Weinberg, {\em {The quantum theory of fields. Vol. 2: Modern applications}}.
\newblock Cambridge University Press,
2013.
\newblock

\bibitem{Banks:1973ps}
T.~Banks, C.~M. Bender, and T.~T. Wu, ``{Coupled anharmonic oscillators. 1.
  Equal mass case},''
\href{http://dx.doi.org/10.1103/PhysRevD.8.3346}{{\em Phys. Rev.} {\bfseries
  D8} (1973) 3346--3378}.

\bibitem{Banks:1974ij}
T.~Banks and C.~M. Bender, ``{Coupled anharmonic oscillators. ii. unequal-mass
  case},''
\href{http://dx.doi.org/10.1103/PhysRevD.8.3366}{{\em Phys. Rev.} {\bfseries
  D8} (1973) 3366--3378}.

\bibitem{Coleman:1977th}
S.~R. Coleman, V.~Glaser, and A.~Martin, ``{Action Minima Among Solutions to a
  Class of Euclidean Scalar Field Equations},''
\href{http://dx.doi.org/10.1007/BF01609421}{{\em Commun.Math.Phys.} {\bfseries
  58} (1978) 211}.

\bibitem{Ivanov:2022osf}
A.~Ivanov, M.~Matteini, M.~Nemevsek, and L.~Ubaldi, ``{Analytic thin wall false
  vacuum decay rate},'' \href{http://dx.doi.org/10.1007/JHEP03(2022)209}{{\em
  JHEP} {\bfseries 03} (2022) 209},
  \href{http://arxiv.org/abs/2202.04498}{{\ttfamily arXiv:2202.04498
  [hep-th]}}.

\bibitem{Lee:1985uv}
K.-M. Lee and E.~J. Weinberg, ``{Tunneling without barriers},''
\href{http://dx.doi.org/10.1016/0550-3213(86)90150-1}{{\em Nucl.Phys.}
  {\bfseries B267} (1986) 181}.

\bibitem{Kusenko:1996bv}
A.~Kusenko, K.-M. Lee, and E.~J. Weinberg, ``{Vacuum decay and internal
  symmetries},'' \href{http://dx.doi.org/10.1103/PhysRevD.55.4903}{{\em Phys.
  Rev. D} {\bfseries 55} (1997) 4903--4909},
  \href{http://arxiv.org/abs/hep-th/9609100}{{\ttfamily arXiv:hep-th/9609100}}.

\bibitem{Fubini:1976jm}
S.~Fubini, ``{A New Approach to Conformal Invariant Field Theories},''
\href{http://dx.doi.org/10.1007/BF02785664}{{\em Nuovo Cim.} {\bfseries A34}
  (1976) 521}.

\bibitem{Lipatov:1976ny}
L.~N. Lipatov, ``{Divergence of the Perturbation Theory Series and the
  Quasiclassical Theory},'' {\em Sov. Phys. JETP} {\bfseries 45} (1977)
  216--223.
[Zh. Eksp. Teor. Fiz.72,411(1977)].

\bibitem{DiLuzio:2015iua}
L.~Di~Luzio, G.~Isidori, and G.~Ridolfi, ``{Stability of the electroweak ground
  state in the Standard Model and its extensions},''
  \href{http://dx.doi.org/10.1016/j.physletb.2015.12.009}{{\em Phys. Lett. B}
  {\bfseries 753} (2016) 150--160},
  \href{http://arxiv.org/abs/1509.05028}{{\ttfamily arXiv:1509.05028
  [hep-ph]}}.

\bibitem{JonaLasinio:1964cw}
G.~Jona-Lasinio, ``{Relativistic field theories with symmetry breaking
  solutions},''
\href{http://dx.doi.org/10.1007/BF02750573}{{\em Nuovo Cim.} {\bfseries 34}
  (1964) 1790--1795}.

\bibitem{Coleman:1973jx}
S.~R. Coleman and E.~J. Weinberg, ``{Radiative Corrections as the Origin of
  Spontaneous Symmetry Breaking},''
\href{http://dx.doi.org/10.1103/PhysRevD.7.1888}{{\em Phys. Rev.} {\bfseries
  D7} (1973) 1888--1910}.

\bibitem{Jackiw:1974cv}
R.~Jackiw, ``{Functional evaluation of the effective potential},''
\href{http://dx.doi.org/10.1103/PhysRevD.9.1686}{{\em Phys. Rev.} {\bfseries
  D9} (1974) 1686}.

\bibitem{Martin:2017lqn}
S.~P. Martin, ``{Effective potential at three loops},''
  \href{http://dx.doi.org/10.1103/PhysRevD.96.096005}{{\em Phys. Rev.}
  {\bfseries D96} no.~9, (2017) 096005},
\href{http://arxiv.org/abs/1709.02397}{{\ttfamily arXiv:1709.02397 [hep-ph]}}.

\bibitem{Collins:1984xc}
J.~C. Collins, {\em {Renormalization}}, vol.~26 of {\em Cambridge Monographs on
  Mathematical Physics}.
\newblock Cambridge University Press, Cambridge,
1986.
\newblock

\bibitem{Peskin:1995ev}
M.~E. Peskin and D.~V. Schroeder, {\em {An Introduction to quantum field
  theory}}.
\newblock Addison-Wesley, Reading, USA,
1995.
\newblock

\bibitem{Bardeen:1978yd}
W.~A. Bardeen, A.~J. Buras, D.~W. Duke, and T.~Muta, ``{Deep Inelastic
  Scattering Beyond the Leading Order in Asymptotically Free Gauge Theories},''
\href{http://dx.doi.org/10.1103/PhysRevD.18.3998}{{\em Phys. Rev.} {\bfseries
  D18} (1978) 3998}.

\bibitem{Chetyrkin:2012rz}
K.~G. Chetyrkin and M.~F. Zoller, ``{Three-loop $\beta$-functions for
  top-Yukawa and the Higgs self-interaction in the Standard Model},''
  \href{http://dx.doi.org/10.1007/JHEP06(2012)033}{{\em JHEP} {\bfseries 06}
  (2012) 033},
\href{http://arxiv.org/abs/1205.2892}{{\ttfamily arXiv:1205.2892 [hep-ph]}}.

\bibitem{Mihaila:2012pz}
L.~N. Mihaila, J.~Salomon, and M.~Steinhauser, ``{Renormalization constants and
  beta functions for the gauge couplings of the Standard Model to three-loop
  order},'' \href{http://dx.doi.org/10.1103/PhysRevD.86.096008}{{\em Phys.
  Rev.} {\bfseries D86} (2012) 096008},
\href{http://arxiv.org/abs/1208.3357}{{\ttfamily arXiv:1208.3357 [hep-ph]}}.

\bibitem{Ford:1992mv}
C.~Ford, D.~R.~T. Jones, P.~W. Stephenson, and M.~B. Einhorn, ``{The Effective
  potential and the renormalization group},''
  \href{http://dx.doi.org/10.1016/0550-3213(93)90206-5}{{\em Nucl. Phys.}
  {\bfseries B395} (1993) 17--34},
\href{http://arxiv.org/abs/hep-lat/9210033}{{\ttfamily arXiv:hep-lat/9210033
  [hep-lat]}}.

\bibitem{Buttazzo:2013uya}
D.~Buttazzo, G.~Degrassi, P.~P. Giardino, G.~F. Giudice, F.~Sala, A.~Salvio,
  and A.~Strumia, ``{Investigating the near-criticality of the Higgs boson},''
  \href{http://dx.doi.org/10.1007/JHEP12(2013)089}{{\em JHEP} {\bfseries 12}
  (2013) 089},
\href{http://arxiv.org/abs/1307.3536}{{\ttfamily arXiv:1307.3536 [hep-ph]}}.

\bibitem{Nielsen:1975fs}
N.~K. Nielsen, ``{On the Gauge Dependence of Spontaneous Symmetry Breaking in
  Gauge Theories},''
\href{http://dx.doi.org/10.1016/0550-3213(75)90301-6}{{\em Nucl. Phys.}
  {\bfseries B101} (1975) 173--188}.

\bibitem{Metaxas:1995ab}
D.~Metaxas and E.~J. Weinberg, ``{Gauge independence of the bubble nucleation
  rate in theories with radiative symmetry breaking},''
  \href{http://dx.doi.org/10.1103/PhysRevD.53.836}{{\em Phys. Rev.} {\bfseries
  D53} (1996) 836--843},
\href{http://arxiv.org/abs/hep-ph/9507381}{{\ttfamily arXiv:hep-ph/9507381
  [hep-ph]}}.

\bibitem{DiLuzio:2014bua}
L.~Di~Luzio and L.~Mihaila, ``{On the gauge dependence of the Standard Model
  vacuum instability scale},''
  \href{http://dx.doi.org/10.1007/JHEP06(2014)079}{{\em JHEP} {\bfseries 06}
  (2014) 079},
\href{http://arxiv.org/abs/1404.7450}{{\ttfamily arXiv:1404.7450 [hep-ph]}}.

\bibitem{Andreassen:2014eha}
A.~Andreassen, W.~Frost, and M.~D. Schwartz, ``{Consistent Use of Effective
  Potentials},'' \href{http://dx.doi.org/10.1103/PhysRevD.91.016009}{{\em Phys.
  Rev.} {\bfseries D91} no.~1, (2015) 016009},
\href{http://arxiv.org/abs/1408.0287}{{\ttfamily arXiv:1408.0287 [hep-ph]}}.

\bibitem{Andreassen:2014gha}
A.~Andreassen, W.~Frost, and M.~D. Schwartz, ``{Consistent Use of the Standard
  Model Effective Potential},''
  \href{http://dx.doi.org/10.1103/PhysRevLett.113.241801}{{\em Phys.Rev.Lett.}
  {\bfseries 113} no.~24, (2014) 241801},
\href{http://arxiv.org/abs/1408.0292}{{\ttfamily arXiv:1408.0292 [hep-ph]}}.

\bibitem{Espinosa:2016nld}
J.~R. Espinosa, M.~Garny, T.~Konstandin, and A.~Riotto, ``{Gauge-Independent
  Scales Related to the Standard Model Vacuum Instability},''
  \href{http://dx.doi.org/10.1103/PhysRevD.95.056004}{{\em Phys. Rev.}
  {\bfseries D95} no.~5, (2017) 056004},
\href{http://arxiv.org/abs/1608.06765}{{\ttfamily arXiv:1608.06765 [hep-ph]}}.

\bibitem{Fujimoto:1982tc}
Y.~Fujimoto, L.~O'Raifeartaigh, and G.~Parravicini, ``{Effective Potential for
  Nonconvex Potentials},''
\href{http://dx.doi.org/10.1016/0550-3213(83)90305-X}{{\em Nucl. Phys.}
  {\bfseries B212} (1983) 268--300}.

\bibitem{Dannenberg:1987fw}
A.~Dannenberg, ``{Dysfunctional Methods and the Effective Potential},''
\href{http://dx.doi.org/10.1016/0370-2693(88)90862-3}{{\em Phys. Lett.}
  {\bfseries B202} (1988) 110--116}.

\bibitem{Sher:1988mj}
M.~Sher, ``{Electroweak Higgs Potentials and Vacuum Stability},''
\href{http://dx.doi.org/10.1016/0370-1573(89)90061-6}{{\em Phys. Rept.}
  {\bfseries 179} (1989) 273--418}.

\bibitem{Brandenberger:1984cz}
R.~H. Brandenberger, ``{Quantum Field Theory Methods and Inflationary Universe
  Models},''
\href{http://dx.doi.org/10.1103/RevModPhys.57.1}{{\em Rev. Mod. Phys.}
  {\bfseries 57} (1985) 1}.

\bibitem{Weinberg:1987vp}
E.~J. Weinberg and A.-q. Wu, ``{Understanding Complex Perturbative Effective
  Potentials},''
\href{http://dx.doi.org/10.1103/PhysRevD.36.2474}{{\em Phys. Rev.} {\bfseries
  D36} (1987) 2474}.

\bibitem{Krive:1976sg}
I.~V. Krive and A.~D. Linde, ``{On the Vacuum Stability Problem in Gauge
  Theories},''
\href{http://dx.doi.org/10.1016/0550-3213(76)90573-3}{{\em Nucl. Phys.}
  {\bfseries B117} (1976) 265--268}.

\bibitem{Krasnikov:1978pu}
N.~V. Krasnikov, ``{Restriction of the Fermion Mass in Gauge Theories of Weak
  and Electromagnetic Interactions},''
{\em Yad. Fiz.} {\bfseries 28} (1978) 549--551.

\bibitem{Maiani:1977cg}
L.~Maiani, G.~Parisi, and R.~Petronzio, ``{Bounds on the Number and Masses of
  Quarks and Leptons},''
\href{http://dx.doi.org/10.1016/0550-3213(78)90018-4}{{\em Nucl. Phys.}
  {\bfseries B136} (1978) 115--124}.

\bibitem{Politzer:1978ic}
H.~D. Politzer and S.~Wolfram, ``{Bounds on Particle Masses in the
  Weinberg-Salam Model},''
  \href{http://dx.doi.org/10.1016/0370-2693(79)90746-9}{{\em Phys. Lett.}
  {\bfseries 82B} (1979) 242--246}.
[Erratum: Phys. Lett.83B,421(1979)].

\bibitem{Hung:1979dn}
P.~Q. Hung, ``{Vacuum Instability and New Constraints on Fermion Masses},''
\href{http://dx.doi.org/10.1103/PhysRevLett.42.873}{{\em Phys. Rev. Lett.}
  {\bfseries 42} (1979) 873}.

\bibitem{Cabibbo:1979ay}
N.~Cabibbo, L.~Maiani, G.~Parisi, and R.~Petronzio, ``{Bounds on the Fermions
  and Higgs Boson Masses in Grand Unified Theories},''
\href{http://dx.doi.org/10.1016/0550-3213(79)90167-6}{{\em Nucl. Phys.}
  {\bfseries B158} (1979) 295--305}.

\bibitem{Linde:1979ny}
A.~D. Linde, ``{Vacuum Instability, Cosmology and Constraints on Particle
  Masses in the {Weinberg-Salam} Model},''
\href{http://dx.doi.org/10.1016/0370-2693(80)90318-4}{{\em Phys. Lett.}
  {\bfseries 92B} (1980) 119--121}.

\bibitem{Lindner:1988ww}
M.~Lindner, M.~Sher, and H.~W. Zaglauer, ``{Probing Vacuum Stability Bounds at
  the Fermilab Collider},''
\href{http://dx.doi.org/10.1016/0370-2693(89)90540-6}{{\em Phys. Lett.}
  {\bfseries B228} (1989) 139--143}.

\bibitem{Arnold:1989cb}
P.~B. Arnold, ``{Can the Electroweak Vacuum Be Unstable?},''
\href{http://dx.doi.org/10.1103/PhysRevD.40.613}{{\em Phys. Rev.} {\bfseries
  D40} (1989) 613}.

\bibitem{Arnold:1991cv}
P.~B. Arnold and S.~Vokos, ``{Instability of hot electroweak theory: bounds on
  m(H) and M(t)},''
\href{http://dx.doi.org/10.1103/PhysRevD.44.3620}{{\em Phys. Rev.} {\bfseries
  D44} (1991) 3620--3627}.

\bibitem{Aad:2012tfa}
{\bfseries ATLAS} Collaboration, G.~Aad {\em et~al.}, ``{Observation of a new
  particle in the search for the Standard Model Higgs boson with the ATLAS
  detector at the LHC},''
  \href{http://dx.doi.org/10.1016/j.physletb.2012.08.020}{{\em Phys. Lett.}
  {\bfseries B716} (2012) 1--29},
\href{http://arxiv.org/abs/1207.7214}{{\ttfamily arXiv:1207.7214 [hep-ex]}}.

\bibitem{Chatrchyan:2012xdj}
{\bfseries CMS} Collaboration, S.~Chatrchyan {\em et~al.}, ``{Observation of a
  new boson at a mass of 125 GeV with the CMS experiment at the LHC},''
  \href{http://dx.doi.org/10.1016/j.physletb.2012.08.021}{{\em Phys. Lett.}
  {\bfseries B716} (2012) 30--61},
\href{http://arxiv.org/abs/1207.7235}{{\ttfamily arXiv:1207.7235 [hep-ex]}}.

\bibitem{Sher:1993mf}
M.~Sher, ``{Precise vacuum stability bound in the standard model},''
  \href{http://dx.doi.org/10.1016/0370-2693(94)91078-2,
  10.1016/0370-2693(93)91586-C}{{\em Phys. Lett.} {\bfseries B317} (1993)
  159--163}, \href{http://arxiv.org/abs/hep-ph/9307342}{{\ttfamily
  arXiv:hep-ph/9307342 [hep-ph]}}.
[Addendum: Phys. Lett.B331,448(1994)].

\bibitem{Altarelli:1994rb}
G.~Altarelli and G.~Isidori, ``{Lower limit on the Higgs mass in the standard
  model: An Update},''
\href{http://dx.doi.org/10.1016/0370-2693(94)91458-3}{{\em Phys. Lett.}
  {\bfseries B337} (1994) 141--144}.

\bibitem{Casas:1994qy}
J.~A. Casas, J.~R. Espinosa, and M.~Quiros, ``{Improved Higgs mass stability
  bound in the standard model and implications for supersymmetry},''
  \href{http://dx.doi.org/10.1016/0370-2693(94)01404-Z}{{\em Phys. Lett.}
  {\bfseries B342} (1995) 171--179},
\href{http://arxiv.org/abs/hep-ph/9409458}{{\ttfamily arXiv:hep-ph/9409458
  [hep-ph]}}.

\bibitem{Espinosa:1995se}
J.~R. Espinosa and M.~Quiros, ``{Improved metastability bounds on the standard
  model Higgs mass},''
  \href{http://dx.doi.org/10.1016/0370-2693(95)00572-3}{{\em Phys. Lett.}
  {\bfseries B353} (1995) 257--266},
\href{http://arxiv.org/abs/hep-ph/9504241}{{\ttfamily arXiv:hep-ph/9504241
  [hep-ph]}}.

\bibitem{Espinosa:2007qp}
J.~R. Espinosa, G.~F. Giudice, and A.~Riotto, ``{Cosmological implications of
  the Higgs mass measurement},''
  \href{http://dx.doi.org/10.1088/1475-7516/2008/05/002}{{\em JCAP} {\bfseries
  0805} (2008) 002},
\href{http://arxiv.org/abs/0710.2484}{{\ttfamily arXiv:0710.2484 [hep-ph]}}.

\bibitem{Ellis:2009tp}
J.~Ellis, J.~R. Espinosa, G.~F. Giudice, A.~Hoecker, and A.~Riotto, ``{The
  Probable Fate of the Standard Model},''
  \href{http://dx.doi.org/10.1016/j.physletb.2009.07.054}{{\em Phys. Lett.}
  {\bfseries B679} (2009) 369--375},
\href{http://arxiv.org/abs/0906.0954}{{\ttfamily arXiv:0906.0954 [hep-ph]}}.

\bibitem{Bezrukov:2012sa}
F.~Bezrukov, M.~{\relax Yu}. Kalmykov, B.~A. Kniehl, and M.~Shaposhnikov,
  ``{Higgs Boson Mass and New Physics},''
  \href{http://dx.doi.org/10.1007/JHEP10(2012)140}{{\em JHEP} {\bfseries 10}
  (2012) 140}, \href{http://arxiv.org/abs/1205.2893}{{\ttfamily arXiv:1205.2893
  [hep-ph]}}.
[,275(2012)].

\bibitem{Degrassi:2012ry}
G.~Degrassi, S.~Di~Vita, J.~Elias-Miro, J.~R. Espinosa, G.~F. Giudice,
  G.~Isidori, and A.~Strumia, ``{Higgs mass and vacuum stability in the
  Standard Model at NNLO},''
  \href{http://dx.doi.org/10.1007/JHEP08(2012)098}{{\em JHEP} {\bfseries 08}
  (2012) 098},
\href{http://arxiv.org/abs/1205.6497}{{\ttfamily arXiv:1205.6497 [hep-ph]}}.

\bibitem{Masina:2012tz}
I.~Masina, ``{Higgs boson and top quark masses as tests of electroweak vacuum
  stability},'' \href{http://dx.doi.org/10.1103/PhysRevD.87.053001}{{\em Phys.
  Rev. D} {\bfseries 87} no.~5, (2013) 053001},
  \href{http://arxiv.org/abs/1209.0393}{{\ttfamily arXiv:1209.0393 [hep-ph]}}.

\bibitem{Bednyakov:2015sca}
A.~V. Bednyakov, B.~A. Kniehl, A.~F. Pikelner, and O.~L. Veretin, ``{Stability
  of the Electroweak Vacuum: Gauge Independence and Advanced Precision},''
  \href{http://dx.doi.org/10.1103/PhysRevLett.115.201802}{{\em Phys. Rev.
  Lett.} {\bfseries 115} no.~20, (2015) 201802},
\href{http://arxiv.org/abs/1507.08833}{{\ttfamily arXiv:1507.08833 [hep-ph]}}.

\bibitem{Iacobellis:2016eof}
G.~Iacobellis and I.~Masina, ``{Stationary configurations of the Standard Model
  Higgs potential: electroweak stability and rising inflection point},''
  \href{http://dx.doi.org/10.1103/PhysRevD.94.073005}{{\em Phys. Rev. D}
  {\bfseries 94} no.~7, (2016) 073005},
  \href{http://arxiv.org/abs/1604.06046}{{\ttfamily arXiv:1604.06046
  [hep-ph]}}.

\bibitem{Andreassen:2017rzq}
A.~Andreassen, W.~Frost, and M.~D. Schwartz, ``{Scale Invariant Instantons and
  the Complete Lifetime of the Standard Model},''
  \href{http://dx.doi.org/10.1103/PhysRevD.97.056006}{{\em Phys. Rev.}
  {\bfseries D97} no.~5, (2018) 056006},
\href{http://arxiv.org/abs/1707.08124}{{\ttfamily arXiv:1707.08124 [hep-ph]}}.

\bibitem{Chigusa:2017dux}
S.~Chigusa, T.~Moroi, and Y.~Shoji, ``{State-of-the-Art Calculation of the
  Decay Rate of Electroweak Vacuum in the Standard Model},''
  \href{http://dx.doi.org/10.1103/PhysRevLett.119.211801}{{\em Phys. Rev.
  Lett.} {\bfseries 119} no.~21, (2017) 211801},
\href{http://arxiv.org/abs/1707.09301}{{\ttfamily arXiv:1707.09301 [hep-ph]}}.

\bibitem{Espinosa:2020qtq}
J.~R. Espinosa, ``{Vacuum Decay in the Standard Model: Analytical Results with
  Running and Gravity},''
  \href{http://dx.doi.org/10.1088/1475-7516/2020/06/052}{{\em JCAP} {\bfseries
  06} (2020) 052}, \href{http://arxiv.org/abs/2003.06219}{{\ttfamily
  arXiv:2003.06219 [hep-ph]}}.

\bibitem{Mihaila:2012fm}
L.~N. Mihaila, J.~Salomon, and M.~Steinhauser, ``{Gauge Coupling Beta Functions
  in the Standard Model to Three Loops},''
  \href{http://dx.doi.org/10.1103/PhysRevLett.108.151602}{{\em Phys. Rev.
  Lett.} {\bfseries 108} (2012) 151602},
\href{http://arxiv.org/abs/1201.5868}{{\ttfamily arXiv:1201.5868 [hep-ph]}}.

\bibitem{Chetyrkin:2013wya}
K.~G. Chetyrkin and M.~F. Zoller, ``{$\beta$-function for the Higgs
  self-interaction in the Standard Model at three-loop level},''
  \href{http://dx.doi.org/10.1007/JHEP04(2013)091,
  10.1007/JHEP09(2013)155}{{\em JHEP} {\bfseries 04} (2013) 091},
  \href{http://arxiv.org/abs/1303.2890}{{\ttfamily arXiv:1303.2890 [hep-ph]}}.
[Erratum: JHEP09,155(2013)].

\bibitem{Kniehl:2015nwa}
B.~A. Kniehl, A.~F. Pikelner, and O.~L. Veretin, ``{Two-loop electroweak
  threshold corrections in the Standard Model},''
  \href{http://dx.doi.org/10.1016/j.nuclphysb.2015.04.010}{{\em Nucl. Phys.}
  {\bfseries B896} (2015) 19--51},
\href{http://arxiv.org/abs/1503.02138}{{\ttfamily arXiv:1503.02138 [hep-ph]}}.

\bibitem{Martin:2013gka}
S.~P. Martin, ``{Three-loop Standard Model effective potential at leading order
  in strong and top Yukawa couplings},''
  \href{http://dx.doi.org/10.1103/PhysRevD.89.013003}{{\em Phys. Rev.}
  {\bfseries D89} no.~1, (2014) 013003},
\href{http://arxiv.org/abs/1310.7553}{{\ttfamily arXiv:1310.7553 [hep-ph]}}.

\bibitem{Martin:2015eia}
S.~P. Martin, ``{Four-Loop Standard Model Effective Potential at Leading Order
  in QCD},'' \href{http://dx.doi.org/10.1103/PhysRevD.92.054029}{{\em Phys.
  Rev.} {\bfseries D92} no.~5, (2015) 054029},
\href{http://arxiv.org/abs/1508.00912}{{\ttfamily arXiv:1508.00912 [hep-ph]}}.

\bibitem{Martin:2018emo}
S.~P. Martin and H.~H. Patel, ``{Two-loop effective potential for generalized
  gauge fixing},'' \href{http://dx.doi.org/10.1103/PhysRevD.98.076008}{{\em
  Phys. Rev.} {\bfseries D98} no.~7, (2018) 076008},
\href{http://arxiv.org/abs/1808.07615}{{\ttfamily arXiv:1808.07615 [hep-ph]}}.

\bibitem{Cline:2018ebc}
J.~M. Cline and J.~R. Espinosa, ``{Axionic landscape for Higgs coupling
  near-criticality},'' \href{http://dx.doi.org/10.1103/PhysRevD.97.035025}{{\em
  Phys. Rev.} {\bfseries D97} no.~3, (2018) 035025},
\href{http://arxiv.org/abs/1801.03926}{{\ttfamily arXiv:1801.03926 [hep-ph]}}.

\bibitem{Branchina:2014usa}
V.~Branchina, E.~Messina, and A.~Platania, ``{Top mass determination, Higgs
  inflation, and vacuum stability},''
  \href{http://dx.doi.org/10.1007/JHEP09(2014)182}{{\em JHEP} {\bfseries 09}
  (2014) 182},
\href{http://arxiv.org/abs/1407.4112}{{\ttfamily arXiv:1407.4112 [hep-ph]}}.

\bibitem{Branchina:2014rva}
V.~Branchina, E.~Messina, and M.~Sher, ``{Lifetime of the electroweak vacuum
  and sensitivity to Planck scale physics},''
  \href{http://dx.doi.org/10.1103/PhysRevD.91.013003}{{\em Phys.Rev.}
  {\bfseries D91} no.~1, (2015) 013003},
\href{http://arxiv.org/abs/1408.5302}{{\ttfamily arXiv:1408.5302 [hep-ph]}}.

\bibitem{Branchina:2015nda}
V.~Branchina and E.~Messina, ``{Stability and UV completion of the Standard
  Model},'' \href{http://dx.doi.org/10.1209/0295-5075/117/61002}{{\em EPL}
  {\bfseries 117} no.~6, (2017) 61002},
\href{http://arxiv.org/abs/1507.08812}{{\ttfamily arXiv:1507.08812 [hep-ph]}}.

\bibitem{Andreassen:2016cvx}
A.~Andreassen, D.~Farhi, W.~Frost, and M.~D. Schwartz, ``{Precision decay rate
  calculations in quantum field theory},''
  \href{http://dx.doi.org/10.1103/PhysRevD.95.085011}{{\em Phys. Rev.}
  {\bfseries D95} no.~8, (2017) 085011},
\href{http://arxiv.org/abs/1604.06090}{{\ttfamily arXiv:1604.06090 [hep-th]}}.

\bibitem{Patel:2017aig}
H.~H. Patel and B.~Radovcic, ``{On the Decoupling Theorem for Vacuum
  Metastability},''
  \href{http://dx.doi.org/10.1016/j.physletb.2017.08.075}{{\em Phys. Lett.}
  {\bfseries B773} (2017) 527--533},
\href{http://arxiv.org/abs/1704.00775}{{\ttfamily arXiv:1704.00775 [hep-ph]}}.

\bibitem{Branchina:2018xdh}
V.~Branchina, F.~Contino, and A.~Pilaftsis, ``{Protecting the stability of the
  electroweak vacuum from Planck-scale gravitational effects},''
  \href{http://dx.doi.org/10.1103/PhysRevD.98.075001}{{\em Phys. Rev. D}
  {\bfseries 98} no.~7, (2018) 075001},
  \href{http://arxiv.org/abs/1806.11059}{{\ttfamily arXiv:1806.11059
  [hep-ph]}}.

\bibitem{Ghosh:2021lua}
J.~K. Ghosh, E.~Kiritsis, F.~Nitti, and L.~T. Witkowski, ``{Revisiting
  Coleman-de Luccia transitions in the AdS regime using holography},''
  \href{http://dx.doi.org/10.1007/JHEP09(2021)065}{{\em JHEP} {\bfseries 09}
  (2021) 065}, \href{http://arxiv.org/abs/2102.11881}{{\ttfamily
  arXiv:2102.11881 [hep-th]}}.

\bibitem{Espinosa:2021tgx}
J.~R. Espinosa, J.~F. Fortin, and J.~Huertas, ``{Exactly solvable vacuum decays
  with gravity},'' \href{http://dx.doi.org/10.1103/PhysRevD.104.065007}{{\em
  Phys. Rev. D} {\bfseries 104} no.~6, (2021) 065007},
  \href{http://arxiv.org/abs/2106.15505}{{\ttfamily arXiv:2106.15505
  [hep-th]}}.

\bibitem{Rajantie:2016hkj}
A.~Rajantie and S.~Stopyra, ``{Standard Model vacuum decay with gravity},''
  \href{http://dx.doi.org/10.1103/PhysRevD.95.025008}{{\em Phys. Rev. D}
  {\bfseries 95} no.~2, (2017) 025008},
  \href{http://arxiv.org/abs/1606.00849}{{\ttfamily arXiv:1606.00849
  [hep-th]}}.

\bibitem{Salvio:2016mvj}
A.~Salvio, A.~Strumia, N.~Tetradis, and A.~Urbano, ``{On gravitational and
  thermal corrections to vacuum decay},''
  \href{http://dx.doi.org/10.1007/JHEP09(2016)054}{{\em JHEP} {\bfseries 09}
  (2016) 054},
\href{http://arxiv.org/abs/1608.02555}{{\ttfamily arXiv:1608.02555 [hep-ph]}}.

\bibitem{Branchina:2016bws}
V.~Branchina, E.~Messina, and D.~Zappala, ``{Impact of Gravity on Vacuum
  Stability},'' \href{http://dx.doi.org/10.1209/0295-5075/116/21001}{{\em EPL}
  {\bfseries 116} no.~2, (2016) 21001},
  \href{http://arxiv.org/abs/1601.06963}{{\ttfamily arXiv:1601.06963
  [hep-ph]}}.

\bibitem{Bentivegna:2017qry}
E.~Bentivegna, V.~Branchina, F.~Contino, and D.~Zappal\`a, ``{Impact of New
  Physics on the EW vacuum stability in a curved spacetime background},''
  \href{http://dx.doi.org/10.1007/JHEP12(2017)100}{{\em JHEP} {\bfseries 12}
  (2017) 100}, \href{http://arxiv.org/abs/1708.01138}{{\ttfamily
  arXiv:1708.01138 [hep-ph]}}.

\bibitem{Franceschini:2022veh}
R.~Franceschini, A.~Strumia, and A.~Wulzer, ``{The collider landscape: which
  collider for establishing the SM instability?},''
  \href{http://arxiv.org/abs/2203.17197}{{\ttfamily arXiv:2203.17197
  [hep-ph]}}.

\bibitem{Espinosa:2015qea}
J.~R. Espinosa, G.~F. Giudice, E.~Morgante, A.~Riotto, L.~Senatore, A.~Strumia,
  and N.~Tetradis, ``{The cosmological Higgstory of the vacuum instability},''
  \href{http://dx.doi.org/10.1007/JHEP09(2015)174}{{\em JHEP} {\bfseries 09}
  (2015) 174}, \href{http://arxiv.org/abs/1505.04825}{{\ttfamily
  arXiv:1505.04825 [hep-ph]}}.

\bibitem{DelleRose:2015bpo}
L.~Delle~Rose, C.~Marzo, and A.~Urbano, ``{On the fate of the Standard Model at
  finite temperature},'' \href{http://dx.doi.org/10.1007/JHEP05(2016)050}{{\em
  JHEP} {\bfseries 05} (2016) 050},
  \href{http://arxiv.org/abs/1507.06912}{{\ttfamily arXiv:1507.06912
  [hep-ph]}}.

\bibitem{griffiths2018introduction}
D.~Griffiths and D.~Schroeter, {\em Introduction to Quantum Mechanics}.
\newblock Cambridge University Press, 2018.

\bibitem{Rajantie}
A.~Rajantie and S.~Stopyra, ``{Standard Model vacuum decay with gravity},''
  \href{http://dx.doi.org/10.1103/PhysRevD.95.025008}{{\em Phys. Rev.}
  {\bfseries D95} no.~2, (2017) 025008},
\href{http://arxiv.org/abs/1606.00849}{{\ttfamily arXiv:1606.00849 [hep-th]}}.

\bibitem{Branchina:2019tyy}
V.~Branchina, E.~Bentivegna, F.~Contino, and D.~Zappal\`a, ``{Direct
  Higgs-gravity interaction and stability of our Universe},''
  \href{http://dx.doi.org/10.1103/PhysRevD.99.096029}{{\em Phys. Rev. D}
  {\bfseries 99} no.~9, (2019) 096029},
  \href{http://arxiv.org/abs/1905.02975}{{\ttfamily arXiv:1905.02975
  [hep-ph]}}.

\end{thebibliography}\endgroup

\end{document}